\documentclass[12pt,fleqn]{article}
\usepackage {verbatim}
\usepackage{amsmath,amssymb,eucal,amsfonts,amsthm,enumerate}

\newcommand{\ra}{\rightarrow}
\newcommand{\bra}{\langle} \newcommand{\ket}{\rangle}
\newcommand{\be}{\begin{equation}}
\newcommand{\ee}{\end{equation}}
\newcommand{\bea}{\begin{eqnarray}}
\newcommand{\eea}{\end{eqnarray}}
\newcommand{\eps}{\epsilon}
\newcommand{\E}{\mbox{e}}

\newcommand{\tfe}{\widetilde{\varphi}}
\newcommand{\ffi}{\varphi}

\newcommand{\ep}{\qquad {\vrule height 10pt width 8pt depth 0pt}}

\newcommand{\grintl}{[\kern-.18em [}
\newcommand{\grintr}{]\kern-.18em ]}

\newcommand{\dist}{\mbox{dist\,}}
\newcommand{\hil}{{\mathcal H}}
\newtheorem{lem}{Lemma}[section]
\newtheorem{prop}{Proposition}[section]
\newtheorem{thm}{Theorem}[section]

\newtheorem{Remark}{Remark}[section]
\usepackage{amsfonts}

\def\un{\hbox{$\mit I$\kern-.77em$\mit I$}}
\def\0{\hbox{$\mit I$\kern-.70em$\mit O$}}
\def\R{{\mathbb R}}
\def\T{{\mathbb T}}
\def\Z{{\mathbb Z}}
\def\N{{\mathbb N}}
\def\C{{\mathbb C}}
\def\E{{\mathbb E}}
\def\P{{\mathbb P}}

\begin{document}

\title{Dynamical Localization for Unitary Anderson Models}
\author{Eman Hamza\footnote{partially supported through MSU New Faculty Grant 07-IRGP-1192.}\\
Michigan State University\\
Department of Mathematics\\ East Lansing, MI 48823\\ U.S.A.
\and
Alain Joye\\
Institut Fourier
\\ Universit\'e de Grenoble, BP 74
\\38402 Saint-Martin d'H\`eres\\
France\\
 \and
G\"unter Stolz\footnote{partially supported through US-NSF grant DMS-0653374}\\
University of Alabama at Birmingham\\
Department of Mathematics
 CH 452\\
Birmingham, AL 35294-1170\\ U.S.A.
}
\date{}
\maketitle

\abstract{
  This paper establishes dynamical localization properties of certain families of unitary random operators on the $d$-dimensional lattice in various regimes. These operators are generalizations of one-dimensional physical models of quantum transport and draw their name from the analogy with the discrete Anderson model of solid state physics.  They consist in a product of a deterministic unitary operator and  a random unitary operator. The deterministic operator has a band structure, is absolutely continuous and plays the role of the discrete Laplacian. The random operator is diagonal with elements given by i.i.d. random phases distributed according to some absolutely continuous measure and plays the role of the random potential. In dimension one, these operators belong to the family of CMV-matrices in the theory of orthogonal polynomials on the unit circle.

We implement the method of Aizenman-Molchanov to prove exponential decay of the  fractional moments of the Green function for the unitary Anderson model in the following three regimes: In any dimension,  throughout the spectrum at large disorder and near the band edges at arbitrary disorder and, in dimension one, throughout the spectrum at arbitrary disorder. We also prove that exponential decay of fractional moments of the Green function implies dynamical localization, which in turn implies spectral localization.

These results complete the analogy with the self-adjoint case where dynamical localization is known to be true in the same three regimes.

}


\setcounter{equation}{0}
\section{Introduction}

The spectral theory of Schr\"odinger operators and other selfadjoint operators $H$ used to model hamiltonians of quantum mechanical systems has a long history. It can be argued that on physical grounds the main motivation for studying spectral properties is their close connection (e.g.\ via the RAGE-Theorem) with dynamical properties of the corresponding time evolution $e^{-iHt}$, i.e.\ the propagation of wave packets under the time-dependent Schr\"odinger equation $i\psi'(t)=H\psi(t)$.

However, the dynamical information following from spectral properties is not very accurate and examples have been found where spectral properties are misleading about the dynamics. In particular, this is the case for operators with singular continuous spectrum or dense point spectrum, spectral types quite common in quantum mechanical models of disordered media such as quasiperiodic or random Schr\"odinger operators, see for instance \cite{gua0}, \cite{simon90}, \cite{dRJLS}.

As a consequence, much of the recent work on hamiltonians governing disordered systems has focused on directly establishing dynamical properties. For example, it has been shown that Anderson-type random hamiltonians exhibit {\it dynamical localization} in various energy regimes, the property that wave packets $\psi(t)$ stay localized in space for all times, see \cite{ks}, \cite{A}, \cite{AG}, \cite{gdb}, \cite{ds}, \cite{gk}, for example.

The central object of interest in understanding dynamics is the unitary group $U(t)=e^{-iHt}$, rather than the hamiltonian $H$ itself. As short time fluctuations will generally not have a major impact on long time dynamics, one may discretize time by choosing a time unit $T>0$ and study $U(nT)=U^n=e^{-inTH}$ as $n\to\infty$, with the fixed unitary propagator $U=U(T)$.

To further stress the role of the propagator as the central object in studies of dynamics, consider hamiltonians $H(t)$ which depend periodically on time, $H(t+T)=H(t)$ for some $T>0$ and all $t$. In this case the large time behavior of solutions of $i\psi'(t)=H(t)\psi(t)$ is governed by the unitary propagator $U(nT,0) =U^n$, where $U=U(T,0)$ is commonly referred to as the monodromy operator of the time-periodic system. Note that in this case $U$ does not have a meaningful representation of the form $e^{iA}$ any more. While such representations with selfadjoint operators $A$ exist for abstract reasons, the operator $A$ may have little to do with the time-dependent hamiltonian $H(t)$. Actually, in the periodic or quasi-periodic cases, the operator $A$ is linked to the so called quasienergy or Floquet operator \cite{h}, \cite{y}, \cite{bel0}, \cite{jl}.

One consequence of this last fact is that it becomes legitimate to study time periodic systems by directly modeling the monodromy operator $U$ based on physical properties.
For example, time dependent analyses of electronic transport in disordered metallic rings have been considered within such a framework, \cite{lv}, \cite{ao}, \cite{abdn}, \cite{ade}, \cite{bb}. Kicked systems, often used in the study of quantum chaos, provide another example of such models, see e.g.\  \cite{cfgv}, \cite{c}, \cite{c2}, \cite{dbf}, \cite{dO}, \cite{b}, \cite{mccawetal}, \cite{rhk}. Similar studies were performed on the dynamical properties of pulsed systems, given by smooth Floquet operators, in  \cite{bel}, \cite{how},  \cite{jl}, \cite{duclosetal}.

Our central goal here is to investigate the dynamics of one such model $U_{\omega}$, which we call the {\it unitary Anderson model}, indicating the presence of disorder by the random parameter $\omega$. The name is chosen by analogy to the selfadjoint Anderson model, which, in its discrete version on $\ell^2(\Z^d)$, takes the form
\be \label{eq:saAnderson}
h_{\omega}= h_0 +V_{\omega}.
\ee
The potential in (\ref{eq:saAnderson}) is given by real-valued i.i.d.\ random variables $\{V_{\omega}(k)\}_{k\in\Z^d}$ and $h_0$ is a deterministic selfadjoint operator, most commonly the discrete Laplacian on $\Z^d$. By comparison, following our previous works \cite{J1, J2, HJS}, for the unitary Anderson model we choose a unitary operator on $\ell^2(\Z^d)$ of the form
\be \label{eq:uniAnderson}
U_{\omega} = D_{\omega} S.
\ee
Here $S$ is a deterministic unitary operator and $D_{\omega}$ a multiplication operator by random phases, i.e.\ for every $\phi \in \ell^2(\Z^d)$ and $k\in \Z^d$,
\be \label{eq:randomphases}
(D_{\omega} \phi)(k) = e^{-i\theta_k^{\omega}} \phi(k),
\ee
with i.i.d.\ random phases $\theta_k^{\omega}$ taking values in $\T:= \R/2\pi\Z$. Note that, despite the formal analogy, there is no simple relation between selfadjoint and unitary Anderson models. In particular, due to non-commutativity, $e^{-i(h_0+V_{\omega})}$ is not a unitary Anderson model and, vice versa, $D_{\omega}S$ can generally not be written in the form $e^{-i(h_0+V_{\omega})}$.

For $S$ we choose what we consider to be a unitary analog of the discrete Laplacian. For $d=1$ it has a five-diagonal structure which finds its roots in some physics models, see \cite{BHJ}. It turns out that $S$ has the same five-diagonal structure as the so-called CMV-matrices, which have recently found much interest in the theory of orthogonal polynomials on the unit circle, where they arise as unitary analogs of Jacobi matrices, see \cite{Simon1,Simon2, Simon4} and references therein. For $d>1$ we define $S$ as a $d$-fold tensor product of its one-dimensional version. For details see Section~\ref{sec:model}. One of the reasons for this choice of $S$ is that one can view the CMV-matrix structure as the simplest non-trivial band structure which a unitary operator on $\ell^2(\Z)$ can have, see e.g.\ \cite{BHJ}, similar to the role of the discrete Laplacian among selfadjoint band matrices. Moreover, we choose $S$ such that it will be invariant under translations by multiples of $2$, yielding ergodicity of the unitary random operator $U_{\omega}$.

Monodromy operators of the form (\ref{eq:uniAnderson}), though not necessarily incorporating Anderson-type randomness, have also been proposed and studied in the physics literature  \cite{lv}, \cite{ao},  \cite{bb} (see also \cite{rhk}). A mathematical investigation of these models was initiated in \cite{BHJ} and continued in \cite{J1, J2, HJS, HS, dOS}. In particular, spectral localization for the unitary Anderson model was established in \cite{HJS} for the one-dimensional model and in \cite{J2} for arbitrary dimension in the presence of large disorder. Here {\it spectral localization} refers to the property that $U_{\omega}$ has pure point spectrum for almost every $\omega$.

We will provide proofs of dynamical localization (as formally defined in Section~\ref{sec:results} below) for the unitary Anderson model in three different regimes: At arbitrary disorder and throughout the spectrum for the one-dimensional model as well as in the large disorder and band-edge regimes in arbitrary dimension (see Section~\ref{sec:results} for a detailed description of these regimes). This coincides with the regimes where localization has been found to hold for selfadjoint Anderson models.

Our approach to localization proofs will be via a unitary version of the fractional moment method, which was initiated as a tool in the theory of selfadjoint Anderson models by Aizenman and Molchanov in \cite{AM}. Dynamical localization will follow as a general consequence of exponential decay of spatial correlations in the fractional moments of Green's function (Section~\ref{sec:dynlocproof}). To complete the proof of dynamical localization in the three regimes described above, the latter property of Green's function will then be established in those regimes.

In fact, in the large disorder regime this has already been done in \cite{J2}. Its proof for the one-dimensional model is one of the main results of the thesis \cite{Hamza}, from where we borrow the proofs presented here (Section~\ref{sec:emanproof}). Some of our general results, in particular the proof that exponential decay of  fractional moments of Green's function implies dynamical localization (Section~\ref{sec:dynlocproof}) and the proof that fractional moments of Green's function are bounded (Section~\ref{sec:fmboundproof}), are also essentially taken from \cite{Hamza}.

The hardest, but possibly also most rewarding part of our work, is the proof of exponential decay of fractional moments of Green's function in the band edge regime, which is carried out in Sections~\ref{sec:Neumann} to \ref{sec:proofofbandedgelocalization}. Several preparatory sections are devoted to building up various mathematical tools which do not seem to be known in the context of unitary operators, such as the Feynman-Hellmann theorem from perturbation theory (Section~\ref{sec:FeymanHellmann}) and Combes-Thomas type bounds on eigenfunctions (Section~\ref{sec:CombesThomas}). Along the way to localization we establish the spectral theoretic precursor of Lifshits tails of the integrated density of states for the unitary Anderson model (Section~\ref{sec:lifshits}) as well as a decoupling procedure required in the iterative proof of exponential decay of fractional moments near the edges of the spectrum (Sections~\ref{sec:splitting} and \ref{sec:decoupling}).

In Section~\ref{dynspecproof} we also include a proof of the general fact that, in the context of the unitary Anderson model, dynamical localization implies spectral localization, as previously known for selfadjoint Anderson models. In the unitary context, this follows from a version of the RAGE-Theorem provided in \cite{EV}, whose proof we reproduce here.

As already mentioned, when $d=1$ and when restricted to $l^2(\N)$ our unitary Anderson matrices $U_\omega$ bear close resemblance with the CMV matrices in the theory of orthogonal polynomials on the unit circle, see \cite{J1}. These polynomials are determined by an infinite set of complex numbers on the unit disc that are  called Verblunsky coefficients. Actually, $U_\omega$ corresponds to a choice of Verblunsky coefficients characterized by constant moduli $r$ and correlated random phases, see \cite{HJS} for details. Other choices of random Verblunsky coefficients have been studied in the literature,  see e.g. \cite{Simon3}, \cite{Simon4}, \cite{stoiciu} and references therein. We note that for i.i.d.\ Verblunsky coefficients in the unit disc with rotation invariant distribution, Simon proves dynamical localization in \cite{Simon3}. While not spelled out explicitly, our results for the one dimensional case show that dynamical localization also holds for the CMV matrices considered in \cite{HJS} with constant moduli Verblunsky coefficients and correlated phases.

Finally, there is an underlying pedagogical goal to our paper: We use the unitary models considered here to give a self-contained presentation of the mathematical theory of Anderson localization via the fractional moment approach. Making use of state-of-the-art techniques from localization theory, we revisit the peculiarities of the one-dimensional case and techniques covering various regimes in the multi-dimensional case within a unitary framework. This requires developing and adapting all necessary background, which we do in a widely self-contained fashion.

\medskip

{\bf Acknowledgments:} E.\ H.\ would like to acknowledge support
through a Junior Research Fellowship at the Erwin Schr\"odinger Institute in Vienna, where
part of this work was done. Also, E.\ H.\ and G.\ S.\ would like to express their gratitude for hospitality at the Institut Fourier
of Universit\'e de Grenoble during visits at crucial stages of this work.

\setcounter{equation}{0}
\section{The Unitary Anderson Model} \label{sec:model}

As the {\it unitary Anderson model} we denote a unitary random operator of the form
\be \label{eq:unAn}
U_{\omega} = D_{\omega} S
\ee
in $\ell^2(\Z^d)$. Motivated by earlier investigations in the physics literature, e.g.\ \cite{bb}, this model was studied mathematically in \cite{BHJ}, \cite{HJS}, \cite{J1}, \cite{J2} and \cite{HS}, from where we take the following definitions and basic results.

A deterministic unitary operator $S$ on $l^2(\Z^d)$, sometimes referred to as the ``free'' unitary operator or ``unitary Laplacian'', is constructed as follows:

Starting with $d=1$, let
$B_1$ and $B_2$ be unitary $2\times2$ matrices given by \be
\label{B1B2}
B_1 = \left( \begin{array}{cc} r & t \\
-t & r \end{array} \right) \quad \mbox{and} \quad B_2 = \left(
\begin{array}{cc} r & -t \\ t & r \end{array} \right), \ee with the
real parameters $t$ and $r$ linked by $r^2+t^2=1$ to ensure
unitarity. Now let $U_e$ be the unitary matrix operator in $l^2(\Z)$ found
as the direct sum of identical $B_1$-blocks with blocks starting at
even indices. Similarly, construct $U_o$ with identical
$B_2$-blocks, where blocks start at odd indices. Define $S_0 = U_e U_o$, which will serve as the operator $S$ in (\ref{eq:unAn}) for dimension $d=1$. The operator $S_0$ is unitary,
with band structure
\be\label{s0}
S_0 = \left( \begin{array}{cccccc}\ddots & rt & -t^2& & & \\
              & r^2 & -rt  & & & \\
              & rt & r^2 & rt & -t^2 & \\
              & -t^2 &-tr & r^2& -rt & \\
              & & & rt &r^2 & \\
              & & & -t^2 & -tr & \ddots \end{array} \right),
\ee
where the position of the origin in $\Z$ is fixed by $\bra
e_{2k-2}|S e_{2k}\ket =-t^2$, with $e_k$ ($k\in\Z$) denoting the
canonical basis vectors in $l^2(\Z)$.  Note also that $S_0$ is invariant under translations by multiples of $2$. Due to elementary unitary equivalences it will suffice to consider $0\leq t,r \leq 1$. Thus $S_0$ is determined by
$t$. We shall sometimes write $S_0(t)$ to emphasize this parameter dependence.
The spectrum of $S_0(t)$ is given by the arc
\be
\sigma(S_0(t))=\Sigma(t)= \{e^{i\vartheta}: \vartheta \in
[-\lambda_0, \lambda_0]\},
\ee
with $\lambda_0 :=  \arccos(r^2-t^2)$. The spectrum is symmetric about
the real axis and grows from the single point $\{1\}$ for $t=0$ to
the entire unit circle for $t=1$. The spectrum is purely absolutely continuous for $t>0$.

To define the multidimensional unitary Laplacian, we follow \cite{J2} in viewing $l^2(\Z^d)$ as $\otimes_{j=1}^d
l^2(\Z)$ so that for all $k\in\Z^d$, $e_k\simeq
e_{k_1}\otimes...\otimes e_{k_d}$. Using $S_0=S_0(t)$ from above we define $S=S(t)$ by
\be
\label{highdimS}
S(t) =\otimes_{j=1}^d S_0(t).
\ee
The spectrum of $S(t)$ is
\be
\sigma(S(t)) = \{e^{i \vartheta}: \vartheta \in [-d\lambda_0,d\lambda_0]\}.
\ee

Throughout this paper $|\cdot|$ will denote the maximum norm on $\Z^d$. Using this norm it is easy to see that $S(t)$ inherits
the band structure of $S_0$ such that
\be
\langle e_k| S(t)
e_l\rangle=0 \qquad \mbox{if } |k-l|>2.
\ee

For the definition of the random phase matrix $D_{\omega}$ in (\ref{eq:unAn}),
introduce the probability space
$(\Omega,\mathcal{F},\mathbb{P})$, where $\Omega =
{\mathbb{T}^{\Z^d}}$ ($\T = \R/2\pi\Z$), $\mathcal{F}$ is the
$\sigma$-algebra generated by cylinders of Borel sets, and
$\mathbb{P}=\bigotimes_{k\in\Z^d}\mu$, where $\mu$ is a non trivial
probability measure on $\mathbb{T}$. The expectation with respect to $\mathbb{P}$ will be denoted by $\E$. We will assume throughout that
$\mu$ is absolutely continuous with bounded density,
\be
\label{eq:mudensity}
d\mu(\theta) = \tau(\theta)\,d\theta, \quad
\tau \in L^{\infty}(\T).
\ee
The random variables $\theta_k$ on
$(\Omega,\mathcal{F},\mathbb{P})$ are defined by
\be
\label{highphases} \theta_k: \Omega \rightarrow \mathbb{T}, \ \
 \ \theta_k^\omega=\omega_{k}, \ \ \ k\in \Z^d.
\ee
In other words, the $\{\theta_k^{\omega}\}_{k\in\Z^d}$ are $\T$-valued i.i.d.\ random variables with common distribution $\mu$.

The diagonal operator $D_\omega$ in $\ell^2(\Z^d)$ is given by
\be \label{highdimeD}
D_\omega e_k=e^{-i\theta^\omega_k} e_k.
\ee

With this choice for $U_{\omega}$ and $S=S(t)$ we define the unitary Anderson model $U_\omega$ via (\ref{eq:unAn}).

This definition and the periodicity of $S$ ensures that the operator
$U_\omega$ is ergodic with
respect to the $2$-shift in $\Omega$. $U_\omega$ also inherits the band
structure of the original operator $S$. The general
theory of ergodic operators, as for example presented in Chapter V of
\cite{CL} for the self-adjoint case, carries over to
the unitary setting. In particular, it follows that the spectrum
of $U_\omega$ is almost surely deterministic, i.e.\ there is a
subset $\Sigma$ of the unit circle such that $\sigma(U_\omega) =
\Sigma$ for almost every $\omega$. The same holds for the
absolutely continuous, singular continuous and pure point parts of
the spectrum: There are $\Sigma_{ac}$, $\Sigma_{sc}$ and
$\Sigma_{pp}$ such that almost surely $\sigma_{ac}(U_{\omega}) =
\Sigma_{ac}$, $\sigma_{sc}(U_\omega) = \Sigma_{sc}$ and
$\sigma_{pp}(U_\omega) = \Sigma_{pp}$. Moreover, we can characterize $\Sigma$ in terms of the support of
$\mu$ and of the spectrum of $S$;
 \be \label{eq:asspectrum}
\Sigma =\exp{(-i\,\mbox{supp}\,\mu)}\,\sigma(S)=
\{e^{i\alpha}:\alpha\in [-d\lambda_0,d\lambda_0]+ \mbox{supp}\,\mu \}.
\ee
Here supp$\,\mu$ denotes
the support of the probability measure $\mu$, defined as
\be
\text{supp}\,\mu:=\{a\;|\;\mu((a-\epsilon,a+\epsilon))>0\text{ for all
}\epsilon>0\}.
\ee
The identity (\ref{eq:asspectrum}) is shown in \cite{J1} for the one-dimensional model, but the argument carries over to arbitrary dimension.

As $t \to0$, $S_0(t)$ tends to the identity
operator, whereas as $t\to 1$, $S_0(t)$ tends to a direct sum of
shift operators. Accordingly,  if $t$ is zero then the operators $U_{\omega}$ are
diagonal and thus trivially have pure point spectrum. On the other
hand, if $t=1$  then it is not hard to see that all
$U_{\omega}$ are purely absolutely continuous (in fact, they are unitarily equivalent to a direct sum of shift operators). Thus, excluding the trivial
special cases, we shall from now on assume that $0<t<1$. For the unitary Anderson model the parameter $t$ takes the role of a disorder parameter with small $t$ corresponding to large disorder, as this means that $U_{\omega}$ is dominated by its diagonal part.


\setcounter{equation}{0}
\section{The Results} \label{sec:results}

As discussed in the introduction, our main goal is to study regimes in which a quantum mechanical system governed by the unitary propagator $U_{\omega}$ is dynamically localized. This can be expressed in terms of the transition amplitudes $\langle e_k|U_{\omega}^n e_l\rangle$, whose squares measure the probability that a system initially in state $e_l$ evolves into state $e_k$ after time $n$. By {\it dynamical localization} we will refer to the property that the expectation of these amplitudes stays exponentially small in the distance of $k$ and $l$, uniformly for all times, i.e.\ the existence of constants $C<\infty$ and $\alpha>0$ such that
\be \label{eq:dl}
\E(\sup_{n\in\Z} |\langle e_k | U_{\omega}^n e_l \rangle|) \le Ce^{-\alpha|k-l|}.
\ee
In fact, what we will prove in several corresponding regimes is the stronger result that
\be \label{eq:dl2}
\E(\sup_f |\langle e_k | f(U_{\omega}) e_l \rangle|) \le Ce^{-\alpha|k-l|},
\ee
where the supremum is taken over all functions $f\in C({\mathbb S})$
with $\|f\|_{\infty}:=\sup_{z\in\mathbb S} |f(z)|\le 1$. Here ${\mathbb S} = \{z\in \C:|z|=1\}$ is the
unit circle.

Also, dynamical localization may only hold in an arc $\{e^{i\theta}: \theta\in [a,b]\}$ of the spectrum of $U_{\omega}$. In this case, the spectral projection $P_{[a,b]}^{\omega}$ of $U_{\omega}$ onto this arc will be applied to the state $e_l$ in (\ref{eq:dl2}), restricting the initial state to the localized part of the spectrum.

Our detailed results, stated in the following two subsections, fall into two categories: We start with results which show that dynamical localization can be established by a unitary version of the fractional moments method, using as a criterion the exponential decay of fractional moments of spatial correlations of the Green function. We also show that dynamical localization generally implies spectral localization, i.e.\ that $U_{\omega}$ almost surely has pure point spectrum in the corresponding part of the spectrum. Finally, we state that dynamical localization implies almost sure finiteness of all quantum moments of the position operator. All these results hold for
arbitrary dimension $d$, for general values of the disorder
parameter $t \in (0,1)$, and without restriction
on the spectral parameter of the unitary operators $U_{\omega}$.

Our second set of results concerns the proof of dynamical localization in three concrete regimes: for the one-dimensional unitary Anderson model, as well as for large disorder and at band edges in arbitrary dimensions. In each case this will be done by verifying exponential decay of the fractional moments of the Green function.

\subsection{Fractional Moment Criteria for Localization}

For $z\in \C$ with $|z|\not= 1$ let
\be \label{eq:resolvent}
G(z) =
G_{\omega}(z) = (U_{\omega}-z)^{-1},
\ee and
\be \label{eq:green}
G(k,l;z) = \langle e_k| G(z)e_l \rangle, \quad k,l \in \Z^d
\ee
be the Green function of $U_{\omega}$ (to use a term from the theory of selfadjoint hamiltonians in the unitary setting).

The Green function becomes singular as $z$ approaches the spectrum of $U_{\omega}$. The first insight which makes the fractional moment method a useful tool in localization theory is that these singularities are fractionally integrable with respect to the random parameters. This means that for $s\in (0,1)$ the fractional moments $\E(|G(k,l;z)|^s)$ have bounds which are uniform for $z$ arbitrarily close to the spectrum. This is the content of our first result. In fact, we will need a somewhat stronger result later, namely that it suffices to average over the random variables $\theta_k$ and $\theta_l$ to get uniform bounds on $|G(k,l;z)|^s$. The role of bounds of this type in localization proofs via the fractional moment method roughly corresponds to the use of {\it Wegner estimates} in the approach to localization by the method of multi scale analysis.

\begin{thm} \label{thm:fmbound}
Assume that the random variables $\{\theta_k\}_{k\in \Z^d}$ are
i.i.d.\ with distribution $\mu$ satisfying (\ref{eq:mudensity}).
Then for every $s\in (0,1)$ there exists $C(s) <\infty$ such that
\be \label{eq:fmbound}
\iint |G(k,l;z)|^s d\mu(\theta_k) d\mu(\theta_l) \le C(s)
\ee
for all $z\in
\C$, $|z|\not=1$, all $k, l \in \Z^d$, and arbitrary values of $\theta_j$, $j\not\in \{k,l\}$.
Consequently,
\be\label{eq:fmbound2}
\E(|G(k,l;z)|^s) \le C(s),
\ee
for all $z\in
\C$, $|z|\not=1$.
\end{thm}
The second statement simply derives from the bound (\ref{eq:fmbound}), which uniform in the random variables $\theta_j$, $j\not\in \{k,l\}$, and the independence of the $\theta_j$.
The proof of (\ref{eq:fmbound}) is given in
Section~\ref{sec:fmboundproof}.

In the next subsection we will identify several situations where the fractional moments $\E(|G(k,l;z)|^s)$ are not just uniformly bounded, but decay exponentially in the distance of $k$ and $l$. The following general result shows that this can be used as a criterion for dynamical
localization of $U_{\omega}$.

\begin{thm} \label{thm:dynamicallocalization}
Assume that the random variables $\{\theta_k\}_{k\in \Z^d}$ satisfy
(\ref{eq:mudensity}) and that for some $s\in (0,1)$, $C<\infty$,
$\alpha>0$, $\varepsilon >0$ and an interval $[a,b]\in \T$,
\be \label{eq:fmexpdecay}
\E(|G(k,l;z)|^s) \le Ce^{-\alpha|k-l|}
\ee
for all $k,l \in \Z^d$ and all $z\in \C$ such that $1-\varepsilon <
|z|<1$ and arg$\,z \in [a,b]$.

Then there exists $\tilde{C}$ such that \be
\label{eq:dynamicallocalization} \E[\sup_{\substack{f\in C({\mathbb
S})\\ \|f\|_{\infty}\le 1}} |\langle e_k|
f(U_{\omega}) P_{[a,b]}^{\omega} e_l \rangle|] \le \tilde{C}e^{-\alpha
|k-l|/4} \ee for all $k,l \in \Z^d$.
\end{thm}

Here, as usual arg$\,z \in \T$ refers to the polar representation $z=|z|\exp{(i\,\mbox{arg}\,z)}$ of a complex number.
We prove Theorem~\ref{thm:dynamicallocalization} in
Section~\ref{sec:dynlocproof}.

\vspace{.5cm}

It has been shown in \cite{J2} that fractional moment bounds of the form (\ref{eq:fmexpdecay}) for a unitary Anderson model imply spectral localization via a spectral averaging technique, following an approach to localization which is due to Simon and Wolff \cite{SW} for the selfadjoint Anderson model. For completeness, we present a direct proof of the fact that
dynamical localization expressed by (\ref{eq:dynamicallocalization})
implies spectral localization in the unitary setup. This follows
from a simple adaptation of arguments borrowed from Enss-Veselic
\cite{EV}, see also \cite{bjlpn}, on the geometric characterization of bound states, i.e.\ a RAGE-type theorem for unitary operators.
We prove

\begin{prop}\label{dynspec}
Assume that for an interval $[a,b]$ there exist constants $C<\infty$ and $\alpha>0$ such that
\be
 \E[\sup_{\substack{f\in C({\mathbb
S})\\ \|f\|_{\infty}\le 1}} |\langle e_k|
f(U_{\omega}) P^\omega_{[a,b]} e_l \rangle|] \le Ce^{-\alpha
|k-l|} \ee for all $k,l \in \Z^d$. Then
\be
(a,b)\cap\Sigma_{cont}=\emptyset,
\ee
where $\Sigma_{cont}=\Sigma_{ sc} \cup \Sigma_{ac}$. In other words, almost surely
$P_{[a,b]}^\omega U_\omega$ has pure point spectrum.
\end{prop}

Another direct consequence of the dynamical localization estimate
(\ref{eq:dynamicallocalization}), is that it prevents the spreading of
wave packets from $P_{[a,b]}^\omega l^2({\mathbb Z}^d)$ under the
discrete dynamics generated by $U_{\omega}$. This dynamical localization
property is measured in terms of the boundedness in time of all
quantum moments of the position operator on the lattice. More
precisely, for $p>0$ let $|X|_e^p$ be the maximal multiplication operator such that
\be
|X|_e^pe_j=|j|_e^p e_j, \ \ \ \mbox{for } j\in{\mathbb Z}^d,
\ee
where $|j|_e$ denotes the Euclidean norm on $\Z^d$. We have

\begin{prop}\label{physloc}
Assume that there exist $C<\infty$ and $\alpha>0$ such that
\be
 \E[\sup_{\substack{f\in C({\mathbb
S})\\ \|f\|_{\infty}\le 1}} |\langle e_k|
f(U_{\omega}) P^\omega_{[a,b]} e_l \rangle|] \le Ce^{-\alpha
|k-l|}
\ee
for all $k,l \in \Z^d$. Then, for any $p\geq 0$ and for any
$\psi$ in $l^2({\mathbb Z}^d)$ of compact support,
\be
\sup_{n\in {\mathbb Z}}\||X|_e^p U_\omega^nP_{[a,b]}^\omega\psi\|<\infty \ \ \mbox{almost surely.}
\ee
\end{prop}

Similar results hold under weaker support conditions on
$\psi$. Our choice is made to keep things simple.
The proofs of Propositions~\ref{dynspec} and \ref{physloc} are given in Section~\ref{dynspecproof}.

\subsection{Localization Regimes}

The other main goal of our work is to identify three different
regimes, where the fractional moment condition (\ref{eq:fmexpdecay})
can be verified, and thus dynamical localization follows by
Theorem~\ref{thm:dynamicallocalization}.

\vspace{.5cm}

\subsubsection{Large Disorder}

As explained at the end of Section~\ref{sec:model}, the parameter $t \in (0,1)$ used in the definition of $S(t)$ and thus, implicitly, also in the definition of $U_{\omega}$,
can be thought of as a measure of the degree of disorder in the
unitary Anderson model. Thus one can expect a
tendency toward localization for small values of $t$. This
was confirmed in \cite{J2} where the following was proven:

\begin{thm} \label{thm:alain}
Suppose that the i.i.d.\ random variables
$\{\theta_k\}_{k\in\Z^d}$ have distribution $\mu$
satisfying (\ref{eq:mudensity}) and let $s\in (0,1)$. Then there
exists $t_0>0$ and $C<\infty$ such that if $t
<t_0$, there exists $\alpha>0$ so that
\be
\label{eq:alain}
 \E(|\langle e_j| U_{\omega}(U_{\omega}-z)^{-1} e_k \rangle|^s) \le Ce^{-\alpha|j-k|}
\ee
for all $j,k\in \Z^d$ and all $z\in \C$ with $|z|\neq 1$.
\end{thm}

In fact, \cite{J2} considers a more general model in which different parameters $t_i$ are chosen in each factor of (\ref{highdimS}) and shows (\ref{eq:alain}) under the condition that $\sum_i t_i$ is sufficiently small.

Using (\ref{eq:greentomodified}) below this implies that
(\ref{eq:fmexpdecay}) holds for all $|z|\not= 1$. Thus in the large disorder regime $t<t_0$
dynamical localization holds on the entire spectrum of $U_{\omega}$ by Theorem~\ref{thm:dynamicallocalization}
(the spectral projection $P^\omega_{[a,b]}$ in
(\ref{eq:dynamicallocalization}) can be dropped).

\vspace{.5cm}

\subsubsection{The One-Dimensional Model}

For the one-dimensional
self-adjoint Anderson model, localization holds throughout the
spectrum, independent of the amount of disorder. The same is true
for the unitary Anderson model, as implied by the following result:

\begin{thm} \label{thm:eman}
Let $d=1$ and suppose that the i.i.d.\ random variables
$\{\theta_k\}_{k\in\Z}$ have distribution $\mu$
satisfying (\ref{eq:mudensity}).
Then for every $t<1$ there exist  $s>0$, $C<\infty$ and $\alpha>0$
such that
\be \label{eq:eman}
\E(|G(k,l;z)|^s) \le C e^{-\alpha|k-l|}
\ee
for all $z\in \C$ such that
$0<||z|-1|<1/2$ and all $k,l \in \Z^d$.
\end{thm}

By Theorem~\ref{thm:dynamicallocalization} this implies dynamical
localization for the one-dimensional unitary Anderson model throughout the spectrum.

\vspace{.3cm}

 Many of the special tools which have been heavily exploited in studies of the one-dimensional self-adjoint Anderson model, are also available for the one-dimensional Unitary model. First of all, there is a transfer-matrix formalism which allows the definition of Lyapunov exponents. In particular, it has been shown in \cite{HS} that under assumption (\ref{eq:mudensity}) (in fact, for much more general distributions) the Lyapunov exponent is positive on the entire spectrum and continuous in the spectral parameter. This is the central ingredient into Lemma~\ref{lemma:ckm} stated in Section~\ref{sec:emanproof} below. For a proof of Lemma~\ref{lemma:ckm}, which is a close analog of a result proven for the selfadjoint one-dimensional Anderson model in \cite{CKM}, we will refer to \cite{Hamza}. In Section~\ref{sec:emanproof} we will then explain in detail how this leads to (\ref{eq:eman}).

\vspace{.5cm}

\subsubsection{Band Edge Localization} \label{sec:beloc}

For notational simplicity and without restriction we assume for the following result that supp$\,\mu \subset [-a,a]$ with $a\in (0,\pi)$ and $-a, a\in \mbox{supp}\,\mu$.
Furthermore, we assume that
\be
a + d\lambda_0 < \pi,
\ee
which by (\ref{eq:asspectrum}) guarantees the existence of a gap in the almost sure spectrum $\Sigma$ of $U_{\omega}$,
\be
\{e^{i\vartheta}: \, \vartheta \in (d\lambda_0+a, 2\pi -d\lambda_0-a)\} \cap \Sigma = \emptyset,
\ee
and that $e^{i(d\lambda_0+a)}$ and $e^{i(2\pi -d\lambda_0-a)}$ are band edges of $\Sigma$.
Our main result is
\begin{thm} \label{thm:bandedgelocalization}
Assume (\ref{eq:mudensity}) and let $0<s<1/3$. There exist $\delta>0$, $\alpha>0$ and $C<\infty$ such that
\be \label{eq:bandedgeloc}
\E(|G(k,l;z)|^s) \le Ce^{-\alpha|k-l|}
\ee
for all $k,l \in \Z^d$ and all $z\in \C$ with $|z|\not= 1$ and arg$\,z \in [d\lambda_0 +a-\delta, d\lambda_0+a] \cup [2\pi-d\lambda_0-a, 2\pi -d\lambda_0-a+\delta]$.
\end{thm}

Note that by Theorem~3.2 this implies dynamical localization for $U_{\omega}$ near the edges $d\lambda_0+a$ and $2\pi-d\lambda_0-a$ of its almost sure spectrum.

\vspace{.3cm}

The strategy of the proof of Theorem~\ref{thm:bandedgelocalization} is the
following.
We control the expectation value of fractional moments of the infinite volume
Green function in terms of the expectation value of fractional moments of the
Green function of a {\it finite} volume restriction of the operator. This
requires addressing several distinct issues.

The first one is the definition
of an appropriate finite volume restriction. We restrict the problem to a finite
 but large box, by
introducing appropriate boundary conditions in Section~8. Our choice of
boundary conditions is governed by the fact that we need monotoncity
properties which are similar to those of Neumann conditions in the
selfadjoint case. The link between the infinite and finite volume resolvents is
provided by a geometric resolvent estimate and a decoupling argument, similar
to the self-adjoint case. Provided one has good estimates on the expectation
value of the fractional moments of the finite volume Green function,  this
allows to lift such estimates to the fractional moments of the infinite volume
resolvent by means of an iteration, for large but fixed size of the box, in a
neighborhood of the band edges. This second step is addressed in
Section~\ref{sec:decoupling}.

To get the sought for estimates on the resolvent of the finite volume
restriction, in the band edge regime, we need to control the probability
that the finite volume restriction of $U_\omega$ has eigenvalues close to
the band edge.
Quantitatively, this amounts to showing that the probability of a small
distance, algebraic in the inverse size of the box,
between the eigenvalues of this finite volume
restriction and the band edges is exponentially small, as the size of the box
increases, see Proposition~\ref{prop:lt}. This is an expression of the fact that
the spectrum close to the band edges is very fluctuating which gives rise to
Lifshits tails in the density of states. To prove this, we follow the
self-adjoint route, see
\cite{Stollmann}. We first study the effect of Neumann
boundary conditions on the spectrum of $S$ in Section~\ref{sec:Neumann}, and we make use of
a unitary version of the
Feynman-Hellmann formula, Proposition~\ref{fehe},
to control this effect on the spectrum of the random operator $U_\omega$.
Then, the Lifshits tail estimate together with a unitary version of the
Combes-Thomas estimate, Proposition~\ref{prop:CT}, allow to show that the
expectation of the moments of the finite volume Green function is
exponentially small in a power of the size of the box,
Proposition~\ref{prop:smallfactor}.

\setcounter{equation}{0}
\section{Boundedness of Fractional Moments} \label{sec:fmboundproof}

In this section we prove Theorem~\ref{thm:fmbound}.

As the bound (\ref{eq:fmbound}) is trivial for
$|z|<1/2$, it suffices to consider $|z|\ge
1/2$ and $|z|\not=1$. Below we prefer to work with the modified resolvent $(U_\omega+z)(U_\omega-z)^{-1}$. Since
\be\label{eq:greentomodified}
(U_\omega-z)^{-1}=\frac{1}{2z}[(U_\omega+z)(U_\omega-z)^{-1}-I],
\ee
it easy to see that the existence of $0<C<\infty$ for
which
\be \label{eq:modfmbound}
\iint |\langle e_k|
(U_\omega+z)(U_\omega-z)^{-1} e_l \rangle|^s \,d\mu(\theta_k) \,d\mu(\theta_l) \leq C,
\ee
for all
$z$ with $|z|\not=1$, all $k,l\in\Z^d$, and uniformly in $\theta_j$, $j\not\in \{k,l\}$, gives the required bound.

Key to the proof of (\ref{eq:modfmbound}) is knowledge of the exact algebraic dependence of the Green function on the two parameters $\theta_k$ and $\theta_l$. Similar formulas for rank one and rank two perturbations of the resolvents of unitary operators have been derived in \cite{Simon1}, Section~4.5, from where we took guidance.

We mostly focus on the proof of (\ref{eq:modfmbound}) for the case $k\neq l$. At the end of the proof we comment on the simpler case $k=l$, where (\ref{eq:modfmbound}) only requires averaging over one parameter.

For $k\neq l$ weborrow an idea from \cite{AENSS} and introduce the change of variables $\alpha = \frac{1}{2}(\theta_k+\theta_l)$, $\beta = \frac{1}{2}(\theta_k-\theta_l)$. This will have the effect of essentially reducing (\ref{eq:modfmbound}) to averaging over the single parameter $\alpha$ (although this still corresponds to a rank two perturbation).

Let $A=\{k,l\}\subset\Z^d$ and define
\bea
\eta_j & = & \left\{\begin{array}{ll} \alpha, & j\in A
\\ 0, & j\notin A
\end{array}\right.
\\ \xi_j & = & \left\{\begin{array}{ll} \beta, & j=k
\\ -\beta, & j=l
\\ 0, & j\notin A
\end{array} \right.
\\ \widehat{\theta}_j^\omega & = & \left\{\begin{array}{ll} 0, & j\in A
\\ \theta^\omega_j, & j\notin A
\end{array}\right.
\eea
 Next, we define the diagonal operators $D_\alpha$, $D_\beta$ and
 $\widehat{D}$ by
 \be
D_\alpha e_j =e^{-i\eta_j} e_j, \quad D_\beta e_j=e^{-i\xi_j} e_j,
\quad \widehat{D}e_j=e^{-i\hat{\theta}_j}e_j.
\ee
Using these
definitions we can write
\be\label{def.V}
U_\omega=D_\alpha
V_\omega,
\ee
with the unitary operator $V_\omega=D_\beta
\widehat{D} S$. In what follows we explore the
relation between the modified resolvents of $U_\omega$ and
$V_\omega$.

Let $P_A$ be the orthogonal projection onto the span of
$\{V_\omega^{-1}e_j:j\in
 A\}$. Using that $\{V_\omega^{-1}e_j:j\in\Z^d\}$ is an orthonormal basis of $l^2(\Z^d)$, simple calculations show that
 $(U_\omega-V_\omega)(I-P_A)=0$ and $V^{-1}_\omega U_\omega
 =e^{-i\alpha}I$ on range $P_A$. In particular, $U_{\omega}-V_{\omega}=(U_{\omega}-V_{\omega})P_A$ is a finite rank operator.
Therefore,
\be\label{uandv}
U_\omega=V_\omega(I-P_A)+e^{-i\alpha}V_\omega P_A.
\ee
 For $z\in \C\setminus \{0\}$ with $|z|\neq 1$, let $F_z=P_A(U_\omega+z)(U_\omega-z)^{-1}P_A$ while
 $\widehat{F}_z=P_A(V_\omega+z)(V_\omega-z)^{-1}P_A$, both viewed as operators on the range of $P_A$ (i.e.\ $2\times 2$-matrices).
 We see that
 \be
 \widehat{F}_z+\widehat{F}_z^*=P_A(2I-2|z|^2)(V_\omega-z)^{-1}[(V_\omega-z)^{-1}]^*P_A.
 \ee
This shows that $\widehat{F}_z+\widehat{F}_z^*$ is invertible and
$\widehat{F}_z+\widehat{F}_z^*<0$ for $|z|>1$. Therefore,
$-i\widehat{F}_z$ is a dissipative operator, {\it i.e.} an operator
$A$ such that $(A-A^*)/2i >0$.
 Similarly,
$-i\widehat{F}_z^{-1}$ is also a dissipative operator. In the case
$|z|<1$, we have that $i\widehat{F}_z$, $i\widehat{F}_z^{-1}$ are
dissipative.

Next we explore the relation between $\widehat{F}_z$ and $F_z$.
Following Section~4.5 of \cite{Simon1} we use the fact that
$(x+z)(x-z)^{-1}=1+2z(x-z)^{-1}$ along with (\ref{uandv}) to obtain
\be
F_z-\widehat{F}_z=-2zP_A(V_\omega-z)^{-1}V_\omega
P_A(e^{-i\alpha}-1)P_A(U_\omega-z)^{-1}P_A.
\ee

Using the definitions of $F_z$ and $\widehat{F}_z$, this can be
rewritten in the form
\be
F_z-\widehat{F}_z=\frac{1}{2}(1+\widehat{F}_z)(e^{-i\alpha}-1)(1-F_z).
\ee

For $\alpha\notin\{0,\pi\}$, let
$m(\alpha)=i\frac{1+e^{-i\alpha}}{1-e^{-i\alpha}}\in\R$. A
straightforward calculation shows that
\be\label{ranktwopert}
F_z=-i(-i\widehat{F}_z+m(\alpha))^{-1}-i(-i\widehat{F}_z^{-1}-m^{-1}(\alpha))^{-1}.
\ee
Note that $\widehat{F}_z$ depends on $\beta$ and the $\widehat{\theta}_j$, but not on $\alpha$.

From  the definitions of $F_z$ and $P_A$ and the fact that
$V^{-1}_\omega U_\omega=e^{-i\alpha}I$ on
span$\{V_\omega^{-1}e_j:j\in\{k,l\}\}$, a simple calculation shows
that  \bea \langle e_k|(U_\omega+z)(U_\omega-z)^{-1}e_l\rangle & = &
\langle  V^{-1}_\omega e_k|F_z V_\omega^{-1}e_l\rangle. \eea

Therefore,
\bea \label{modifiedprojection}
\lefteqn{\iint |\langle e_k|(U_\omega+z)(U_\omega-z)^{-1}e_l\rangle|^s \,d\mu(\theta_k)\, d\mu(\theta_l)} \nonumber \\
& \leq &
||\tau||_\infty^2  \int_0^{2\pi}\int_0^{2\pi}|\langle
V^{-1}_\omega e_k|F_z V_\omega^{-1}e_l\rangle|^s \,d\theta_k\,
d\theta_l \nonumber \\
& \leq & \|\tau\|_{\infty}^2 \int_0^{2\pi} \int_0^{2\pi} \|F_z\|^s\,d\theta_k\,d\theta_l \nonumber \\
& \leq & 2 ||\tau||_\infty^2 \int_{-\pi}^{\pi} \left( \int_0^{2\pi} \|F_z\|^s\,d\alpha \right)\,d\beta,
\eea
where we have changed to the variables $\alpha$ and $\beta$ and slightly enlarged the integration domain into the rectangle $0\le \alpha \le 2\pi$, $-\pi \le \beta \le \pi$.

We split the $\alpha$-integral according to (\ref{ranktwopert}),
\be \label{eq:alphasplit}
\int_0^{2\pi} \|F_z\|^s\,d\alpha \le \int_0^{2\pi} \|(-i\widehat{F}_z+m(\alpha))^{-1}\|^s\,d\alpha + \int_0^{2\pi} \|(-i\widehat{F}_z^{-1}-m^{-1}(\alpha))^{-1}\|^s\,d\alpha.
\ee

Recalling that $m(\alpha)$ has singularities at $\alpha\in\{0,2\pi\}$, we make the change of variables $x=m(\alpha)$,
\bea
\label{intestimate}
\lefteqn{\int_0^{2\pi}  ||(-i\widehat{F}_z+m(\alpha))^{-1}||^s  d\alpha} \nonumber \\ & = &
\lim_{R\to\infty}\int_{-R}^{R} \frac{2}{x^2+1} ||(-i\widehat{F}_z+x)^{-1}||^s\,dx \nonumber
\\ & = & 2\sum_{n\in\Z}\int_{n}^{n+1} \frac{1}{x^2+1} ||(-i\widehat{F}_z+x)^{-1}||^s\,dx \nonumber
\\ & \leq & 2 \sum_{n\in\Z}\frac{1}{(|n|-1)^2+1}\int_{n}^{n+1} ||(-i\widehat{F}_z+x)^{-1}||^s\,dx.
\eea
We can now complete the proof of (\ref{eq:modfmbound}) by treating the cases $|z|>1$ and $|z|<1$ separately. If $|z|>1$ then $-i\widehat{F}_z$ is dissipative and Lemma~\ref{lem:2by2dis} shows boundedness of the integral on the right of (\ref{intestimate}), uniform in $n$, $|z|>1$, $\beta$ and $\widehat{\theta}_j$. Thus, after summation, $\int_0^{2\pi}  ||(-i\widehat{F}_z+m(\alpha))^{-1}||^s \le C(s)$. The second term on the right of (\ref{eq:alphasplit}) can be bounded in a similar way, using that $-i\widehat{F}_z^{-1}$ is dissipative as well. Inserting these bounds into (\ref{modifiedprojection}) makes the $\beta$-integration trivial and completes the proof of (\ref{eq:modfmbound}) for the case $|z|>1$.

If $|z|<1$, then $-i\widehat{F}_z$ and $-i\widehat{F}_z^{-1}$ and anti-dissipative. As Lemma~\ref{lem:2by2dis} obviously also holds for anti-dissipative matrices $A$, the proof of (\ref{eq:modfmbound}) goes through with the same argument.

\vspace{.3cm}

The proof of (\ref{eq:modfmbound}) for the case $k=l$ is similar but simpler. We don't need a change of variables, but directly work with one of the parameters $\theta_l$ instead of $\alpha$, leading to rank one perturbations. The objects corresponding to $F_z$ and $\widehat{F}_z$ become scalars and we only have to use the trivial scalar version of Lemma~\ref{lem:2by2dis} to conclude.

\vspace{.5cm}

We finally provide an elementary proof of the following Lemma which was used above. A much more general result of this form is given as Lemma~3.1 in \cite{AENSS}.

\begin{lem} \label{lem:2by2dis}
For every $s\in (0,1)$ there exists $C(s)<\infty$ such that
\be \label{eq:disbound}
\int_E \|(A+xI)^{-1}\|^s\,dx \le C(s)
\ee
for every dissipative $2\times 2$-matrix $A$ and every unit interval $E$.
\end{lem}

{\bf Proof:} First observe that a general dissipative $2\times 2$-matrix $A$ is unitarily equivalent to an upper triangular dissipative matrix (choose as unitary transformation any matrix whose first column is given by a normalized eigenvector of $A$). Thus we may assume that
\be \label{eq:Aform}
A = \left( \begin{array}{cc} a_{11} & a_{12} \\ 0 & a_{22} \end{array} \right),
\ee
which implies
\be \label{eq:disinvers}
(A+xI)^{-1} = \left( \begin{array}{cc} \frac{1}{a_{11}+x} & - \frac{a_{12}}{(a_{11}+x)(a_{22}+x)} \\ 0 & \frac{1}{a_{22}+x} \end{array} \right).
\ee
The bound (\ref{eq:disbound}) follows if we can establish a corresponding fractional integral bound for the absolute value of each entry of (\ref{eq:disinvers}) separately. For the diagonal entries this is obvious.

We bound the upper right entry of (\ref{eq:disinvers}) by
\bea \label{eq:upperright}
\left| \frac{a_{12}}{(a_{11}+x)(a_{22}+x)}\right| & \le & \frac{|a_{12}|}{|\mbox{Im}\,((a_{11}+x)(a_{22}+x))|} \nonumber \\
& = & \frac{1}{\left| x \frac{\mbox{\footnotesize Im}\,a_{11} +\mbox{\footnotesize Im}\,a_{22}}{|a_{12}|} + \frac{\mbox{\footnotesize Im}(a_{11}a_{22})}{|a_{12}|} \right|}.
\eea
The positive matrix
\be
\mbox{Im}\,A = \frac{1}{2i}(A-A^*) = \left( \begin{array}{cc} \mbox{Im}\,a_{11} & \frac{1}{2i}a_{12} \\ -\frac{1}{2i}\bar{a}_{12} & \mbox{Im}\,a_{22} \end{array} \right)
\ee
has positive determinant, i.e.\ det Im$\,A = \mbox{Im}\,a_{11} \mbox{Im}\,a_{22} -|a_{12}|^2/4$.
We thus get
\be
\left| \frac{\mbox{Im}\,a_{11} + \mbox{Im}\,a_{22}}{a_{12}}\right|^2 \ge \frac{2\mbox{Im}\,a_{11} \mbox{Im}\,a_{22}}{|a_{12}|^2} \ge \frac{1}{2}.
\ee
The latter allows to conclude the required integral bound for (\ref{eq:upperright}).
\ep

\setcounter{equation}{0}
\section{Dynamical Localization via Green's Function} \label{sec:dynlocproof}

Here we will prove Theorem~\ref{thm:dynamicallocalization}, i.e.\ that exponential decay of fractional moments of Green's function implies dynamical localization.

Our proof uses an idea which in the context of selfadjoint Anderson models is due to Graf \cite{Graf}, namely that second moments of an Anderson model's Green function can be bounded in terms of its fractional moments (including, however, a scalar factor which becomes singular as the spectral parameter approaches the spectrum). While the details of the proof are more involved than in the selfadjoint case, we find a bound of this form for unitary Anderson models in Section~\ref{sec:secmoment}.

Another tool we use is the integral formula (\ref{f(u)}), which expresses operator functions $f(U)$ in terms of the resolvent of $U$. In Section~\ref{sec:poissonrep} we provide a proof of this formula, which combines the spectral theorem for unitary operators with the representation of Borel measures on $\T$ by Poisson integrals.

Equipped with these tools we complete the proof of dynamical localization in Section~\ref{sec:completeproof}.

\subsection{A Second Moment Estimate} \label{sec:secmoment}

We start by a bound of second moments of Green's function in terms
of its fractional moments, which holds pointwise in the spectral
parameter.

\begin{prop} \label{prop:2ndmoment}
Assume that the $\{\theta_k^{\omega}\}_{k\in\Z^d}$ satisfy
(\ref{eq:mudensity}). Then for every $s\in (0,1)$ there exists
$C(s)<\infty$ such that
\be \label{eq:2ndmomentbound}
\E((1-|z|^2)|G(k,l;z)|^2) \le C(s) \sum_{|m-k|\le 4}
\E(|G(m,l;z)|^s) \ee for all $|z|<1$ and $k,l \in \Z^d$.
\end{prop}

\vspace{.5cm}

 {\bf Proof:} Throughout the proof we will assume $z\not=0$. The bound
(\ref{eq:2ndmomentbound}) carries over to $z=0$ by continuity.

For $\delta \in\T$, let
$\eta_k=e^{-i(\theta_k^{\omega}+\delta)}-e^{-i\theta_k^{\omega}}$. Then
define $D_\omega^\delta=D_\omega+\eta_k P_k$, where $P_k$ is the
orthogonal projection into the span of $e_k$. Let
$U_\omega^\delta=D_\omega^\delta S$. Using the resolvent identity we
have
\be
(U^\delta_\omega-z)^{-1}-(U_\omega-z)^{-1}=-(U^\delta_\omega-z)^{-1}\eta_k P_k
S(U_\omega-z)^{-1},
\ee
for all $z\in\C$ such that $0<|z|<1$. Letting
$F(z)=S(U_\omega-z)^{-1}$ and
$F_\delta(z)=S(U^\delta_\omega-z)^{-1}$, the last equation takes the
form
\be
F_\delta(z)-F(z)=-F_\delta(z)\eta_k P_k F(z).
\ee
Denoting $F(i,j,z)=\bra e_i| F(z)e_j\ket$ it is easy to see that
\be
F_\delta(z)-F(z)=-\eta_k \frac{F(z)P_k F(z)}{1+\eta_k F(k,k,z)}.
\ee
Therefore, for all $l\in\Z^d$
\be \label{realtionofperturbed}
F_\delta(k,l,z)=\frac{F(k,l,z)}{1+\eta_k F(k,k,z)}.
\ee

On the other hand, we also have that
\bea
|F_\delta(k,l,z)|^2 & \leq
& \sum_{y\in\Z}|F_\delta(k,y,z)|^2 \nonumber
\\ & = & \bra e_k|
S(U^\delta_\omega-z)^{-1}[(U^\delta_\omega-z)^{-1}]^*S^* e_k\ket.
\eea
Since $U^\delta_\omega$ is a unitary operator, the following
identity holds
\be\label{from*}
[(U^\delta_\omega-z)^{-1}]^*=\frac{-1}{\bar{z}}(U^\delta_\omega-\frac{1}{\bar{z}})^{-1}U^\delta_\omega.
\ee
Thus, it follows that
\bea
|F_\delta(k,l,z)|^2 & \leq &
\frac{-1}{\bar{z}}\bra e_k|
S(U^\delta_\omega-z)^{-1}(U^\delta_\omega-\frac{1}{\bar{z}})^{-1}D_\omega^\delta
e_k\ket \nonumber
\\ & = & \frac{-e^{-i(\theta^\omega_k+\delta)}}{\bar{z}}\bra e_k|
S(U^\delta_\omega-z)^{-1}(U^\delta_\omega-\frac{1}{\bar{z}})^{-1}
e_k\ket.
\eea
Again using the resolvent identity, we see that
\be
(U^\delta_\omega-z)^{-1}(U^\delta_\omega-\frac{1}{\bar{z}})^{-1}
=\frac{\bar{z}}{|z|^2-1}\{(U^\delta_\omega-z)^{-1}-(U^\delta_\omega-\frac{1}{\bar{z}})^{-1}\}.
\ee
Hence,
\be
|F_\delta(k,l,z)|^2 \leq
\frac{e^{-i(\theta^\omega_k+\delta)}}{1-|z|^2}\{\bra e_k|
S(U^\delta_\omega-z)^{-1}e_k\ket-\bra
e_k|S(U^\delta_\omega-\frac{1}{\bar{z}})^{-1} e_k\ket\}.
\ee
From (\ref{from*}), the definition of $U_\omega^\delta$ and the fact that
$(U^\delta_\omega-z)^{-1}=-\frac1z[I-U_\omega^\delta(U^\delta_\omega-z)^{-1}]$,
it follows that
\bea
\bra e_k|S(U^\delta_\omega-\frac{1}{\bar{z}})^{-1} e_k\ket & = &
-\bar{z}e^{i(\theta_k^\omega+\delta)} \bra
e_k|[(U^\delta_\omega-z)^{-1}]^* e_k\ket \nonumber
\\ & = & -\bar{z}e^{i(\theta_k^\omega+\delta)}\{1- \bra
U_\omega^\delta(U^\delta_\omega-z)^{-1} e_k|e_k\ket\} \nonumber
\\ & = & e^{i(\theta_k^\omega+\delta)}\{1-e^{i(\theta^\omega_k+\delta)}
\overline{F_\delta(k,k,z)}\}.
\eea
Therefore, we now obtain that
\bea
|F_\delta(k,l,z)|^2 & \leq &
\frac{1}{1-|z|^2}\{2\Re{[e^{-i(\theta^\omega_k+\delta)}F_\delta(k,k,z)]}-1\} \nonumber
\\ & = & \frac{1}{1-|z|^2}\{|F_\delta(k,k,z)|^2-|e^{i(\theta^\omega_k+\delta)}-F_\delta(k,k,z)|^2\},
\eea
 since $|x-y|^2=|x|^2+|y|^2-2\Re[{\bar{x}y}]$. Using
(\ref{realtionofperturbed}), to rewrite $F_\delta(k,k,z)$ in terms
of elements of $F$, along with the definition of $\eta_k$ we get
\be
|F_\delta(k,l,z)|^2 \leq
\frac{1}{1-|z|^2}\{\frac{|F(k,k,z)|^2-|e^{i\theta^\omega_k}-F(k,k,z)|^2}{|1
+\eta_kF(k,k,z)|^2}\}.
\ee
This inequality gives, in particular,
that $F(k,k,z)\neq0$. Therefore
\be \label{firstestimatesquare}
(1-|z|^2)|F_\delta(k,l,z)|^2 \leq \frac{1-|1-
e^{i\theta_k}F(k,k,z)^{-1}|^2}{|\eta_k +F(k,k,z)^{-1}|^2}.
\ee
Finally note that the last inequality allows us to
conclude that $|1- e^{i\theta_k}F(k,k,z)^{-1}|\leq1$.

 One can also use the fact that
\be
|F_\delta(k,k,z)| \leq  ||(U_\omega^\delta-z)^{-1}| \leq
\frac{1}{1-|z|},
\ee
 for all $\delta\in\T$, to get a different
upper bound on $(1-|z|^2)|F_\delta(k,l,z)|^2$. Since
(\ref{realtionofperturbed}) can be rewritten as
\be \label{realtion2}
F_\delta(k,l,z) =\frac{1}{\eta_k
+F(k,k,z)^{-1}}\frac{F(k,l,z)}{F(k,k,z)},
\ee
it follows that
$1-|\eta_k +F(k,k,z)^{-1}|\leq|z|$. Then by choosing $\delta$ such
that $e^{-i\delta}=\frac{1- e^{i\theta_k}F(k,k,z)^{-1}}{|1-
e^{i\theta_k}F(k,k,z)^{-1}|}$, we see that
\be
|1-e^{i\theta_k}F(k,k,z)^{-1}|\leq |z|.
\ee
Using this along with
(\ref{realtion2}) we obtain the following upper bound
\be \label{secondestimatesqaure}
(1-|z|^2)|F_\delta(k,l,z)|^2\leq
\frac{1-|1- e^{i\theta_k}F(k,k,z)^{-1}|^2}{|\eta_k
+F(k,k,z)^{-1}|^2}\frac{|F(k,l,z)|^2}{|F(k,k,z)|^2}.
\ee

Combining the two estimates (\ref{firstestimatesquare}) and
(\ref{secondestimatesqaure}) and using that for $0<s<1$ we have
$\min(1,|x|^2)\leq |x|^s$, it follows that
\be
(1-|z|^2)|F_\delta(k,l,z)|^2\leq \frac{1-|1-
e^{i\theta_k}F(k,k,z)^{-1}|^2}{|e^{-i\delta}-(1
-e^{i\theta_k}F(k,k,z)^{-1})|^2}\frac{|F(k,l,z)|^s}{|F(k,k,z)|^s}.
\ee
Letting $y=1- e^{i\theta_k}F(k,k,z)^{-1}$, this can be rewritten
as
\be
(1-|z|^2)|F_\delta(k,l,z)|^2\leq
\frac{(1-|y|^2)|1-y|^s}{|e^{-i\delta}-y|^2}|F(k,l,z)|^s.
\ee

Since the expectations of $F$ and $F_\delta$ are related by
\be
\E[|F(k,l,z)|^2]=\E[\int d\mu(\theta_k+\delta)|F_\delta(k,l,z)|^2],
\ee
 it follows that
\bea
\lefteqn{\E[(1-|z|^2)|F(k,l,z)|^2]} \nonumber \\ & \leq &
||\tau||_\infty\E\Big[|F(k,l,z)|^s
\sup_{\{y\in\C:|y|<1\}}\int_0^{2\pi}d\delta
\frac{(1-|y|^2)|1-y|^s}{|e^{-i\delta}-y|^2}\Big] \nonumber
\\ & \leq & 2^s||\tau||_\infty\E\Big[|F(k,l,z)|^s
\sup_{\{y\in\C:|y|<1\}}\int_0^{2\pi}d\delta
\frac{(1-|y|^2)}{|e^{-i\delta}-y|^2}\Big].
\eea
Next we evaluate the
integral
\bea
\int_0^{2\pi}d\delta
\frac{(1-|y|^2)}{|e^{-i\delta}-y|^2} & = & \int_0^{2\pi}d\delta
\Re{[\frac{e^{-i\delta}+y}{e^{-i\delta}-y}]} \nonumber
\\ & = & \Re\int_0^{2\pi}d\delta
[\frac{2e^{-i\delta}}{e^{-i\delta}-y}-1].
\eea
The latter integral
can be easily evaluated using Cauchy integral formula, by simply
substituting $z=e^{-i\delta}$ and gives
\be
\int_0^{2\pi}d\delta
\frac{(1-|y|^2)}{|e^{-i\delta}-y|^2} =2\pi.
\ee
Therefore,
\be \label{from2tos}
\E[(1-|z|^2)|F(k,l,z)|^2] \leq
2^{s+1}\pi||\tau||_\infty\E[|F(k,l,z)|^s].
\ee
Since $S$ is a
unitary operator with band structure, we see that for $s\in(0,1)$
\bea \label{eq:1stbound}
\E[|F(k,l,z)|^s] & = & \E[|\bra
S^{*}e_k| (U_\omega-z)^{-1}e_l\ket|^s] \nonumber
\\ & \leq & C_1(s)\sum_{|m-k|\leq 2}\E[|\bra e_m|
(U_\omega-z)^{-1}e_l\ket|^s].
\eea

 Finally, in order to get the required bound of the second moment of elements of $(U_\omega-z)^{-1}$
  we use again that $S$ is a unitary operator with
band structure. Therefore, for $k,l\in\Z^d$
\bea \label{eq:2ndbound}
\E[(1-|z|^2)|\bra e_k| (U_\omega-z)^{-1}e_l\ket|^2] & = &
\E[(1-|z|^2)|\bra Se_k| S(U_\omega-z)^{-1}e_l\ket|^2] \nonumber
\\ & \leq & C_2(s) \sum_{|m-k|\leq 2 }\E[(1-|z|^2)|F(m,l,z)|^2].
 \eea
Combining (\ref{eq:2ndbound}), \eqref{from2tos} and
(\ref{eq:1stbound})  gives (\ref{eq:2ndmomentbound}). \ep

\vspace{.5cm}

\subsection{An Integral Representation for $f(U)$} \label{sec:poissonrep}

We will reduce bounds for $f(U)$ to bounds on resolvents by the
formula
\be \label{f(u)}
f(U)=w-\lim_{r\ra
1^-}\frac{1-r^2}{2\pi}\int_0^{2\pi}(U-re^{i\theta})^{-1}(U^{-1}-re^{-i\theta})^{-1}f(e^{i\theta})d\theta
\ee
for $f\in C(\mathbb S)$ and $U$ a unitary operator. This is a
simple consequence of the representation of Borel measures on
$\mathbb T$ by Poisson integrals:

Let $\ffi\in{\mathcal H}$  be normalized and consider the non
negative spectral measure $d\mu_\ffi$ on $\mathbb T$ such that
\be
\bra \ffi | U \ffi\ket=\int_{\mathbb T}e^{i\alpha}d\bra\ffi |
E(\alpha)\ffi \ket=\int_{\mathbb T}e^{i\alpha}d\mu_\ffi(\alpha),
\ee
where $E(\cdot)$ denotes the spectral family of $U$. We can thus
rewrite for $0\leq r<1$
\bea\label{pi}
(1-r^2)\bra \ffi
|(U-re^{i\theta})^{-1}(U^{-1}-re^{-i\theta})^{-1}\ffi\ket&=&
\int_{\mathbb
T}\frac{1-r^2}{|e^{i\alpha}-re^{i\theta}|^2}d\mu_\ffi(\alpha) \nonumber \\
=\int_{\mathbb
T}\frac{1-r^2}{1+r^2-2r\cos(\theta-\alpha)}d\mu_\ffi(\alpha)&=&u(r,\theta),
\eea
which coincides with the Poisson integral of the measure
$d\mu_\ffi$. As a function of $z=x+iy=re^{i\theta}$, with the
standard abuses of notations, the RHS of (\ref{pi}) is non negative
and harmonic in the open unit disc (\cite{Rudin} Sect. 11.17), but
it is not bounded as $r\ra 1^-$ at the atoms of $d\mu_\ffi$. Now,
Theorem 3.9.8 in \cite{Pinsky} on the representation of non negative
Borel measures on $\mathbb T$ says that for any $f\in C(\mathbb S)$
we have
\be
\lim_{r\ra
1^-}\int_0^{2\pi}u(r,\theta)f(e^{i\theta})\frac{d\theta}{2\pi}=\int_{\mathbb
T}f(e^{i\alpha})d\mu_\ffi(\alpha)\equiv \bra \ffi | f(U)\ffi\ket.
\ee
Considering the matrix element $\bra\ffi | \cdot \psi\ket$ for
arbitrary $\ffi$ and $\psi$ in ${\mathcal H}$, we get the equivalent
result  by polarization, which proves (\ref{f(u)}).\medskip

If $P_{[a,b]}$ denotes the spectral projector of $U$ on the arc
$[a,b]\subset {\mathbb T}$ and $\ffi  \in{\mathcal H} $ is
normalized, then
\be
d\bra\ffi |P_{[a,b]}E(\alpha)\ffi\ket\equiv
d\mu_\ffi^{[a,b]}(\alpha)=d\mu_\ffi(\alpha)|_{[a,b]}.
\ee

By the same argument with $P_{[a,b]}\ffi$ in place of $\ffi$, i.e.
with $d\mu_\ffi^{[a,b]}$ in place of $ d\mu_\ffi$, we have for any
$f\in C({\mathbb S})$
\bea
P_{[a,b]} f(U) &=&w-\lim_{r\ra 1^-}\frac{1-r^2}{2\pi}\int_{a}^{b}(U-re^{i\theta})^{-1}(U^{-1}-re^{-i\theta})^{-1}f(e^{i\theta})d\theta\nonumber\\
&\equiv& w-\lim_{r\ra 1^-}f_r(U).
\eea

\subsection{Conclusion of the Proof of Theorem~\ref{thm:dynamicallocalization}} \label{sec:completeproof}

Let $f\in C(\mathbb S)$ such that $\|f\|_\infty\leq 1$ and consider
\be
f_r(U)=\frac{1-r^2}{2\pi}\int_{a}^{b}(U-re^{i\theta})^{-1}(U^{-1}-re^{-i\theta})^{-1}f(e^{i\theta})d\theta.
\ee
We compute
\bea
\lefteqn{|\bra e_k|f_r(U)e_j\ket|} \nonumber \\ & \leq & \frac{1-r^2}{2\pi}\int_a^{b}|\bra e_k|(U-re^{i\theta})^{-1}(U^{-1}-re^{-i\theta})^{-1}e_j\ket||f(e^{i\theta})|d\theta \nonumber\\
& \leq & \frac{1-r^2}{2\pi}\int_a^{b}\sum_{l\in{\mathbb Z}^d}|\bra
e_k|(U-re^{i\theta})^{-1}e_l\ket||\bra
e_j|(U-re^{i\theta})^{-1}e_l\ket|d\theta,
\eea
which is independent
of $f$. Hence, using  Fatou's Lemma and Fubini's Theorem and taking
the supremum over all $f\in C({\mathbb S})$ such that
$\|f\|_{\infty}\le 1$,
\bea
\lefteqn{{\mathbb E}[\sup_{f}|\bra
e_k|P_{[a,b]}f(U_\omega)e_j\ket|]} \nonumber \\ & = &
{\mathbb E}[\sup_{f} \lim_{r\ra 1^-}|\bra e_k|f_r(U)e_j\ket|] \nonumber \\
& \leq & {\mathbb E}[ \liminf_{r\ra 1^-} \sup_{\|f\|_\infty\leq 1} |\bra e_k|f_r(U)e_j\ket|] \nonumber \\
&\leq & \liminf_{r\ra
1^-}\frac{1-r^2}{2\pi}\int_a^{b}\sum_{l\in{\mathbb Z}^d}{\mathbb
E}[|\bra e_k|(U-re^{i\theta})^{-1}e_l\ket||\bra
e_j|(U-re^{i\theta})^{-1}e_l\ket|]|d\theta.
\eea
By the
Cauchy-Schwarz inequality and the second moment bound
(\ref{eq:2ndmomentbound}), we have
\bea
\lefteqn{{\mathbb E}[|(1-r^2)\bra e_k|(U-re^{i\theta})^{-1}e_l\ket||\bra e_j|(U-re^{i\theta})^{-1}e_l\ket|]} \nonumber \\ & \leq &
({\mathbb E}[|(1-r^2)\bra
e_k|(U-re^{i\theta})^{-1}e_l\ket|^2])^{1/2}
({\mathbb E}[|(1-r^2)\bra e_j|(U-re^{i\theta})^{-1}e_l\ket|^2])^{1/2} \nonumber \\
& \leq & C_1 \left( \sum_{m:|m-k|\le 4} {\mathbb E}(G(m,l;re^{i\theta})|^s)\right)^{1/2} \left( \sum_{m:|m-j|\le 4} {\mathbb E}(G(m,l;re^{i\theta})|^s)\right)^{1/2} \nonumber \\
& \leq & C_2 e^{-\alpha |k-l|/2} e^{-\alpha |j-l|/2},
\eea
where
ultimately the assumption (\ref{eq:fmexpdecay}) on exponential decay
of fractional moments of Green's function was used. Since there
exists a finite $c$ such that
\be
\sum_{l\in{\mathbb
Z}^d}e^{-\alpha(|k-l|+|j-l|)/2}\leq c e^{-\alpha |k-j|/4}
\ee
we
finally get
\be\label{norm}
{\mathbb E}[\sup_f |\bra
e_k|P_{[a,b]}f(U_\omega)e_j\ket|]\leq  \tilde{C} e^{-\alpha|k-j|/4},
\ee
with $\tilde{C}=c\,C_2$. This completes the proof of
Theorem~\ref{thm:dynamicallocalization}.

\setcounter{equation}{0}
 \section{Dynamical Localization implies  Spectral Localization}  
\label{dynspecproof}

We start by recalling the geometrical concepts and results
introduced in \cite{EV} in a general framework.
\medskip

Let $U$ be a unitary operator in a separable Hilbert space ${\mathcal H}$ and
let  ${\mathcal H}_{pp}(U)$, respectively  ${\mathcal H}_{c}(U)$ denote  the pure point, respectively continuous, spectral subspace of $U$.
We will  denote the spectral resolution of $U$  by $E_U$.

 A practical characterization of the vectors of ${\mathcal H}_{pp}(U)$ and  ${\mathcal H}_{c}(U)$ in terms of their behaviour under the discrete dynamics with respect to families of finite dimensional subspaces of ${\mathcal H}$ is provided in \cite{EV} and reads as follows:

\medskip

Let $P=\{P_r\}_{r\geq 0}$ be a family of projections on ${\mathcal H}$
such that
\be\label{fam}
P_r=P_r^2=P_r^*, \ \ \mbox{Rank } P_r
<\infty, \ \ \mbox{and}\ s-\lim_{r\ra\infty}P_r=I.
\ee
We define
the set of bounded trajectories with respect to the family $P$ by
\be
{\mathcal M}_U^b(P)=\{\psi\in {\mathcal H} \ |\
\lim_{r\ra\infty}\sup_{n\in\mathbb Z}\|(I -P_r)U^n\psi\|=0 \}.
\ee
Similarly, the set of propagating trajectories with respect to $P$ is defined by
 \be
 {\mathcal M}^p_U(P)=\{\psi\in {\mathcal H}
\ |\ \lim_{N\ra\infty}\frac{1}{2N+1}\sum_{n=-N}^N\|P_rU^n\psi\|=0, \
\forall r\geq 0\},
\ee
Both ${\mathcal M}_U^b(P)$ and ${\mathcal M}_U^p(P)$ are easily seen to be closed subspaces of ${\cal H}$.

Then the following  holds.

\begin{lem}\label{ev} For any unitary operator $U$ on a separable Hilbert space ${\mathcal H}$, and any family $P$ as in (\ref{fam}),
\be
{\mathcal M}_U^b(P)={\mathcal H}_{pp}(U)\ \ \ \mbox{and} \ \ \
{\mathcal M}^p_U(P)={\mathcal H}_{c}(U)
\ee
\end{lem}

\vspace{.3cm}

While the definition above suffices for our purpose, more general families of operators than $P$ can be considered, see \cite{EV}. For completeness, we provide a detailed proof of this Lemma, which is a slight adaptation of the argument in \cite{EV}.

\vspace{.2cm}

{\bf Proof of Lemma~\ref{ev}} in three steps:

First,  ${\mathcal H}_{pp}(U)\subset {\mathcal M}_U^b(P)$:

If $U\ffi=e^{i\alpha}\ffi$, then
\be
\|(I -P_r)U^n\ffi\|=\|(I -P_r)\ffi\|\ra 0, \ \ \mbox{uniformly in $n$ as $r\ra \infty$}.
\ee
We conclude using that ${\mathcal H}_{pp}(U)=\overline{\{\ffi \ | \ U\ffi=e^{i\alpha}\ffi\}}$ and the fact that ${\mathcal M}_U^b(P)$ is closed.

Second, ${\mathcal M}_U^b(P)\perp{\mathcal M}_U^p(P)$:

Let $\ffi\in {\mathcal M}_U^p(P)$ and
$\psi\in {\mathcal M}_U^b(P)$. Then
\bea
\bra \ffi |\psi\ket&=&\frac{1}{2N+1}\sum_{n=-N}^N\bra\ffi |\psi\ket=\frac{1}{2N+1}\sum_{n=-N}^N\bra U^n\ffi |U^n\psi\ket\nonumber\\
&=&\frac{1}{2N+1}\sum_{n=-N}^N\bra U^n\ffi |(I-P_r)U^n\psi\ket+\bra U^n\ffi |P_r U^n\psi\ket.
\eea
By our choice of $\psi$, for any $\eps>0$ there exists $r(\eps)$, uniform in $n$,  such that $|\bra U^n\ffi |(I-P_{r(\eps)})U^n\psi\ket|<\eps$. And our choice of $\ffi$ and the selfadjointness of $P_{r(\eps)}$ imply
\be
\left|\frac{1}{2N+1}\sum_{n=-N}^N\bra U^n\ffi |P_{r(\eps)} U^n\psi\ket\right|\leq\frac{1}{2N+1}\sum_{n=-N}^N\|P_{r(\eps)}U^n\ffi\|\|\psi\|\ra 0 \ \ \mbox{as $N\ra\infty,$}
\ee
which shows that $\bra \ffi |\psi\ket=0$.

Third, ${\mathcal H}_{c}(U)\subset {\mathcal M}_U^p(P)$:

Let $\psi\in {\mathcal H}_{c}(U)=P_c(U){\mathcal H}$, where $P_c(U)$ is the spectral projector of $U$ on the continuous spectral subspace of $U$. Since the orthogonal projector $P_r$ is finite dimensional for any $r$, there exists vectors $\ffi_i\in$ Ran~$P_r$ such that we can write for some $m<\infty$
\be
P_r=\sum_{i=1}^{m}|\ffi_i\ket\bra \ffi_i |, \ \ \mbox{where} \ \ \ \bra\ffi_i|\ffi_j\ket=\delta_{ij}.
\ee
Hence, $\|P_rU^n\psi\|^2=\sum_{i=1}^{m}|\bra\ffi_i|U^n\psi\ket|^2$.  Now, by the Spectral Theorem we have
\be
f_{\ffi_i,\psi}(n):=\bra\ffi_i|U^n\psi\ket=\int_{\mathbb T}e^{ixn}d\mu_{\ffi_i, \psi}(x),
\ee
where $d\mu_{\ffi, \psi}(x)=d\bra P_c(U)\ffi_i|E_U(x)P_c(U)\psi\ket$ is a continuous complex valued measure. Thus, by Wiener's Theorem,
\be
\lim_{N\ra\infty}\frac{1}{2N+1}\sum_{n=-N}^N|f_{\ffi_i,\psi}(n)|^2=\sum_{x\in {\mathbb T}}|\mu_{\ffi_i,\psi}(\{x\})|^2=0.
\ee
Consequently, making use of Cauchy-Schwarz,
\bea
\frac{1}{2N+1}\sum_{n=-N}^N\|P_rU^n\psi\| & \leq & \left\{\frac{1}{2N+1}\sum_{n=-N}^N\|P_rU^n\psi\|^2\right\}^{1/2} \nonumber \\
& = & \left\{\sum_{i=1}^m\frac{1}{2N+1}\sum_{n=-N}^N|f_{\ffi_i,\psi}(n)|^2\right\}^{1/2}
\eea
which tends to zero as $N\ra 0$.

Finally the decomposition  ${\mathcal H}={\mathcal H}_{pp}(U)\oplus {\mathcal H}_{c}(U)$ leads to the conclusion.\ep

\medskip

{\bf Proof of Proposition \ref{dynspec}}

Coming back to the random situation at hand, we construct a suitable
family of projectors $P=\{P_r\}_{r\geq 0}$ by  means of
\be
P_r=\sum_{\substack{k\in {\mathbb Z}^d\\|k|\leq r}} |e_k\ket\bra
e_k|, \ \ r\geq 0,
\ee
and consider the vectors of the form
$P^\omega_{[a,b]}e_j$, $j\in {\mathbb Z}^d$. We have for all
$n\in\mathbb Z$
\be
\|(I -P_r)U_\omega^n P^\omega_{[a,b]}e_j\| = \{\sum_{\substack{k\in {\mathbb Z}^d \\
|k| > r}}|\bra e_k |P^\omega_{[a,b]}U_\omega^n e_j\ket|^2\}^{1/2}.
\ee
Therefore, since $|\bra e_k |P^\omega_{[a,b]}U_\omega^n
e_j\ket|\leq 1$,
\bea
\sup_n\|(I -P_r)U_\omega^n P^\omega_{[a,b]}e_j\| & = & \{\sup_n \sum_{\substack{k\in {\mathbb Z}^d \\
|k| > r}}|\bra e_k |P^\omega_{[a,b]}U_\omega^n e_j\ket|^2\}^{1/2} \nonumber\\
&\leq&\{\sup_n\sum_{\substack{k\in {\mathbb Z}^d  \\
|k| > r}}|\bra e_k |P^\omega_{[a,b]}U_\omega^n e_j\ket|\}^{1/2}.
\eea
Thus, by Fatou's Lemma and Cauchy-Schwarz inequality,
\bea
\lefteqn{{\mathbb E}(\lim_{r\ra\infty}\sup_n\|(I -P_r)U_\omega^n
P^\omega_{[a,b]}e_j\|)} \nonumber \\ &\leq& \liminf_{r\ra\infty}\, {\mathbb
E}(\{\sup_n\sum_{\substack{k\in {\mathbb Z}^d |k| > r}}|\bra e_k
|P^\omega_{[a,b]}U_\omega^n e_j\ket|\}^{1/2}) \nonumber \\  &\leq&
\liminf_{r\ra\infty}\, \{{\mathbb E}(\sup_n\sum_{\substack{k\in {\mathbb Z}^d\\
|k| > r}}|\bra e_k |P^\omega_{[a,b]}U_\omega^n e_j\ket|)\}^{1/2}.
\eea
Now, by Fubini's Theorem,
\bea
{\mathbb E}(\sup_n\sum_{\substack{k\in {\mathbb Z}^d \\
|k| > r}}|\bra e_k |P^\omega_{[a,b]}U_\omega^n e_j\ket|) & \leq & {\mathbb E}(\sum_{\substack{k\in {\mathbb Z}^d\\
|k| > r}}\sup_n |\bra e_k |P^\omega_{[a,b]}U_\omega^n e_j\ket|) \nonumber \\
& = & \sum_{\substack{k\in {\mathbb Z}^d\\
|k| > r}}{\mathbb E}(\sup_n |\bra e_k |P^\omega_{[a,b]}U_\omega^n
e_j\ket|) \nonumber \\
& \leq & \tilde{C} \sum_{\substack{k\in {\mathbb Z}^d\\
|k| > r}}e^{-\alpha |k-j|/4} \leq  \tilde{C} e^{\alpha
|j|/4} \sum_{\substack{k\in {\mathbb Z}^d\\ |k| > r}}e^{-\alpha
|k|/4},
\eea
where the right hand side decays exponentially fast to
zero as $r\ra\infty$. As a consequence,
\be
{\mathbb E}(\lim_{r\ra\infty}\sup_n\|(I -P_r)U_\omega^n
P^\omega_{[a,b]}e_j\|)=0,
\ee
so that there exists a set
$\Omega_j\subset \Omega$ such that ${\mathbb P}(\Omega_j)=1$ and for
all $\omega\in \Omega_j $
\be
\lim_{r\ra\infty}\sup_n\|(I
-P_r)U_\omega^n P^\omega_{[a,b]}e_j\|=0.
\ee
In other words, $P^\omega_{[a,b]}e_j\in {\mathcal H}_{pp}(U_\omega), \ \ \forall
\omega\in \Omega_j. $ Then, $\tilde \Omega=\cap_{j\in {\mathbb
Z}^d}\Omega_j$ is a set of probability one such that for all $\omega
\in\tilde \Omega$, $ P^\omega_{[a,b]}e_j\in {\mathcal
H}_{pp}(U_\omega), \ \ \forall j\in {\mathbb Z}^d. $ Hence,
\be
P^\omega_{[a,b]} l^2({\mathbb Z}^d)\subset{\mathcal H}_{pp}(U_\omega),
\ \ \mbox{a.s.}
\ee
\ep

\medskip
{\bf Proof of Proposition \ref{physloc}}\\
By assumption, $\psi=\sum_k \psi_k e_k$ satisfies $\psi_k=0$ if $|k|>R$, for some $R>0$. Hence, by Cauchy-Schwarz
\bea
\||X|_e^pU_\omega^nP_{[a,b]}^\omega\psi\|^2 & = & \sum_j |\bra e_j||X|_e^pU_\omega^nP_{[a,b]}^\omega\psi\ket|^2 \nonumber \\
& = & \sum_j |j|_e^{2p}\left| \sum_k\bra e_j|U_\omega^nP_{[a,b]}^\omega e_k\ket\psi_k \right|^2 \nonumber \\
& \leq & \sum_{j}\sum_{|k|\leq R} |j|_e^{2p}|\bra e_j|U_\omega^nP_{[a,b]}^\omega e_k\ket|^2\|\psi\|^2 \nonumber \\ & \leq & \sum_{j}\sum_{|k|\leq R} |j|_e^{2p}|\bra e_j|U_\omega^nP_{[a,b]}^\omega e_k\ket|\|\psi\|^2,
\eea
since $|\bra e_j|U_\omega^nP_{[a,b]}^\omega e_k\ket|\leq 1$. By the same steps as those performed in the previous proof, one gets that
\be
{\mathbb E}(\sup_{n\in {\mathbb Z}^d}\||X|_e^pU_\omega^nP_{[a,b]}^\omega\psi\|)<\infty \ \ \mbox{if } \ \
 \sum_{j}\sum_{|k|\leq R} |j|_e^{2p}{\mathbb E}(\sup_{n}|\bra e_j|U_\omega^nP_{[a,b]}^\omega e_k\ket|)<\infty.
\ee
That the latter sum is finite follows from (\ref{eq:dynamicallocalization}), which ends the proof.\ep

\setcounter{equation}{0}
\section{One-Dimensional Localization}\label{sec:emanproof}

In this section we prove Theorem~\ref{thm:eman}.

\subsection{Basic Properties of the One-Dimensional Model} \label{sec:basic1D}

The proof of Theorem~\ref{thm:eman} uses a number of tools, which are specific to the one-dimensional model, where the operators studied here can be considered as unitary analogs of Jacobi matrices. We start this section by briefly presenting the properties which we will need. While we have a somewhat different point of view, much of this is related or equivalent to facts on CMV-matrices, which can be found throughout \cite{Simon1, Simon2}.

Due to the specific band structure of $U_{\omega}$, the generalized eigenvectors can be
studied using complex $2\times 2$-transfer matrices. Specifically, generalized eigenvectors are
solutions (not necessarily in $l^2$) of the eigenvalue equation
\be
U_{\omega}\psi=z\psi, \quad \psi=\sum_{k\in\Z} \psi_k e_k,
\ee
for $z\in\C\backslash\{0\}$ and characterized by
the relations
\be \label{eigeneq}
\begin{pmatrix} \psi_{2k+1} \\
\psi_{2k+2} \end{pmatrix}= T_z(\theta_{2k}^\omega,\theta_{2k+1}^\omega)
\begin{pmatrix} \psi_{2k-1} \\ \psi_{2k} \end{pmatrix},
\ee
for all $k\in\Z$. Here the transfer matrices
$T_z:\mathbb{T}^2\rightarrow GL(2,\C)$ are defined by
\be
\label{transfer matrices}
T_z(\theta, \eta) = \begin{pmatrix}
-\dfrac{e^{-i\eta}}{z} &
\dfrac{r}{t}\left(e^{i(\theta-\eta)} -\dfrac{e^{-i\eta}}{z}\right) \\
\dfrac{r}{t}\left(1-\dfrac{e^{-i\eta}}{z}\right) &
-\dfrac{ze^{i\theta}}{t^2}
+\dfrac{r^2}{t^2}\left(1+e^{i(\theta-\eta)}-\dfrac{e^{-i\eta}}{z}\right)
\end{pmatrix}.
\ee
Note that det\,$T_z(\theta_{2k}^\omega,\theta_{2k+1}^\omega)=e^{i(\theta_{2k}^\omega-\theta_{2k+1}^\omega)}$
has modulus one and is independent of $z$. We have for any $n\in\N$
\be \label{cocycle}
\begin{pmatrix} \psi_{2n-1} \cr
\psi_{2n} \end{pmatrix}  = T_z(\theta_{2(n-1)}^\omega,\theta_{2(n-1)+1}^\omega)\cdots
T_z(\theta_{0}^\omega,\theta_{1}^\omega)\begin{pmatrix} \psi_{-1} \cr
\psi_{0} \end{pmatrix}
 \equiv T_z(\omega,n)\begin{pmatrix} \psi_{-1} \cr \psi_{0} \end{pmatrix},
 \ee
 \be
\begin{pmatrix} \psi_{-2n-1} \cr
\psi_{-2n}\end{pmatrix} =T_z(\theta_{-2n}^\omega,\theta_{-2n+1}^\omega)^{-1}\cdots
T_z(\theta_{-2}^\omega,\theta_{-1}^\omega)^{-1}\begin{pmatrix} \psi_{-1}
\cr \psi_{0} \end{pmatrix} \equiv T_{z}(\omega,-n)\begin{pmatrix} \psi_{-1}
\cr \psi_{0} \end{pmatrix}.
\ee
 We also set
$T_z(\omega,0)=I$.

The transfer matrix formalism allows to introduce the Lyapunov exponent
$\gamma(z)$;
\be
\gamma(z)=\lim_{n\to\infty}{\dfrac{\E(\ln{||T_z(\omega,n)||})}{n}}\;.
\ee
Positivity and continuity of the Lyapunov exponent for
all values of $z$ under the current assumptions was proven in
\cite{HS}. A consequence of these properties of $\gamma(z)$
is the following unitary version of Lemma 5.1 of~\cite{CKM}.

\begin{lem}\label{lemma:ckm}
Assume that the random variables $\{\theta_k\}_{k\in \Z^d}$ satisfy
(\ref{eq:mudensity}), then for each compact subset $\Lambda$ of $\C$ there exist $\alpha=\alpha(\Lambda)>0$
and $0<\delta=\delta(\Lambda)<1$ and $C=C(\Lambda)<\infty$ such that
\be
\E[||T_z(\omega,n)v||^{-\delta}]\leq Ce^{-\alpha n}
\ee
for all $z\in \Lambda$, $n\geq 0$ and unit vector $v\in \C^2$.
\end{lem}

We omit the proof of this lemma which is very similiar to the one given for the
self-adjoint Anderson model in~\cite{CKM}, see Appendix~A in
\cite{Hamza} for details.

\vspace{0.3cm}

We will frequently work with restrictions of the infinite matrices $U_\omega$ to discrete finite intervals $[a,b]$ or half-lines $[a,\infty)$ and $(-\infty,b]$, respectively. Here we slightly abuse notation and write, for example, $[a,b]$ for $[a,b]\cap \Z$. To guarantee that the restrictions remain unitary we have to choose suitable boundary conditions. At each finite endpoint these boundary conditions can be labeled by a parameter in $\T$.

For $a\in \Z$ and $\eta\in\T$, the unitary operator
$S_\eta^{[a,\infty)}$ on $l^2([a,\infty))$ is constructed as follows:
If $a=2n$ is even, let the $2\times
2$ matrices $B_1$ and $B_2$ be defined as in \eqref{B1B2}, then let
$U_e^{[2n,\infty)}$ be the unitary operator in $l^2([2n,\infty))$
found as the direct sum of identical $B_1$-blocks with blocks
starting at $2n$. On the other hand construct $U_o^{[2n,\infty)}$ starting
with a single $1\times1$ block $e^{i\eta}$, then identical $B_2$-blocks
starting at $2n+1$. Now let
$S_\eta^{[2n,\infty)}=U_e^{[2n,\infty)}U_o^{[2n,\infty)}$. The
operator $S_\eta^{[2n,\infty)}$ on $l^2([2n,\infty))$ will have a
band structure
\be
S_\eta^{[2n,\infty)}={\begin{pmatrix} re^{i\eta}
& rt & -t^2& \cr
               -te^{i\eta} & r^2& -rt& \cr
               & rt & r^2 & rt & -t^2& \cr
              & -t^2 &-tr & r^2& -rt& \cr
              & & & rt &r^2 & \cr
              & & & -t^2& -tr&\ddots \end{pmatrix}}.
\ee
The parameter $\eta$ can be thought of as a
boundary condition at $2n$.

Similarly, $U_e^{(-\infty,2n+1]}$ is
found as the direct sum of identical $B_1$-blocks with blocks
starting at even indices, while $U_o^{(-\infty,2n+1]}$ has
identical $B_2$-blocks starting at odd indices, with
$(U_o^{(-\infty,2n+1]})(2n+1,2n+1)=e^{i\eta}$. Thus,
\be
S_\eta^{(-\infty,2n+1]}=  {\begin{pmatrix}\ddots & rt & -t^2& & &
\cr
              & r^2& -rt  & &  \cr
              & rt & r^2 & rt & -t^2\cr
              & -t^2 &-rt & r^2& -rt\cr
              & & & rt &r^2 & te^{i\eta}\cr
              & & & -t^2& -rt & re^{i\eta}\end{pmatrix}}.
\ee

To define $S_\eta^{(-\infty,2n]}$ and $S_\eta^{[2n+1,\infty)}$ we slightly modify this construction,
this time filling up $U_e^{[2n+1,\infty)}$ and $U_e^{(-\infty,2n]}$ with a $1\times 1$-block $e^{i\eta}$, respectively, yielding
\be
S_\eta^{(-\infty,2n]}=
{\begin{pmatrix}\ddots & rt & -t^2& &  \cr
              & r^2& -rt  & & \cr
              & rt & r^2 & rt & -t^2\cr
              & -t^2 &-rt & r^2& -rt\cr
              & & & te^{i\eta} &re^{i\eta}\end{pmatrix}},
\ee
while
\be
S_\eta^{[2n+1,\infty)}={\begin{pmatrix}  re^{i\eta} &
-te^{i\eta}& \cr
               rt & r^2 & rt & -t^2& \cr
               -t^2 &-rt & r^2& -rt& \cr
               & & rt &r^2 & \cr
              & & -t^2& -rt&\ddots \end{pmatrix}}.
\ee

In similar fashion, for integers $-\infty< a<b< \infty$ we can
construct the unitary operator $S_{\eta_a,\eta_b}^{[a,b]}$ with
$\eta_a$ boundary condition at $a$ and $\eta_b$ boundary condition
at $b$, for example we have
\be S_{\eta_{2n},\eta_{2m}}^{[2n,2m]}= {\begin{pmatrix}re^{i\eta_{2n}} & rt &
-t^2& & \cr
             -te^{i\eta_{2n}} & r^2& -rt  & & \cr
              & rt & r^2 & rt & -t^2 & \cr
              & -t^2 &-rt & r^2& -rt & \cr
              & & & & \ddots\cr
  & & & & te^{i\eta_{2m}} & re^{i\eta_{2m}}\end{pmatrix}}.\ee
 Finally, we define
 \begin{align}\label{Ufinite} U^{[a,b]}_{\omega,\eta_a,\eta_b}=D^{[a,b]}_\omega
 S_{\eta_a,\eta_b}^{[a,b]},\end{align}
 where the diagonal operator $D^{[a,b]}_\omega$ on $l^2([a,b])$ is defined as in \eqref{highdimeD}. Similarly, we define $U_{\omega,\eta}^{[a,\infty)}$ and $U_{\omega,\eta}^{(-\infty,b]}$.

As before, the generalized eigenvectors of $U^{[a,b]}_{\omega,\eta_a,\eta_b}$, $U_{\omega,\eta}^{[a,\infty)}$ and $U_{\omega,\eta}^{(-\infty,b]}$ are characterized by
the relations \eqref{eigeneq} are supplemented with appropriate relations to reflect the boundary conditions, see~\cite{Hamza} for details. In the proof of Theorem~\ref{thm:eman} we only use $\eta=0$, so for the rest of this section we write
$U^{[a,b]}_{\omega,0,0}=U^{[a,b]}$, $U_{\omega,0}^{[a,\infty)} = U^{[a,\infty)}$ and $U_{\omega,0}^{(-\infty,b]} = U^{(-\infty,b]}$ for simplicity, also frequently leaving the $\omega$-dependence implicit. Other boundary conditions will be used later, see Section~\ref{sec:Neumann}.

\vspace{.3cm}

In the following discussion we use the notation $U^{[a,b]}$ for general $-\infty \le a < b \le \infty$, i.e.\ we write $U^{[-\infty,\infty]}$, $U^{[a,\infty]}$ and $U^{[-\infty,b]}$ for $U$, $U^{[a,\infty)}$ and $U^{(-\infty,b]}$.

Another feature of the one-dimensional model is that Green's function
$G(k,l;z)$ can be expressed in terms of two generalized eigenfunctions to $z$ which, separately at each endpoint $a$, $b$, are square-summable or satisfy the boundary condition at the endpoint.

For a solution $\ffi$ of $(U-z)\ffi=0$, we define $\tfe$ by
\be\label{tfeproperty}
\begin{pmatrix} \tfe_{2n}\\
\tfe_{2n+1}\end{pmatrix}=
\begin{pmatrix}t^2&rt\\ rt&r^2-ze^{i\theta_{2n}}\end{pmatrix}
\begin{pmatrix}  \ffi_{2n-1}\\ \ffi_{2n}\end{pmatrix}.
\ee
 A straightforward calculation shows that $\tfe$
 is characterized by the relations
\be\label{tfetransfer}
\begin{pmatrix} \tfe_{2k}\\
\tfe_{2k+1}\end{pmatrix}= \widetilde{T_z}(\theta_{2k-1},\theta_{2k})
\begin{pmatrix}  \tfe_{2k-2}\\
\tfe_{2k-1}\end{pmatrix}, \ee for all $k\in\Z$. Here the transfer
matrices $\widetilde{T_z}:\T^2\to GL(2,\C)$ are defined by
\be\label{tfetransfermatrix}
\widetilde{T_z}(\theta,\eta)=T^t_z(\eta,\theta),
\ee
where $T_z(\eta,\theta)$ is given by \eqref{transfer matrices} and $T^t$ denotes the transpose of $T$.

For $-\infty \le a < b \le \infty$ let
\be\label{finiteresolvent}
G^{[a,b]}(z) = (U^{[a,b]}-z)^{-1}.
\ee
To any $z$ not in the spectrum of $U^{[a,b]}$ choose generalized eigenvectors $\varphi^a$ and $\varphi^b$ as follows:

If $a$ is finite, then $\varphi^a$ is the unique solution of $(U^{[a,\infty)}-z)\varphi^a=0$ with $\varphi^a(a)=1$, i.e.\ a generalized eigenvector to $z$ which satisfies the boundary condition at $a$. If $a=-\infty$, then for $\varphi^a$ we choose a non-trivial solution of $(U-z)\varphi^a=0$, which is square-summable at $-\infty$. In the latter case $\varphi^a$ is determined up to a constant (one can construct if from the tail of $(U-z)^{-1} e_0$ at $-\infty$ and the fact that the transfer matrices (\ref{transfer matrices}) have determinant of modulus one shows that there can't be two linearly independent solutions which are square-summable at $-\infty$). Similar we choose $\varphi^b$ with prescribed boundary behavior at $b$.

The following proposition
gives an expression of the elements of $G^{[a,b]}(z)$ in terms of
$\varphi^a$ and $\varphi^b$ and the corresponding $\tfe^a$, $\tfe^b$
defined as in \eqref{tfeproperty}.

\begin{prop}\label{thm:greeninsol}
For all finite $k$, $l$ with $a\leq k,l\leq b$, if $l=2n$ or $l=2n+1$,
\be\label{eq:greeninsol}
G^{[a,b]}(k,l;z) = \begin{cases}
c_l \tfe^b_l\ffi^a_k& \text{if } k<l \text{ or } k=l \text{ are even},
\\ c_l \tfe^a_l\ffi^b_k& \text{if } k>l\text{ or } k=l \text{ are odd},
\end{cases}
\ee where
$c_l=\dfrac{e^{i\theta_l}}{\tfe^a_{2n+1}\tfe^b_{2n}-\tfe^a_{2n}\tfe^b_{2n+1}}$.
\end{prop}

{\bf Proof:} A straightforward, if rather tedious, calculation shows that the matrix
whose entries are given by the right hand side of \eqref{eq:greeninsol} is indeed the
inverse of $U_{\omega}^{[a,b]}-z$, see \cite{Hamza} for details. \ep

\vspace{.3cm}

We conclude this section by proving the following lemma that is used later in the proof of Theorem~\ref{thm:eman}.

\begin{lem}\label{lem:boundbynorm}
 For $z\in \C$ with $0<||z|-1|<1/2$, $a\in\Z$ and $s\in(0,1)$ there exists $0<C_\mu(t,s)<\infty$ such that
\be \int_0^{2\pi}
d\mu(\theta_{2m})\frac{1}{|t\ffi^a_{2m-1}+(r-ze^{i\theta_{2m}})\ffi^a_{2m}|^s}
\leq C_\mu(t,s)\Big\|\begin{pmatrix}\ffi^a_{2m-1}\\\ffi^a_{2m}
\end{pmatrix}\Big\|^{-s},\ee
for all $m\geq a+2$.
\end{lem}
{\bf Proof:}
First note that both $\ffi^a_{2m-1}$ and $\ffi^a_{2m}$ are
 independent of $\theta_{2m}$ and can
not vanish simultaneously. This follows from the fact that the transfer matrices needed to construct them via (\ref{eigeneq}) from $\varphi^a(a)=1$ only contain $\theta_a, \ldots, \theta_{2m-1}$.

Therefore we have the following cases:

\textbf{Case 1}: $\ffi^a_{2m}=0$, using that
$\Big\|\begin{pmatrix}\ffi^a_{2m-1}\\\ffi^a_{2m}
\end{pmatrix}\Big\|\leq 2|\ffi^a_{2m-1}|+2|\ffi^a_{2m}|$,  the bound follows directly.

\textbf{Case 2}:  $\ffi^a_{2m}\neq0$. In this case
\bea
  \lefteqn{\int_0^{2\pi}
d\mu(\theta_{2m})\frac{1}{|t\ffi^a_{2m-1}+(r-ze^{i\theta_{2m}})\ffi^a_{2m}|^s}} \nonumber \\
& = & \frac{1}{|\ffi^a_{2m}|^s}\int_0^{2\pi}
d\mu(\theta_{2m})\frac{1}{|t\frac{\ffi^a_{2m-1}}{\ffi^a_{2m}}+(r-ze^{i\theta_{2m}})|^s}.
\eea
 Let $M=\sup\{|r-ze^{i\theta_{2m}}|:\theta\in[0,2\pi],
 0<\big||z|-1\big|<1/2\}<\infty$ and distinguish between
 two subcases;

\textbf{Case 2a}: If
$t\Big|\frac{\ffi^a_{2m-1}}{\ffi^a_{2m}}\Big|>2M$, it follows
that
\be
\Big|t\frac{\ffi^a_{2m-1}}{\ffi^a_{2m}}+(r-ze^{i\theta_{2m}})\Big|
>\frac{t}{2}\Big|\frac{\ffi^a_{2m-1}}{\ffi^a_{2m}}\Big|.
\ee
On the other hand
\be
\Big\|\begin{pmatrix}\ffi^a_{2m-1}\\\ffi^a_{2m}
\end{pmatrix}\Big\|^s\leq (2+\frac1M)^s|\ffi^a_{2m-1}|^s.
\ee
Using these estimates we conclude that
\be
  \int_0^{2\pi}
d\mu(\theta_{2m})\frac{1}{|t\ffi^a_{2m-1}+(r-ze^{i\theta_{2m}})\ffi^a_{2m}|^s}\leq
\frac{2^{s+1}\pi||\tau||_\infty}{t^s}(2+\frac1M)^s\Big\|\begin{pmatrix}\ffi^a_{2m-1}\\\ffi^a_{2m}
\end{pmatrix}\Big\|^{-s}.
\ee

\textbf{Case 2b}: Assume that
$t\Big|\frac{\ffi^a_{2m-1}}{\ffi^a_{2m}}\Big|\leq 2M$.
For all $0<s<1$ there
exists $0<C_\mu(s)<\infty$ such that for all $\beta\in\C$
\be \label{decoupling}
\int_\T d\mu(\theta)\frac{1}{|(e^{\pm
i\theta}-\beta)|^{s}}\leq C_\mu(s),
\ee
see e.g.\ \cite{J2}. Using this we get the bound
\be
  \int_0^{2\pi}
d\mu(\theta_{2m})\frac{1}{|t\ffi^a_{2m-1}+(r-z e^{i\theta_{2m}})\ffi^a_{2m}|^s}
\leq \frac{2^s C_\mu^{(1)}(s)}{|\ffi^a_{2m}|^s}.
\ee
 However, under the current
assumption we have
\be
\Big\|\begin{pmatrix}\ffi^a_{2m-1}\\\ffi^a_{2m}
\end{pmatrix}\Big\|^s\leq (4\frac{M}{t}+2)^s|\ffi^a_{2m}|^s,
\ee
which gives the required bound. \ep

\subsection{Proof of Theorem~\ref{thm:eman}}

We now turn to the proof of Theorem \ref{thm:eman}.
The proof is done in three steps: first we prove that it suffices to
deal with even matrix elements of Green's function, i.e.\ that we only
need to show \eqref{eq:eman} for the case that $k=2n$, $l=2m$.
This mainly serves the purpose of avoiding to cover four separate sub-cases.
Then we show that
proving the bound \eqref{eq:eman} for an element $(2n,2m)$
 of $(U_\omega-z)^{-1}$ can be reduced to
proving the same bound for the element $(2n,2m)$ of the resolvent
of the finite volume operator $U^{[2n,2m]}_\omega$ at $z$. Finally,
we prove that expectation of fractional moments of
$G^{[2n,2m]}(2n,2m;z)$ decays exponentially. Based on the Green's function formula (\ref{eq:greeninsol}) this
will be found by combining Lemma~\ref{lem:boundbynorm} with Lemma~\ref{lemma:ckm}.


\vspace{0.5cm}

{\bf Step One:}
The following Lemma shows that the expectation of a fractional moment
of any element $G(k,l;z)$ can be reduced to the expectation of
fractional moments of even matrix elements of $G(z)$. This comes at the cost of enlarging the fractional exponent $s$ due to the use of H\"older's inequality.

\begin{lem} \label{reductiontoeven}
Let $s\in(0,1/4)$ and $k,l\in \Z$ such that $|k-l|>4$. Choose $n,m\in\Z$ such that $k\in\{2n,2n+1\}$, $l\in\{2m,2m+1\}$.
There exists
$\kappa(t,s,\mu)<\infty$ such that
\be
\E[|G(k,l;z)|^s]\leq \kappa(t,s,\mu)
\sum_{i,j=0}^1\big(\E[|G(2n+2i,2m-2j;z)|^{4s}]\big)^{1/4},
\ee
for all
$z\in\C$ with $0<||z|-1|<1/2$.
\end{lem}

{\bf Proof:} Using the definition of $n,m$ and that $|k-l|>4$ one
has $|n-m|\geq2$. Since $\tfe^{\pm\infty}$, defined by \eqref{tfeproperty},
satisfies \eqref{tfetransfer} and \eqref{tfetransfermatrix}, a
straightforward calculation shows that
\bea \tfe^{\pm\infty}_{2m+1} & = & \frac{1}{rt(e^{i\theta^\omega_{2m-1}}-1/z)} \big\{ [r^2(e^{i\theta^\omega_{2m}}+e^{i\theta^\omega_{2m-1}}-\frac1z) \nonumber \\ & & -ze^{i(\theta^\omega_{2m}+\theta^\omega_{2m-1})}]\tfe^{\pm\infty}_{2m}
-t^2e^{i\theta^\omega_{2m}}\tfe^{\pm\infty}_{2m-2}\big\}.
\eea
Using Theorem~\ref{thm:greeninsol} along with \eqref{tfetransfer} it follows that
for $k\notin\{2m-1,2m\}$
\bea
G(k,2m+1;z) & = & \frac{e^{i\theta^\omega_{2m+1}}}{rt(e^{i\theta^\omega_{2m-1}}-1/z)} \big\{ [r^2(1+e^{-i(\theta^\omega_{2m}-\theta^\omega_{2m-1})}-\frac{e^{-i\theta^\omega_{2m}}}{z})-ze^{i\theta^\omega_{2m-1}}] \nonumber \\ & & G(k,2m;z)
-t^2e^{i(\theta^\omega_{2m-1}-\theta^\omega_{2m-2})}G(k,2m-2;z)\big\}.
\eea
 By H\"{o}lder's inequality and \eqref{decoupling} it follows that for
$s\in(0,1/4)$ there exists $0<C_\mu^{(1)}(s,r)<\infty$ such that
\bea \label{secondodd}
\lefteqn{\E[|G(k,2m+1;z)|^s]} \nonumber \\ & \leq &
C_\mu^{(1)}(s,r) \left((\E[|G(k,2m;z)|^{2s}])^{1/2} +(\E[|G(k,2m-2;z)|^{2s}])^{1/2}\right).
\eea
 Similarly, using
\eqref{eigeneq} and \eqref{transfer matrices} we obtain
\be
\varphi^{\pm\infty}_{2n+1}=\frac{-t}{r(e^{i\theta^\omega_{2n+1}}-1/z)}\big\{\frac1z\varphi^{\pm\infty}_{2n+2}+e^{i\theta^\omega_{2n}}\varphi^{\pm\infty}_{2n}\big\}.
\ee
Thus, for $l\notin\{2n,2n+1\}$,  $s\in(0,1/2)$ and all $z\in\C$ with $0<||z|-1|<1/2$,
\begin{eqnarray}\label{firstodd}
\lefteqn{\E[|G(2n+1,l;z)|^s]} \nonumber \\ & \leq &
C_\mu^{(2)}(s,r) \left((\E[|G(2n+2,l;z)|^{2s}])^{1/2}
+(\E[|G(2n,l;z)|^{2s}])^{1/2}\right).
\end{eqnarray}
It readily follows
from \eqref{secondodd} and \eqref{firstodd} that for
$|n-m|\notin\{0,1\}$ and all $s\in(0,1/4)$
\be \label{bothodd}
\E[|G(2n+1,2m+1;z)|^s] \leq
\tilde{\kappa}_\mu^{(1)}(s,r)\sum_{i,j=0}^1\big(\E[|G(2n+2i,2m-2j;z)|^{4s}]\big)^{1/4}.
\ee
This proves the Lemma for the case $k=2n+1$, $l=2m+1$. The other cases are more direct.
\ep

\vspace{0.5cm}

{\bf Step Two:}
Let $\Big|[k,l]\Big|$ denote the
interval $\big[\min\{k,l\},\max\{k,l\}\big]$. In what follows we
show that the expectation of fractional moments of $G(2n,2m;z)$ can
be reduced to that of $G^{\big|[2n,2m]\big|}(2n,2m;z)$.

\begin{lem}\label{lm:reductiontofinitevol}
For $s\in(0,1/3)$ and $n,m\in\Z$ with $|n-m|\geq 2$, we have
\be \label{toexpoffinite}
\E[|G(2n,2m;z)|^s] \leq C_\mu(t,s)(\E[|G^{|[2n,2m]|}(2n,2m;z)|^{3s}])^{1/3},
\ee
for all $z\in \C$ with $0<||z|-1|<1/2$.
\end{lem}

{\bf Proof:} For definiteness, assume that $m\geq n+2$, the case $n\geq m+2$ being similar.
 Using the definition of $U^{[x,y]}_\omega$ \eqref{Ufinite},
we see that
\be
U_\omega=U^{(-\infty,2n-1]}_\omega\oplus U^{[2n,\infty)}_\omega+\Gamma^e_n,
\ee
where $\Gamma^e_n$ is given by
\be
\Gamma^e_n(k,l)=\begin{cases}(rt-t)e^{-i\theta_{2n-2}}, &
k=2n-2,l=2n-1
\\-t^2e^{-i\theta_{2n-2}},& k=2n-2,l=2n
\\ (r^2-r)e^{-i\theta_{2n-1}}, & k=2n-1,l=2n-1
\\-rte^{-i\theta_{2n-1}}, &k=2n-1,l=2n
\\rte^{-i\theta_{2n}}, & k=2n, l=2n-1
\\(r^2-r)e^{-i\theta_{2n}},& k=2n, l=2n
\\ -t^2e^{-i\theta_{2n+1}}, & k=2n+1,l=2n-1
\\(-rt+t)e^{-i\theta_{2n+1}},& k=2n+1,l=2n
\\0, & \text{otherwise}.
\end{cases}
\ee
Denote $G^{n}(z)=G^{(-\infty,2n-1]}(z)\oplus G^{[2n,\infty)}(z)$. By
the first resolvent identity, we have
\be
G(z)-G^{n}(z)=-G(z)\Gamma^e_n G^{n}(z).
\ee
Therefore, it follows for all $m\geq n+2$ that
\bea \label{semifinite}
G(2n,2m;z) & = & \{1+t^2e^{-i\theta_{2n-2}}G(2n,2n-2;z)+rte^{-i\theta_{2n-1}}G(2n,2n-1;z)\nonumber
\\ & & -(r^2-r)e^{-i\theta_{2n}}G(2n,2n;z)-(t-rt)e^{-i\theta_{2n+1}}G(2n,2n+1;z)\} \nonumber \\ & & G^{[2n,\infty)}(2n,2m;z).
\eea

A similar application of the first resolvent identity, this time to the difference $G^{[2n,\infty)}(z)-G^{[2n,2m]}(z) \oplus G^{[2m+1,\infty)}(z)$, allows to express $G^{[2n,\infty)}(2n,2m;z)$ in terms of $G^{[2n,2m]}(2n,2m;z)$ and ultimately leads to
\bea \label{reductiontofinite}
G(2n,2m;z) & = & \Big\{1+t^2e^{-i\theta_{2n-2}}G(2n,2n-2;z)+rte^{-i\theta_{2n-1}}G(2n,2n-1;z)\nonumber
\\ & & -(r^2-r)e^{-i\theta_{2n}}G(2n,2n;z)
-(t-rt)e^{-i\theta_{2n+1}}G(2n,2n+1;z)\Big\}\nonumber
\\ & & \times \Big\{ 1-e^{-i\theta_{2m}}[(rt-t)G^{[2n,\infty)}(2m-1,2m;z)+ \nonumber \\ & & (r^2-r)G^{[2n,\infty)}(2m,2m;z)
+rtG^{[2n,\infty)}(2m+1,2m;z) \nonumber \\ & & -t^2G^{[2n,\infty)}(2m+2,2m;z)] \Big\} G^{[2n,2m]}(2n,2m;z).
\eea
If $A$ and $B$ denote the two $\{\cdot\}$-factors on the right hand side of (\ref{reductiontofinite}), then it follows from $s<1/2$ and Theorem~\ref{thm:fmbound} that
\be \label{eq:twofactors}
\E(|A|^{3s}) \le C, \quad \E(|B|^{3s}) \le C
\ee
uniformly in $|z|\not= 1$. Here we are using that Theorem~\ref{thm:fmbound} remains true with identical proof for the Green function of $U_{\omega}^{[2n,\infty)}$. An application of H\"older's inequality to (\ref{reductiontofinite}) yields
\be
\E[|G(2n,2m;z)|^s] \le (\E[|A|^{3s}])^{1/3} (\E[|B|^{3s}])^{1/3} \left( \E[|G^{[2n,2m]}(2n,2m;z)|^{3s}] \right)^{1/3},
\ee
which gives (\ref{toexpoffinite}) when combined with (\ref{eq:twofactors}).
  \ep


\vspace{0.5cm}

{\bf Step Three:}
Now that we have reduced the problem to dealing with fractional
moments of elements of the form $G^{|[2n,2m]|}_z(2n,2m)$, we show
that expectations of such objects decay exponentially, in particular
we show;
\begin{lem}\label{decayoffinitemom}
Assume that $\{\theta^\omega_k\}_{k\in\Z}$ are i.i.d. with
probability measure $d\mu(\theta)=\tau(\theta)d\theta$, where
$\tau\in L^\infty(\T)$. There exist $s_0\in(0,1)$, $0<C_1<\infty$,
$\alpha_1>0$ such that
\be
\E[|G^{|[2n,2m]|}(2n,2m;z)|^{s_0}] \leq
C_1 e^{-\alpha_1|m-n|},
\ee
for all $z\in\C$ with $0<||z|-1|<1/2$, and all
 $m,n\in\Z$ such that $|m-n|\geq 2$.
\end{lem}
 {\bf Proof:}
For $m\geq n+2$, let $\ffi^{2n}$ and $\ffi^{2m}$ be
two solutions that satisfy the boundary conditions at
$2n$ and $2m$, respectively, such that $\ffi^{2n}_{2n}=1$ and
$\ffi^{2m}_{2m}=1$. Using \eqref{eq:greeninsol}, we have
\be
G^{[2n,2m]}(2n,2m;z)=\frac{e^{i\theta_{2m}}}{\tfe^{2n}_{2m+1}\tfe^{2m}_{2m}-\tfe^{2n}_{2m}\tfe^{2m}_{2m+1}}\tfe^{2m}_{2m}.
\ee
Since $\ffi^{2m}$ satisfies the boundary condition at $2m$, it follows that
\be
\tfe^{2m}_{2m}=tze^{i\theta^\omega_{2m}},
\ee
\be
\tfe^{2m}_{2m+1}=(r-1)ze^{i\theta^\omega_{2m}}.
\ee
 Using this along with the definition of $\tfe^{2n}$, we obtain
\be
G^{[2n,2m]}(2n,2m;z)
=\frac{e^{i\theta_{2m}}}{t\ffi^{2n}_{2m-1}+(r-ze^{i\theta^\omega_{2m}})\ffi^{2n}_{2m}}.
\ee
Now, for $s\in(0,1)$ the expectation of the s-moment of
$G^{[2n,2m]}(2n,2m;z)$ is given by
\be\label{exptoint}
\E[|G^{[2n,2m]}(2n,2m;z)|^s]=\widehat{\E}\left[
\int_0^{2\pi}
d\mu(\theta_{2m})\frac{1}{|t\ffi^{2n}_{2m-1}+(r-ze^{i\theta^\omega_{2m}})\ffi^{2n}_{2m}|^s}\right],
\ee
where $\widehat{\E}$ is the expectation with respect to the random
variables $\{\theta^\omega_k\}_{k\in\Z\backslash\{2m\}}$. By Lemma \ref{lem:boundbynorm}, we have
\be
\E(|G^{[2n,2m]}_z(2n,2m)|^s)\leq C_\mu(s,t)\E\bigg[\bigg\|T_z(\omega,m-n)\begin{pmatrix}\ffi^{2n}_{2n-1}\\\ffi^{2n}_{2n}
\end{pmatrix}\bigg\|^{-s}\bigg].
\ee

Using Lemma \ref{lemma:ckm},
it follows that there exist $\alpha_1>0$ and
 $s_0\in(0,1)$ and $\widetilde{C}^{(1)}_\mu(s_0,t)<\infty$ such that
\be
 \E[|G^{[2n,2m]}_z(2n,2m)|^{s_0}] \leq
\widetilde{C}^{(1)}_\mu(s_0,t)e^{-\alpha_1(m-n)},
\ee
for all $z\in \C$ with $0<||z|-1|<1/2$, $0<\epsilon<1/2$
and $m,n\in\Z$ such that $m-n\geq 2$.

In the case $n\geq m+2$ let
$\psi^{2n}$ and $\psi^{2m}$ be solutions of $(U-z)\psi=0$ that
satisfy the boundary condition at $2n$ and $2m$,
respectively, with $\psi^{2n}_{2n}=\psi^{2m}_{2m}=1$. Using
\eqref{eq:greeninsol}, \eqref{tfeproperty}, we obtain that
\be
G^{[2m,2n]}(2n,2m;z)=\frac{1}{z(t\psi^{2n}_{2m-1}+(1-r)\psi^{2n}_{2m})}.
\ee
In order to bound the expectation we first integrate with respect to $\theta^\omega_{2m-1}$ and use the same procedure as before. \ep

\vspace{0.5 cm}

{\bf Proof of Theorem \ref{thm:eman}:} Without restriction we may assume $|k-l|>4$ (as (\ref{eq:eman}) for $|k-l|\le 4$ only requires the a-priori bound (\ref{eq:fmbound})). Pick $m,n\in\Z$ such that
$k\in\{2n,2n+1\}$, $l\in\{2m,2m+1\}$ and $|m-n|>1$. Thus using the
results of Lemma \ref{reductiontoeven} and Lemma~\ref{lm:reductiontofinitevol}, there exists $0<
\kappa(t,s,\mu)<\infty$ such that
\be
\E[|G(k,l;z)|^s]
\leq\kappa(t,s,\mu)
\sum_{i,j=0}^1\big(\E[|G^{|[2n+2i,2m-2j]|}(2n+2i,2m-2j;z)|^{12s}]\big)^{1/12}.
\ee

 Next the result of Lemma \ref{decayoffinitemom} gives that there exist $s_0\in(0,1)$,
$0<\widetilde{C}_1<\infty$, $\alpha>0$ such that
\be
\E[|G(k,l;z)|^{s_0}]  \leq \widetilde{C}^{(1)}\kappa(t,s_0,\mu)
\sum_{i,j=0}^1e^{-\alpha|2n-2m+2i+2j|}.
\ee
Using the definition of $m,n$ this yields (\ref{eq:eman}).
\ep

\setcounter{equation}{0}
\section{Neumann Boundary Conditions} \label{sec:Neumann}

In this section we study in detail the properties of one of the finite volume restrictions of the ``free'' unitary operator $S$ introduced in Section~\ref{sec:basic1D}. This particular restricted operator will share many of the properties of selfadjoint Neumann Laplacians. We will therefore think of them as restrictions of $S$ to finite volume with Neumann boundary conditions.

\subsection{Neumann Boundary Conditions for $d=1$} \label{sec:neumann1d}

Let $e^{i\eta}=r+it$, $L\in \N$, and on $l^2([0,2L-1])$
define
\be
\label{nbc}
S_N^{[0,2L-1]}= \left(\begin{array}{cccccccccc} re^{i\eta} &
rt & -t^2 & & & & & & & \\
             -te^{i\eta} & r^2& -rt  & & & & & & & \\
              & rt & r^2 & rt & -t^2 & & & & & \\
              & -t^2 &-rt & r^2& -rt & & & & & \\
              & & & & \ddots & & & & & \\
  & & & & & rt & r^2 & rt & -t^2 & \\
            & & &  & & -t^2 &-rt & r^2& -rt & \\
             & & & & & & & rt &r^2 & te^{i\eta} \\
              & & & & & & & -t^2& -rt &
              re^{i\eta}\end{array} \right),
\ee
which is a special case of the restrictions of $S_0$ to finite intervals defined in Section~\ref{sec:basic1D}. To characterize the spectrum of $S_N^{[0,2L-1]}$ define $\lambda_k = \arccos(r^2-t^2 \cos(k\pi/L))$, $k=0,1,\ldots, n$, i.e.\ in particular $\lambda_L=0$ and $\lambda_0= \arccos(r^2-t^2)$, the latter coinciding with $\lambda_0$ as introduced in Section~\ref{sec:model} and giving the band edges of $S_0$ as $e^{\pm i \lambda_0}$.

\begin{prop}\label{prop:spectrumofN} The $2L$ eigenvalues of $S_N^{[0,2L-1]}$ are non degenerate and given by
\be
\sigma(S_N^{[0,2L-1]})=
\{e^{i\lambda_0},e^{i\lambda_L}=1\}\cup\{e^{\pm i\lambda_k} :  k=1,..,L-1\}.
\ee
In particular,
$\sigma(S_N^{[0,2L-1]})\subset \sigma (S)$.
Moreover,
\be
\ffi_0=(i, 1, i, 1, \cdots, i, 1)^t
\ee
is an eigenvector associated with $e^{i\lambda_0}$ and
\be
\ffi_L=(i, 1, -i, -1, \cdots, (-1)^{L+1}i, (-1)^{L+1})^t
\ee
is an eigenvector associated with $1$.
\end{prop}

To prove Proposition~\ref{prop:spectrumofN} we will use the transfer matrix formalism, i.e.\ solutions of $U_{\omega}\psi=z\psi$ are
characterized by the relations
\be
\left( \begin{array}{c} \psi_{2k+1} \\
\psi_{2k+2}\end{array} \right) =
T_z(\theta_{2k}^\omega,\theta_{2k+1}^\omega)
\left( \begin{array}{c}\psi_{2k-1} \\ \psi_{2k}\end{array} \right),
\ee
for all $k\in\Z$, where the transfer matrices
$T_z:\T^2\rightarrow GL(2,\C)$ are defined by \eqref{transfer matrices}.
In the ``free'' case $S_0\psi=e^{i\lambda}\psi$, the transfer matrix
takes the simple form
\be
T(\lambda) := T_{e^{i\lambda}}(0,0) = \left( \begin{array}{cc}
-e^{-i\lambda}&
\frac{r}{t}\left(1-e^{-i\lambda}\right) \\
\frac{r}{t}\left(1-e^{-i\lambda}\right) &
-\frac{e^{i\lambda}}{t^2}
+\frac{r^2}{t^2}\left(2-e^{-i\lambda}\right)
\end{array} \right).
\ee

\medskip

The following lemma will be used in the proof of Proposition~\ref{prop:spectrumofN}.
\begin{lem}
The vector $(i,1)^t$ is an eigenvector of
$T(\lambda)^L$ if and only if $\lambda\in\{0,\arccos(r^2-t^2)\}
\cup\{\lambda:\cos(\lambda)= r^2-t^2\cos(k\pi/L), k=1,..,L-1\}$.
 \end{lem}

\noindent {\bf Proof:} In order to simplify the analysis we distinguish between two cases:

(i) $(i,1)^t$ is an eigenvector of
$T(\lambda)$.
A straightforward calculation shows that this is only true for $\lambda\in \{0,\arccos(r^2-t^2)\}$ with corresponding eigenvalues $\{-1, 1\}$ respectively.

(ii) The second case is when $(i,1)^t$ is
an eigenvector of $T(\lambda)^L$ but not of $T(\lambda)$, i.e.
\be
 T(\lambda)^L \left( \begin{array}{c} i\\ 1 \end{array} \right) =a \left( \begin{array}{c} i\\ 1 \end{array} \right),
 \ee
while $v=T(\lambda)(i,1)^t$ is linearly independent of $(i,1)^t$. Then it follows that
\be
 T(\lambda)^Lv= T(\lambda)^{L+1} \left( \begin{array}{c} i\\ 1 \end{array} \right) = aT(\lambda) \left( \begin{array}{c} i\\ 1 \end{array} \right)= av.
 \ee
Also since $v$ and $(i,1)^t$ are linearly independent it follows that $T(\lambda)^L=a I$. Finally using that the determinant of $T(\lambda)$ is one, we see that $a^2=1$.

Since the eigenvalues of $T(\lambda)$ are given by $e^{\pm i
\arccos((r^2-cos(\lambda))/t^2)}$, we have that $a$ is an
eigenvalue of  $T(\lambda)^L$ if and only if $e^{\pm i
2L\arccos((r^2-cos(\lambda))t^2)}=1$, i.e.
\bea
\lambda & \in &\{\cos(\lambda)=
r^2-t^2\cos(k\pi/L), k=1,..,2L-1\} \nonumber
\\ & = & \{\cos(\lambda)=
r^2-t^2\cos(k\pi/L), k=1,..,L\}.
\eea
 Combining the two cases gives the result.
\ep

\medskip

\noindent {\bf Proof of Proposition~\ref{prop:spectrumofN}:}
In light of the previous Lemma, it suffices to show that
$e^{i\lambda}\in \sigma (S_N^{[0,2L-1]})$ if and only if
$(i,1)^t$ is an eigenvector of
$T(\lambda)^L$.

First it is not hard to see that $e^{i\lambda}\in \sigma
(S_N^{[0,2L-1]})$ means the existence of $\psi\in l^2([0,2L-1])$ such
that
\be
\left( \begin{array}{c}\psi_{2m+1} \\
\psi_{2m+2}\end{array} \right) = T(\lambda)
\left( \begin{array}{c} \psi_{2m-1} \\ \psi_{2m} \end{array}\right),
\ee
for $m\in\{1,..,L-2\}$, while
\bea \label{eq:upperboundary}
 \left( \begin{array}{c} \psi_{1} \\
\psi_{2}\end{array}\right) = \psi_{0}
\left( \begin{array}{c} \frac{1}{t}(r-e^{i(\eta-\lambda)})
\\
\frac{1}{t^2}(r^2-re^{i(\eta-\lambda)}-e^{i\lambda}+re^{i\eta}) \end{array} \right),
\eea
and
\bea \label{eq:lowertboundary}
\left( \begin{array}{c} \psi_{2L-3} \\
\psi_{2L-2} \end{array} \right)= \psi_{2L-1} \left(\begin{array}{c}
\frac{1}{t^2}(r^2-re^{i(\eta-\lambda)}-e^{i\lambda}+re^{i\eta})
\\ \frac{-1}{t}(r-e^{i(\eta-\lambda)}) \end{array} \right).
\eea
Define $\tilde{\psi}_{-1}$ such that
\be
T(\lambda)^{-1} \left( \begin{array}{c}\psi_{1} \\
\psi_{2} \end{array} \right)=
\left( \begin{array}{c} \tilde{\psi}_{-1} \\ \psi_{0} \end{array} \right).
\ee
Using (\ref{eq:upperboundary}) and that $e^{i\eta}=r+it$, one can see
that this definition is equivalent to having $\tilde{\psi}_{-1}=i\psi_0$.
Similarly, defining $\tilde{\psi}_{2L}$ such that
\be
\left( \begin{array}{c} \psi_{2L-1} \\
\tilde{\psi}_{2L} \end{array} \right) = T(\lambda)
\left( \begin{array}{c} \psi_{2L-3} \\ \psi_{2L-2} \end{array} \right),
\ee
we see that $\tilde{\psi}_{2L}=-i\psi_{2L-1}$. Also by definition,
\bea
T(\lambda)^L
\left( \begin{array}{c} \tilde{\psi}_{-1} \\ \psi_{0} \end{array} \right) & = & \left( \begin{array}{c} \psi_{2L-1} \\
\tilde{\psi}_{2L} \end{array} \right) \\
T(\lambda)^L \left( \begin{array}{c} i \\ 1 \end{array} \right) & = & \frac{-i\psi_{2L-1}}{\psi_0} \left( \begin{array}{c} i \\
1 \end{array} \right).
\eea
which shows the required assertion. The last two claims of Proposition~\ref{prop:spectrumofN} follow from the above together with $T(0)(1,i)^t =-(1,i)^t$ and $T(\lambda_0)(1,i)^t = (1,i)^t$.
\ep

\medskip

We conclude this subsection with several remarks:

(i) $e^{i\arccos(r^2-t^2)}\in \sigma (S_N^{[0,2L-1]})$, while
$e^{-i\arccos(r^2-t^2)}$ is not.

(ii) For $\lambda_0=\arccos(r^2-t^2)=\arccos(1-2t^2)$,  $S_N^{[0,2L-1]}\varphi_0=e^{i\lambda_0}\varphi_0$
has solutions with $|\varphi_{0}(k)|=1$ for all $k\in [0,2L-1]$.

(iii) There is a gap between the upper edge of the spectrum of $S_N^{[0,2L-1]}$, given by
$e^{i\arccos(r^2-t^2)}$, and the next closest eigenvalue. In particular,
for any $k\in\{1,\cdots, L-1\}$ we have
\bea\label{ggap}
|e^{i\arccos(r^2-t^2)}-e^{i\arccos(r^2-t^2\cos(\pi k/L))}|&>& t^2(1-\cos({\pi k}/{L}))=2t^2\sin^2(\pi k/2L)\nonumber\\
&\geq& t^2\left(\frac{\pi k}{L}\right)^2\frac{(4-\pi)^2}{32},
\eea
using the property $\sin(x)\geq x(1-\pi/4)$ if $x\in(0,\pi/2).$

(iv) The previous remarks as well as the ``Neumann-bracketing'' property to be found in Section~\ref{sec:splitting} below will make the operators $S_N^{[0,2L-1]}$ a suitable tool when studying properties of finite volume restrictions of the unitary Anderson model near the upper edge $e^{i\lambda_0}$ of the spectrum of $S_0$. To get the corresponding results also at the lower edge $e^{-i\lambda_0}$ of the spectrum of $S_0$ one needs to modify the definition of $S_N^{[0,2L-1]}$ by setting $e^{i\eta}=r-it$ in (\ref{nbc}). In this case we use, in particular, that the vector $(-i, 1)^t$ is an eigenvector of $T(-\lambda_0)$, leading to $e^{-i\lambda_0}$ becoming an eigenvalue of the restricted operator. One gets properties similar to Proposition~\ref{prop:spectrumofN} and Remarks (i), (ii) and (iii) above.

\subsection{Neumann Boundary Conditions for $d> 1$}

\medskip

For a box $\Lambda := [2l_1,2m_1-1] \times \ldots \times [2l_d,2m_d-1] \subset \Z^d$ define
\be \label{eq:Neumannmultidim}
S_N^{\Lambda} = \otimes_{j=1}^d S_N^{[2l_j,2m_j-1]} \quad \mbox{on} \quad \otimes_{j=1}^d l^2([l_j,2m_j-1]) = l^2(\Lambda).
\ee
Note here that the discussion of Section~\ref{sec:neumann1d} applies with obvious modifications to intervals of the form $[2l_j,2m_j-1]$ and integers $l_j<m_j$.
We will be particularly interested in the case of cubic boxes $\Lambda_L := [-2L,2L+1]^d$ for $L\in\N$. The spectrum of $S_N^{\Lambda_L}$ is given by the $|\Lambda_L| = (4L+2)^d$ eigenvalues
\be
\sigma(S_N^{\Lambda_L})=\left\{\Pi_{j=1}^d e^{i\sigma_{j}\lambda_{k_j}}\right\}
\ee
where
\bea
& k_j \in \{0, 1,2,\cdots, 2L+1\}^{d} \ \ \mbox{for} \ \ j=1,\ldots,d, \nonumber \\
&\sigma_j\in\{+1,-1\} \ \ \mbox{for}\ \ k_j=1,\dots, 2L, \ \ \sigma_j=1 \ \ \mbox{for} \ \ k_j\in\{0,2L+1\}.
\eea

Under the assumption $d\arccos(r^2-t^2)<\pi$, the upper edge of the spectrum of $S$, $e^{id\arccos(r^2-t^2)}=e^{id\lambda_0}$, belongs to $\sigma(S_N^{\Lambda})$ and is non degenerate. An eigenvector corresponding to $e^{id\lambda_0}$ is
$\ffi_0^{(d)}=\otimes_{1}^d\ffi_0$, whose components all have modulus one.

Moreover, there exists $c_0$, a numerical constant, such that
\be\label{gap}
\dist (e^{id\lambda_0},\sigma(S_N^{\Lambda_L})\setminus\{e^{id\lambda_0}\})>\frac{c_0t^2}{|\Lambda_L|^{2/d}}.
\ee
Indeed, the closest eigenvalue to $e^{id\lambda_0}$ is  $e^{i((d-1)\lambda_0+\arccos(r^2-t^2\cos(\pi/(2L+1))))}$, which is $d$-fold degenerate, so that the distance (\ref{gap}) equals
\bea
|e^{id\lambda_0}-e^{i((d-1)\lambda_0+\arccos(r^2-t^2\cos(\pi/(2L+1))))}|&=&
|e^{i\lambda_0}-e^{i\arccos(r^2-t^2\cos(\pi/(2L+1)))}| \nonumber \\
&>& \frac{t^2}{(4L+2)^2} c_0= \frac{c_0t^2}{|\Lambda_L|^{2/d}},
\eea
where $c_0=(\pi(4-\pi))^2/8$, see (\ref{ggap}).

For later study of spectral properties near the lower edge $e^{-id\lambda_0}$ of the spectrum of $S$, we use the modified version of $S_N^{[2l_j,2m_j-1]}$ from the fourth remark at the end of the previous subsection in the definition (\ref{eq:Neumannmultidim})

\setcounter{equation}{0}
\section{The Feynman-Hellmann Formula} \label{sec:FeymanHellmann}

The Feynmann-Hellmann formula provides, on the level of first order perturbation theory, the change of an isolated simple eigenvalue of a selfadjoint operator under an additive perturbation. Here we will need a corresponding result for multiplicative perturbations of unitary operators. We prove such a formula in an analytic framework, which will suffice for our purpose.

\begin{prop}\label{fehe}
Let $ I\subset \R$ be an open interval containing zero, $\hil$ be a separable Hilbert space and $I\ni \alpha \mapsto U(\alpha)$ an analytic map with values in the set of unitary operators on $\hil$. Assume  $\beta(0)\in \mathbb S$ is an isolated simple eigenvalue of $U(0)$ with normalized corresponding eigenvector $\ffi(0)\in\hil$. Then, there exists an open disc centered at $0$ of radius $\alpha_0>0$, $D(0,\alpha_0)\subset \C$, and two analytic maps  $D(0,\alpha_0)\ni \alpha\mapsto \beta(\alpha)\in\C$ and $D(0,\alpha_0)\ni \alpha\mapsto \ffi(\alpha)\in \hil$ such that
\bea \label{triv}
&&U(\alpha)\ffi(\alpha)=\beta(\alpha)\ffi(\alpha) \ \ \ \forall \alpha \in
D(0,\alpha_0), \nonumber \\ \nonumber
&&\dist (\beta(\alpha), \sigma(U(\alpha)\setminus\{\beta(\alpha)\})>0\\
&&\|\ffi(\alpha)\|=1 \ \ \mbox{if} \  \alpha\in D(0,\alpha_0)\cap I.
\eea
Moreover, for all $\alpha\in D(0,\alpha_0)\cap I$,
\be\label{hf}
\beta'(\alpha)=\bra\ffi(\alpha)|U'(\alpha)\ffi(\alpha)\ket.
\ee
\end{prop}

\begin{Remark}

i) For a given $\alpha$, the last formula is of course  true for any choice of normalized eigenvector of $U(\alpha)$, corresponding to $\beta(\alpha)$.

ii) If $I\ni\alpha\mapsto U(\alpha)$ is analytic and takes its values in the set of unitary  {\em finite} matrices, all its eigenvalues and spectral projectors admit analytic extensions in a complex neighborhood of $I$, even at the values of $\alpha$ where eigenvalues of $U(\alpha)$ may cross, see \cite{kato}. Consequently, an analytic choice of normalized eigenvectors can be made for all $\alpha\in I$.
\end{Remark}
\noindent {\bf Proof:}
  By the general theory of analytic perturbations of operators, see e.g.~\cite{kato}, the operator $U(\alpha)$ admits an isolated simple eigenvalue $\beta(\alpha)$,  for small enough values of $|\alpha|$, say in $D(0,\alpha_0)$. Also, the analytic rank one spectral projector on $\beta(\alpha)$, $P(\alpha)$, given by the Riesz formula is analytic for all $\alpha\in D(0,\alpha_0)$.

By definition, for all $\alpha\in D(0,\alpha_0)$,
\be\label{pu}
P(\alpha)U(\alpha)=U(\alpha)P(\alpha)=\beta(\alpha)P(\alpha),
\ee
and since $U(\alpha)$ is unitary on the real axis, $P(\alpha)$ is self-adjoint
for real $\alpha$'s.
Now define the analytic operator $W(\alpha)$ as the unique solution to the ODE
\be
W'(\alpha)=[P'(\alpha),P(\alpha)]W(\alpha), \ \ \ W(0)= I, \ \ \alpha\in D(0,\alpha_0).
\ee
It is a well known property (\cite{kato}) that the following intertwining property
holds for all  $\alpha\in D(0,\alpha_0)$,
\be\label{inter}
P(\alpha)W(\alpha)=W(\alpha)P(0).
\ee

Note that $W(\alpha)$ is unitary on the real axis, since its generator is easily seen to be anti self-adjoint there. We define an analytic vector by
\be
\ffi(\alpha)=W(\alpha)\ffi(0).
\ee
Identities (\ref{inter}) and (\ref{pu}) show that
\be
U(\alpha)\ffi(\alpha)=\beta(\alpha)\ffi(\alpha)
\ee
and $\ffi$ is normalized on the real axis, since $W$ is unitary there. By differentiation of the previous identity and application of $P(\alpha)$ to the result, we obtain
\be
P(\alpha)U'(\alpha)\ffi(\alpha)+P(\alpha)U(\alpha)\ffi'(\alpha)=
\beta'(\alpha)P(\alpha)\ffi(\alpha)+\beta(\alpha)P(\alpha)\ffi'(\alpha)
\ee
which reduces to
\be
P(\alpha)U'(\alpha)\ffi(\alpha)=\beta'(\alpha)\ffi(\alpha)
\ee
due to (\ref{pu}) and (\ref{inter}). Since for all $\alpha\in I\cap D(0,\alpha_0)$ we can write
\be
P(\alpha)=|\ffi(\alpha)\ket\bra \ffi(\alpha)|=W(\alpha)|\ffi(0)\ket\bra \ffi(0)|W^{-1}(\alpha),
\ee
the result follows.
\ep

\medskip

As a specific application, let us consider the family of analytic unitary matrices
\be \label{eq:defualpha}
U^\Lambda(\alpha)=D(\alpha) S^\Lambda_N = \mbox{diag}\{e^{-i\alpha\theta_k}\} S^\Lambda_N,
\ee
where $S^\Lambda_N$ is the Neumann restriction of $S$ to a $d$-dimensional box $\Lambda$ introduced in Section~\ref{sec:Neumann}, $\alpha\in \R$, and
$\theta_k\in \mathbb T$ for all $k\in\Lambda$. $U^\Lambda(\alpha)$ interpolates between $S^\Lambda_N$ and $\mbox{diag}\{e^{-i\theta_k}\} S^\Lambda_N$, at $\alpha=0$ and $\alpha=1$, respectively.
Introducing the self-adjoint matrix $H^\Lambda=\sum_{k\in\Lambda}\theta_k\,|e_k\ket\bra e_k|$ on $l^2(\Lambda)$, we can rewrite
\be
U^\Lambda(\alpha)=e^{-i\alpha H^\Lambda}S^\Lambda_N, \ \ \ \alpha\in\R.
\ee

 \begin{lem} \label{lll}
 If $e^{i\lambda(0)}$ is a discrete non-degenerate eigenvalue of $S^\Lambda_N$
with normalized eigenstate $\varphi(0)$, then, for all $\alpha\in\R$,
there exist analytic eigenvalues $e^{i\lambda(\alpha)}$ of $ U^\Lambda(\alpha)$ with analytic
normalized eigenvectors $\ffi(\alpha)$ such that
    \be
    \frac{d}{d\alpha} e^{i\lambda(\alpha)} = -ie^{i\lambda(\alpha)}
\sum_{k\in\Lambda}\theta_k|\bra e_k|\varphi(\alpha)\ket|^2.
    \ee
In particular, for all $\alpha\in\R$,
$\lambda'(\alpha)=-\sum_{k\in\Lambda}\theta_k\,|\bra e_k|\varphi(\alpha)\ket|^2$.
\end{lem}

\noindent {\bf Proof:}

The existence of analytic eigenvalues $e^{i\lambda(\alpha)}$ and analytic eigenvectors $\ffi(\alpha)$ of $U^{\Lambda}(\alpha)$, $\alpha \in \R$, follows from Proposition~\ref{fehe} and the remark following it.

 We compute
\be
(U^{\Lambda})'(\alpha)=-iH_\Lambda U^{\Lambda}(\alpha), \ \ \ U^{\Lambda}(0)=S^\Lambda_N
\ee
and
\be
\bra\ffi(\alpha)|-iH^\Lambda U^{\Lambda}(\alpha)\ffi(\alpha)\ket=
-i\sum_{k\in\Lambda}\theta_k\, |\bra e_k|\ffi(\alpha)\ket|^2e^{i\lambda(\alpha)}
\ee
and apply (\ref{hf}).
\ep

\setcounter{equation}{0}
\section{Splitting Boxes by Neumann Boundary Conditions} \label{sec:splitting}

Throughout this section we will assume that boxes $\Lambda \subset \Z^d$ are compatible with Neumann boundary conditions as defined in Section~\ref{sec:Neumann}, $S_N^{\Lambda}$ is given by (\ref{eq:Neumannmultidim}) and
\be
U^{\Lambda} = D S_N^{\Lambda} = \mbox{diag}(e^{-i\theta_k}) S_N^{\Lambda}.
\ee

For notational simplicity we will assume in this section that the box $\Lambda$ has a vertex at the origin, which does not cause a restriction.
We first deal with dimension $d=1$.\\

Consider a one dimensional box $\Lambda_0$ consisting of two disjoint adjacent boxes $\Lambda_1$ and $\Lambda_2$:
\be
\Lambda_1=[0,2l-1], \ \ \Lambda_2=[2l,2(l+n)-1], \ \  \Lambda_0=[0,2(l+n)-1],
\ee
with $n, l\geq 2$ (to avoid the special case $S_N^{[0,1]}$). We note that
$U^{\Lambda_0}$ and $U^{\Lambda_1}\oplus U^{\Lambda_2}$ are both defined on $l^2([0,2(l+n)-1])$.

\medskip

We want to show that the eigenvalues of $U^{\Lambda_1}\oplus U^{\Lambda_2}$ are closer to the upper band edge
$e^{i(\lambda_0+a)}$ of the almost sure spectrum $\Sigma$ of $U$ than those of $U^{\Lambda_0}$. Recall here that in Section~\ref{sec:beloc} we have assumed that $|\theta_k|\le a$ and $\lambda_0+a <\pi$. This is the analog of the well known property $H^{\Lambda_1}\oplus H^{\Lambda_2}\leq H^{\Lambda_0}$, where $H^\Lambda$ is the Neumann restriction to a box $\Lambda$ of the discrete Schr\"odinger operator.

\medskip

The following simple observation is the starting point of the analysis. Splitting a box by imposing Neumann boundary conditions is a rank one perturbation:
\begin{lem}
Let $S_N^{\Lambda_j}$, $j=0,1,2$ be defined as above. Then,
\be
S_N^{\Lambda_0}=S_N^{\Lambda_1}\oplus S_N^{\Lambda_2}+|\psi\ket\bra \ffi|
\ee
where
\bea \label{psiffi}
\psi&=& -t e_{2l-2} -re_{2l-1}-ire_{2l} +it e_{2l+1}\nonumber\\
\ffi&=& t(-i e_{2l-1} +e_{2l})
\eea
\end{lem}

The proof is an easy computation. This leads us to using the following fact about rank one perturbations of unitary operators which return a unitary operator.

\begin{lem} \label{r1}
Let $U$ a unitary operator on a Hilbert space $\mathcal H$ and $f,g\in {\mathcal H}\setminus \{0\}$. If
\be \label{pertU}
V=U+|f\ket\bra g|
\ee
is unitary, then there exists $\beta\in (-\pi,\pi]$ such that $e^{i\beta} = 1+\bra Ug|f \ket$ and
\be \label{UVrel}
V=e^{i\beta|\hat{f}\ket\bra\hat{f}|}U,
\ee
where $\hat{f}=f/\|f\|$.
\end{lem}

\noindent {\bf Proof:}

The identity $VV^*=\mathbb I$ implies that
\be\label{isom}
|U g \ket\bra f|+|f\ket\bra U g|+\|g\|^2|f\ket\bra f|=0.
\ee
Applying this to $f$ shows that $Ug$ is proportional to $f$, so that
\be \label{Ugfprop}
U g =\frac{\bra f|U g\ket}{\| f\|^2} f.
\ee
With this it follows from (\ref{pertU}) that
\bea
V & = & (I+|f\ket \bra Ug|)U \nonumber \\
& = & (I+ \bra Ug,f\ket |\hat{f}\ket \bra \hat{f} |)U \nonumber \\
& = & (I- |\hat{f} \ket \bra \hat{f}| + \mu |\hat{f} \ket \bra \hat{f}|)U,
\eea
where $\mu=1+\bra Ug|f \ket$. Thus $I- |\hat{f} \ket \bra \hat{f}| + \mu |\hat{f} \ket \bra \hat{f}| = VU^*$ is unitary, which shows that $|\mu|=1$, i.e.\ $\mu = e^{i\beta}$ for $\beta \in (-\pi,\pi]$, and $I- |\hat{f} \ket \bra \hat{f}| + \mu |\hat{f} \ket \bra \hat{f}| = e^{i\beta |\hat{f} \ket \bra \hat{f}|}$.
\ep

\vspace{.3cm}

Taking into account the random phases, we apply the previous lemma to our case with
\be
U^{\Lambda_0}=U^{\Lambda_1}\oplus U^{\Lambda_2}+|D\psi\ket\bra \ffi|
\ee
and $\psi$, $\ffi$ from (\ref{psiffi}). We compute
$\|D\psi\|=\|\psi\|=\sqrt{2}$, i.e.\ $\hat\psi = \psi/\|\psi\| =\psi/\sqrt{2}$, and
\be
e^{i\beta}= 1+ \bra (U^{\Lambda_1} \oplus U^{\Lambda_2}) \ffi| D\psi \ket  = 1+ \bra (S_N^{\Lambda_1} \oplus S_N^{\Lambda_2}) \ffi|\psi \ket = e^{-i\arccos(r^2-t^2)}=e^{-i\lambda_0}
\ee
so that
\be
U^{\Lambda_0}=e^{-i\lambda_0|D\hat{\psi}\ket\bra D\hat{\psi}|}\,U^{\Lambda_1}\oplus U^{\Lambda_2}=
De^{-i\lambda_0|\hat{\psi}\ket\bra \hat{\psi}|}\,S_N^{\Lambda_1}\oplus S_N^{\Lambda_2}.
\ee

Let us introduce an analytic family of unitary operators defined in $I$, a complex neighborhood of $[0,1]$,
by
\be
I\ni \alpha\mapsto U(\alpha)=e^{-i\alpha\lambda_0|D\hat{\psi}\ket\bra D\hat{\psi}|}\,U^{\Lambda_1}\oplus U^{\Lambda_2},
\ee
such that $U(0)=U^{\Lambda_1}\oplus U^{\Lambda_2}$ and $U(1)=U^{\Lambda_0}$. By Lemma~\ref{lll} we immediately get the following

\begin{prop} \label{prop:neumannsplit}
Let $\alpha\in I$ and  $e^{i\lambda(\alpha)}$ denote any analytic eigenvalue of $U(\alpha)$, which is isolated except at a finite set of values of $\alpha$. Then
\be \label{argmon}
\arg (\lambda(1))\leq \arg(\lambda(0)).
\ee
\end{prop}

\begin{Remark}
In other words,  when a Neumann boundary condition is introduced to split $\Lambda_0$ into $\Lambda_1\cup \Lambda_2$, the eigenvalues of $U_{\Lambda_1}\oplus U_{\Lambda_2}$ are closer to $e^{i(\lambda_0+a)}$ than those of $U_{\Lambda_0}$.
\end{Remark}

Let us generalize now to dimension $d\geq 1$. Consider a box $\Lambda_0$ of the form
\be
\Lambda_0=\Lambda_0(1)\times\Lambda(2)\times\dots\Lambda(d)
\ee
where
\be
\Lambda_0(1)=[0,2(l+n)-1], \ \ \Lambda(j)=[0,2l_j-1], \ j=2, \cdots, d,
\ee
which we split by a Neumann boundary condition perpendicular to the first axis as
$\Lambda_0=\Lambda_1\cup\Lambda_2$
with
\be
\Lambda_k=\Lambda_k(1)\times\Lambda(2)\times\dots\Lambda(d), \ k=1,2
\ee
and
\be
\Lambda_0(1)=\Lambda_1(1)\cup\Lambda_2(1)=[0,2l-1]\cup [2l,2(l+n)-1].
\ee
By the previous results, the corresponding operators $U^{\Lambda_k}$, $k=0,1,2$ are related by
\be
U^{\Lambda_0}=e^{-i\lambda_0|D\hat\psi\ket\bra D\hat\psi|\otimes I\otimes\cdots\otimes I}\,U^{\Lambda_1}\oplus U^{\Lambda_2}.
\ee
Applying Lemma~\ref{lll} again, with $H_\Lambda$ replaced by the non negative operator $|D\hat\psi\ket\bra D\hat\psi|\otimes {\mathbb I}\otimes\cdots\otimes{\mathbb I} $, shows that the spectra of
$U^{\Lambda_0}$ and $U^{\Lambda_1}\oplus U^{\Lambda_2}$ are related in the same way as in the one dimensional case, e.g.\ by (\ref{argmon}). Clearly, the splitting by Neumann boundary conditions can be done perpendicular to any of the $d$ coordinate axes and can also be iterated. Thus we get the above form of spectral monotonicity also, for example, when spitting $U^{\Lambda_L}$ over the cube $[-2L,2L+1]^d$ into a direct sum of $U_{\Lambda_i}$ for $((2L+1)/l)^d$ cubes $\Lambda_i$ of sidelength $2l$.

\setcounter{equation}{0}
\section{A Combes-Thomas Estimate} \label{sec:CombesThomas}

Combes-Thomas bounds, originating from \cite{Combes/Thomas}, have become the standard tool in Schr\"o\-dinger operator theory to show exponential decay of eigenfunctions to eigenvalues which lie outside of the essential spectrum. They also provide a key step in localization proofs for random Schr\"odinger operators in the band edge regime, see e.g.\ \cite{Stollmann}. Here we provide a Combes-Thomas type estimate for unitary operators with band structure.

Let $U$ be unitary on $l^2(\Z^d)$. We say that $U$ has band structure of width $w>0$, if it can be written as
\be
U=D+O \ \ \mbox{with}\ \ \bra e_k |De_j\ket=\delta_{kj} \bra e_k |De_k\ket \ \
\mbox{and} \ \  \bra e_k |O e_j\ket =0 \ \ \mbox{if}\ \ |j-k| > w.
\ee

\begin{prop}[Combes-Thomas type estimate] \label{prop:CT}
For a unitary
operator $U$ on $l^2(\Z^d)$ with band structure of width $w$, there exist $0< B<\infty$ which depends on $U$ only, such that
\be \label{eq:CT}
|\langle e_j|(U-z)^{-1} e_k\rangle|\leq \frac{2}{{\rm dist}(z,\sigma(U))} e^{-{\rm dist}(z,\sigma(U))|j-k|B}.
\ee
\end{prop}

\begin{Remark}
i) The same result holds for $U$ defined on a finite dimensional Hilbert space $l^2(\Lambda)$, $\Lambda \subset \Z^d$, with constants independent of $\Lambda$.

ii) Actually, our proof works  more generally for bounded normal operators $U$ with band
structure. Results of this type are known in the literature, e.g.\ \cite{Demko}.

\end{Remark}

\noindent {\bf Proof:}

Let $x=(x_1,\cdots, x_d)$, where $x_n$ is the self-adjoint  multiplication operator acting
on $e_k$, with $k=(k_1,\ldots, k_d)$,
 as $x_ne_k=k_ne_k$ and defined on its natural domain. We introduce the vector $\alpha=(\alpha_1,\cdots, \alpha_d) \in\R^d$ and construct the self-adjoint operator
\be
e^{\alpha x}\ \ \mbox{acting as}\ \ e^{\alpha x}e_k=e^{\alpha k}e_k
\ee
on
\be
{\mathcal D}_\alpha=\{\psi\in l^2(\Z^d)\ \mbox{s.t.}\ \sum_{k\in\Z^d} |\bra e_k|\psi\ket|^2e^{\alpha k}<\infty\}.
\ee
Here $\alpha k = \sum_n \alpha_n k_n$. Consider the operator
\be
U_\alpha:=e^{\alpha x}Ue^{-\alpha x}=e^{\alpha x}De^{-\alpha x}+e^{\alpha x}Oe^{-\alpha x}=D_\alpha + O_\alpha
\ee
defined {\it a priori} on the dense set
\be
c_0=\{\psi\in l^2(\Z^d)\ \mbox{s.t.}\ \bra e_k|\psi\ket =0 \ \mbox{for $|k|$ large enough}\}.
\ee
The operator $U_\alpha$ is bounded because for any $\psi \in c_0$
\be
O_\alpha\psi=\sum_{j\in F}\sum_{\substack{k\neq j\\ |k-j|\leq w}}e^{\alpha(k-j)}\bra e_k | O e_j\ket
\bra e_j|\psi\ket e_k
\ee
where the set $F$ is finite and $e^{\alpha(k-j)}\leq e^{|\alpha|w}$. Moreover,
$D_\alpha=D$ which shows that $\|U_\alpha\|\leq C_1(\alpha)\leq C_1<\infty$ on $c_0$, for $\alpha$ in a bounded set. Similarly,
\be
\|U-U_\alpha\|=\|O-O_\alpha\|\leq C_2(\alpha)\leq C_3|\alpha|,
\ee
for $|\alpha|$ small enough.
From the resolvent identity, if $z\in\rho(U)$ and $ C_3|\alpha|/\dist(z,\sigma(U))<1/2$,
\be
(U_\alpha-z)^{-1}=(U-z)^{-1}({\mathbb I}+(U_\alpha-U)(U-z)^{-1})^{-1}
\ee
with
\be
\|(U_\alpha-z)^{-1}\|\leq 2\|(U-z)^{-1}\|\leq 2/\dist(z,\sigma(U)).
\ee
Finally, by the formula
\be
(U_\alpha-z)^{-1}=e^{\alpha x}(U-z)^{-1}e^{-\alpha x},
\ee
we derive
\be
\bra j|e^{\alpha x}(U-z)^{-1}e^{-\alpha x}k\ket=e^{-\alpha(k-j)}\bra j | (U-z)^{-1}k\ket = \bra j | (U_\alpha-z)^{-1}k\ket,
\ee
from which we get
\be
|\bra j | (U-z)^{-1}k\ket|\leq 2\frac{e^{+\alpha(k-j)}}{\dist(z,\sigma(U))}.
\ee
Choosing the components of $\alpha$ and their sign  in such a way that $|\alpha_n|=|\alpha|>0$ and
\be
\alpha(j-k)=\sum_{n=1}^d\alpha_n(j-k)_n\geq|\alpha||j-k|,
\ee
we obtain the result, with $|\alpha|=\dist(z,\sigma(U))/4C_3$,
and $B=1/4C_3$.
\ep

\setcounter{equation}{0}
\section{The Genesis of Lifshits Tails} \label{sec:lifshits}

After having introduced some tools in the previous two sections we will now start with the actual proof of Theorem~\ref{thm:bandedgelocalization}. Throughout this proof we will focus on localization at the upper band edge $e^{i(d\lambda_0+a)}$ of the almost sure spectrum $\Sigma$ of $U_{\omega}$. The proof at the lower band edge is completely analogous. It uses the alternate form of Neumann boundary conditions discussed in Section~\ref{sec:Neumann} (setting $e^{i\eta}=r-it$ in (\ref{nbc}) rather than $r+it$) and a corresponding adjustment of the results on splitting boxes in Section~\ref{sec:splitting}.

We find it convenient to rotate the upper band edge of $\Sigma$ to be identical with $e^{id\lambda_0}$, the upper band edge of $S$. This is achieved by replacing the original $U_{\omega}$ by $e^{-ia} U_{\omega}$. In other words, setting $\theta_M = 2a$ we now assume
\be \label{sec:newmu}
\mbox{supp}\,\mu \subset [0,\theta_M] \quad \mbox{with} \quad 0\in \mbox{supp}\,\mu \quad \mbox{and}\quad 2d\lambda_0 + \theta_M<2\pi.
\ee
The latter means that $\Sigma$ has the gap $\{e^{i\vartheta}: d\lambda_0 < \vartheta < 2\pi-(d\lambda_0+\theta_M)\}$.

\medskip

As in earlier sections we will frequently drop the subscript $\omega$ from our notation.
\medskip

We will first establish a Lifshits tail estimate for the spectrum near the band edge $e^{id\lambda_0}$. At the root of this is the following proposition which we prove by following the steps of Stollmann \cite{Stollmann}. As in Section~\ref{sec:Neumann}, for $L\in\N$ we set $\Lambda_L = [-2L,2L+1]^d$.

\begin{prop}\label{prop:lt}
Let $e^{i\lambda(U^{\Lambda_L})}$, respectively  $e^{id\lambda_0}$, be the eigenvalue of largest argument of $U^{\Lambda_L}$, respectively $S^{\Lambda_L}_N$. Then $\lambda(U^{\Lambda_L})\leq d\lambda_0$ and there exist $b>0$ and $\gamma>0$, independent of $L$ and $d$, such that
 \be
\P \left( |e^{i\lambda(U^{\Lambda_L})}-e^{id\lambda_0}|\leq \frac{b}{L^2}
\right) \leq e^{-\gamma L^d},
\ee
for $L$ large enough.
 \end{prop}

Let us first give an easy Corollary of Lemma~\ref{lll}. Recall that $U^{\Lambda_L}(\alpha)$ is defined by (\ref{eq:defualpha}).

\begin{lem} \label{mono}
Consider a fixed realization of $U^{\Lambda_L}(\alpha)$ in dimension $d\geq 1$. Then the analytic continuation of any eigenvalue $e^{i\lambda(\alpha)}$ of $U^{\Lambda_L}(\alpha)$ is such that
$\lambda(\alpha)$ is non increasing. Consequently,
$\left|e^{i\lambda(\alpha)}-e^{i\lambda(0)}\right|$ is a non decreasing
function of $\alpha\geq 0$, as long as $\lambda(0)-\lambda(\alpha)<\pi$.

Moreover, the eigenvalue $e^{id\lambda_0}$ of $S_N^{\Lambda_L}$ and its analytic continuation $e^{i\lambda_0(\alpha)}$ satisfy
\be \label{eq:alphaderiv}
\frac{d}{d\alpha}\lambda_0(\alpha)|_{\alpha=0}= -\frac{1}{(4L+2)^d}\sum_{k\in\Lambda_L}\theta_k.
\ee
\end{lem}

\noindent {\bf Proof:}

The first statement follows from Lemma~\ref{lll} and from
\be
\frac{d}{d\alpha}\left|e^{i\lambda(\alpha)}-e^{i\lambda(0)}\right|^2=2\sin(\lambda(\alpha)-\lambda(0))\lambda'(\alpha).
\ee
The second statement makes use of the fact that the components of the eigenvector $\ffi_0^{(d)}$ all have equal modulus.
\ep

Recall the following standard large deviation estimate whose proof can be found, e.g., in Lemma~2.1.1 of \cite{Stollmann}.

\begin{lem}\label{probablityestimate}
For non-trivial and non-negative i.i.d.\ random variables $\theta_k$ and $s_0=\gamma_0=-\frac{1}{2}\ln(\mathbb{E}(e^{-\theta_0}))>0$, we have
\be
\mathbb{P}\left(\frac{1}{|\Lambda|}\sum_{i\in\Lambda}\theta_i\leq s_0\right)\leq e^{-\gamma_0|\Lambda|}.
\ee
 \end{lem}

Let us consider the small $\alpha$ behavior of $e^{i\lambda_0(\alpha)}$.

 \begin{lem}\label{perturbationofeigenvalues}
 There exist $c_1>0$ and $c_2>0$, independent of $d$ and $L$, such that for $L$ sufficiently large
 \be
 \left|e^{i\lambda_0(\alpha)}-e^{id\lambda_0}-\alpha \left(\frac{d}{d\alpha}e^{i\lambda_0(\alpha)}\right)_{\alpha=0}\right|\leq
 c_1\alpha^2 L^2, \; 0\leq\alpha\leq \frac{c_2}{L^2}.
 \ee
 \end{lem}

\noindent {\bf Proof:}

Expanding  $e^{id\lambda_0(\alpha)}$ in terms of $\alpha\in\R$, we
get that
\be
e^{i\lambda_0(\alpha)}-e^{id\lambda_0}-\alpha
\frac{d}{d\alpha}e^{i\lambda_0(0)}=
\frac{\alpha^2}{2}\frac{d^2}{d^2\alpha}e^{i\lambda_0(\tilde\alpha)}
\ee
for some $0<\tilde\alpha<\alpha$. Next we use Cauchy's integral formula to
bound
\be
\frac{d^2}{d^2\alpha}e^{i\lambda_0(\tilde\alpha)}= \frac{2!}{2 \pi i}\int_{|z-\tilde\alpha|=r}\frac{e^{i\lambda_0(z)} }{(z-\tilde\alpha)^3} dz= \frac{2!}{2 \pi i}\int_{|z-\tilde\alpha|=r}\frac{e^{i\lambda_0(z)}-e^{id\lambda_0} }{(z-\tilde\alpha)^3} dz
\ee
for $\tilde\alpha$ small enough and suitable $r>0$. Thus we need to control
$e^{i\lambda_0(z)}$ for $z$ complex.
 Since
\be
U^{\Lambda_L}(\alpha)-U^{\Lambda_L}(0)=(e^{-i\alpha H^{\Lambda_L}}-I)S_N^\Lambda,
\ee
where
\be
\|e^{-i\alpha H^{\Lambda_L}}- I\|\leq e^{|\alpha|\theta_M}-1,
\ee
we have, by the second resolvent identity,
\be
(U^{\Lambda_L}(\alpha)-z)^{-1}=(S_N^{\Lambda_L}-z)^{-1}(I -
(e^{-i\alpha H^{\Lambda_L}}-I)S_N^\Lambda(U^{\Lambda_L}(\alpha)-z)^{-1})
\ee
for $z\not\in \sigma( U^{\Lambda_L}(\alpha))\cup \sigma(S_N^{\Lambda_L})$. Hence,
\be
\dist (z,\sigma(S_N^{\Lambda_L}))>e^{|\alpha|\theta_M}-1 \ \ \Longrightarrow \ \ z\in \rho(U^{\Lambda_L}(\alpha)).
\ee
Now (\ref{gap}) says
\be
\dist (e^{id\lambda_0},\sigma(S_N^{\Lambda_L})\setminus\{e^{id\lambda_0}\})>\delta=\frac{t^2c_0}{|\Lambda_L|^{2/d}}.
\ee
Thus, if $|\alpha|<\alpha_0:=\frac{\ln(1+\delta/2)}{\theta_M}$,
\be
\{z\ | \ |z-e^{id\lambda_0}|=\delta/2\}\subset \rho(U^{\Lambda_L}(\alpha)).
\ee
We now take $\alpha\in (-\alpha_0/2,\alpha_0/2)$ so that $\tilde\alpha<\alpha_0/2$ and $r=\alpha_0/2$ so that
\be
\{z\ | \ |z-\tilde\alpha|=\alpha_0/2\}\subset \{z \ | \ |z|\leq \alpha_0\}
\ee
and for such $z$'s $|e^{i\lambda_0(z)}-e^{id\lambda_0}|< \delta/2$.
Using $\delta/4<\ln(1+\delta/2)<\delta/2$ if $\delta<1/2$, one gets that
if $0\leq\alpha<\frac{\delta}{8\theta_M}$,
\be
\left|\frac{\alpha^2}{2}\frac{d^2}{d^2\alpha}e^{i\lambda_0(\tilde\alpha)}\right|\leq \alpha^2\frac{\delta}{2}\left(\frac{2}{\alpha_0}\right)^2\leq
\alpha^2 32\frac{\theta_M^2}{\delta}.
\ee
We get the announced result with
\be
c_1=\frac{128 \theta_M^2}{t^2 c_0}, \ \ c_2=\frac{t^2 c_0}{32\theta_M},
\ee
provided $|\Lambda_L|^{2/d} = (4L+2)^2 >2t^2c_0$, i.e.\ for $L$ sufficiently large.
\ep

\vspace{.5cm}

\noindent {\bf Proof of Proposition \ref{prop:lt}:}
Assume that $|e^{i\lambda(U^{\Lambda_L})}-e^{id\lambda_0}|\leq b/L^2$, with $b$ to be determined later. Using the monotony in $\alpha$  (Lemma~\ref{mono}) and Lemma~\ref{perturbationofeigenvalues}, we have for $0\leq\alpha\leq c_2/L^2$ and $L$ large enough
\bea
\left|\alpha \left(\frac{d}{d\alpha}e^{i\lambda_0(\alpha)}\right)_{\alpha=0}\right| & \leq &
\left|e^{i\lambda_0(\alpha)}-e^{id\lambda_0}\right|+c_1\alpha^2 L^2 \nonumber
\\ & \leq & \left|e^{i\lambda(U^\Lambda)}-e^{id\lambda_0}\right|+c_1\alpha^2
L^2 \nonumber
\\ & \leq & \frac{b}{L^2}+c_1\alpha^2 L^2.
\eea
Dividing by $\alpha$ and then choosing $\alpha=c_4/L^2$ such that
$c_4\leq c_2$ and $c_1 c_4\leq s_0/2$, we obtain that
\be
\left|
\left(\frac{d}{d\alpha}e^{i\lambda_0(\alpha)}\right)_{\alpha=0}\right| \leq
b/c_4+s_0/2.
\ee
Next we choose $b$ such that $b/c_4\leq s_0/2$ to get that
\be
\left|
\left(\frac{d}{d\alpha}e^{i\lambda_0(\alpha)}\right)_{\alpha=0}\right|\leq
s_0.
\ee
Note that $b$ is thus independent of $d$ and $L$.
On the other hand we have from (\ref{eq:alphaderiv}) that
\be
    \left|\left(\frac{d}{d\alpha} e^{i\lambda_0(\alpha)}\right)_{_{\alpha=0}}\right| = \frac{1}{(4L+2)^d}\sum_{k\in\Lambda_L}\theta_k.
    \ee
In probabilistic terms
\be
\left\{\omega\ |\  |e^{i\lambda(U^{\Lambda_L})}-e^{id\lambda_0}|\leq b/L^2 \right\}\subset
\left\{ \omega \ |\  \frac{1}{(4L+2)^d}\sum_{k\in\Lambda_L}\theta_k\leq s_0\right\}.
\ee
Finally an application of Lemma~\ref{probablityestimate} ends the
proof with $\gamma=2^d\gamma_0$.
\ep

\vspace{.5cm}

The Lifshits tail estimate of Proposition~\ref{prop:lt} and the properties of the Neumann boundary conditions, Proposition~\ref{prop:neumannsplit}, allow to prove the following result, which is based on an equivalent result for Schr\"odinger operators provided in
\cite{Stollmann}.

\begin{prop}\label{prop:lbeta}
Let $\beta\in (0,1)$. There exist finite positive constants $\bar \gamma$, $C$ and a sequence of positive integers $L_k$ with $L_k\to\infty$ such that
for any $k$ and any $z\in\C$ with
$1<|z|<2 \ \ \mbox{and} \ \ \ d\lambda_0-1/L_k^\beta\leq \arg z \leq d\lambda_0$,
\be
\P(\mbox{\em dist }(z,\sigma(U^{\Lambda_{L_k}}))\leq 1/L_k^\beta)\leq CL_k^{d(1-\beta/2)}e^{-\bar\gamma L_k^{d\beta/2}}.
\ee
\end{prop}

{\bf Proof:}  Let $\beta\in (0,1)$ and $b>0$ the constant found in Proposition~\ref{prop:lt}. Fix a constant $C>1$.

We claim that for each sufficiently large $k\in \N$ there exists $L_k\in\N$ which is a multiple of $k$ and such that
\be\label{scale}
\frac{b}{Ck^2} \le \frac{2}{L_k^{\beta}} \le \frac{b}{k^2}.
\ee
To see this, note that (\ref{scale}) is equivalent to
\be
L_k \in \left[ (2k^2/b)^{1/\beta}, (2k^2/b)^{1/\beta} C^{1/\beta} \right].
\ee
As $\beta<1$, for $k$ sufficiently large, this interval has length larger than $k$, allowing for a choice of $L_k$ as required.

We now show that (\ref{scale}) holds for these $L_k$. Split the box $\Lambda_{L_k} = [-2L_k,2L_k+1]^d$ into $M=|\Lambda_{L_k}|/|\Lambda_k| = (L_k/k)^d$ disjoint boxes as
\be
\Lambda_{L_k}=\cup_{j=1}^M\Lambda_k(j),
\ee
where  $\Lambda_k(j)$ denotes the suitably translated box $\Lambda_k+c(j)$. Consider now
\be
U_N=U^{\Lambda_k(1)}\oplus U^{\Lambda_k(2)}\oplus \cdots \oplus U^{\Lambda_k(M)}
\ee
on $l^2(\Lambda_{L_k})=\oplus_{j=1}^M l^2(\Lambda_k(j))$, where each $ U^{\Lambda_k(j)}$ is provided with Neumann boundary conditions. Proposition~\ref{prop:neumannsplit} shows that passing from $U^{\Lambda_{L_k}}$ to $U_N$ by
introducing Neumann boundary conditions makes the eigenvalues come closer to the upper band edge, i.e.
\be
\lambda(U^{\Lambda_{L_k}})\leq \lambda(U^{\Lambda_k(j_0)}) \ \ \mbox{for some $j_0\in \{1,2, \cdots, M\}$},
\ee
where $e^{i\lambda(U)}$ denotes the eigenvalue of largest argument of $U$. As a consequence, taking  the stochastic independence of the $(U^{\Lambda_k(j)})$ into account together with the relation  (\ref{scale})
\bea
\P \left( |e^{i\lambda(U^{\Lambda_{L_k}})}-e^{id\lambda_0}|\leq 2/L_k^{\beta} \right) &\leq &
\P \left( |e^{i\lambda(U^{(\Lambda_k(j_0))})}-e^{id\lambda_0}|\leq b/k^{2} \:\mbox{for some $j_0$}\right)\nonumber\\
 &\leq&
M\P \left( |e^{i\lambda(U^{(\Lambda_k(1))})}-e^{id\lambda_0}|\leq b/k^{2} \right).
\eea
By applying Proposition~\ref{prop:lt} to the box $\Lambda_k(1)$ we see that the latter is bounded by
\be
\frac{L_k^d}{k^d} e^{-\gamma k^d} \leq C L_k^{d(1-\beta/2)}e^{-\bar \gamma L_k^{d\beta/2}}.
\ee
Finally, since dist$(z,\sigma(U^{\Lambda_{L_k}}))\leq 1/L_k^\beta$ for $z $ such that $|z-e^{id\lambda_0}|\leq 1/L_k^\beta$ implies  $|e^{i\lambda(U^{\Lambda_{L_k}})}-e^{id\lambda_0}|\leq 2/L_k^{\beta}$, we get the result.
\ep

\setcounter{equation}{0}
\section{Towards an Iterative Proof of Exponential Decay} \label{sec:decoupling}

The proof of Theorem~\ref{thm:bandedgelocalization}, to be completed in Section~\ref{sec:proofofbandedgelocalization}, will proceed as follows: To prove exponential decay of $\E(|G(k,l;z)|^s)$ we will join the two sites $k$ and $l$ by a chain of boxes of side length $L$. For a suitable choice of $L$ and arg$\,z$ close to the edge of $\Sigma$, the Lifshits tail and Combes-Thomas estimates will show that the fractional moment of the finite volume Green function $G^{(L)}(k,j;z)$ is small (think ``less than one'' even if this is only true up to some factors which can be controlled). Here $G^{(L)}$ is the resolvent of a restriction of $U$ to a box of side length $L$ centered at $k$ and $j$ is a boundary site of this box. To turn this into a proof of exponential decay of the infinite volume Green function, we need two more tools: (i) a factorization of the infinite volume Green function into finite volume factors, often referred to as a {\it geometric resolvent identity}, (ii) a {\it decoupling argument} which allows to factorize the fractional moments in the geometric resolvent identity. These two remaining tools will be provided in this section.

As explained at the beginning of Section~\ref{sec:lifshits} we will continue to focus on the localization proof at the upper band edge and continue to assume (\ref{sec:newmu}), so that the upper edge of $\Sigma$ is $e^{id\lambda_0}$.

\subsection{A Geometric Resolvent Identity} \label{sec:gri}

Due to the specific structure of our operators (in particular their ergodicity with respect to translations by two) it is of advantage to cut up $\Z^d$ into cubes of side length two. Thus, for $n=(n_1,\ldots,n_d) \in \Z^d$ let
\be
 C_n := [2n_1, 2n_1+1] \times \ldots \times [2n_d, 2n_d+1]
 \ee
and $\chi_n := \chi_{C_n}$ the characteristic function of $C_n$. For $L\in \N$ let
\be
\Lambda_L = \bigcup_{|n|\le L} C_n = [-2L,2L+1]^d.
\ee

We will work with restrictions $U_{\omega}^{\Lambda_L}$ and $U_{\omega}^{\Lambda_L^c}$ of $U_{\omega}$ to $\Lambda_L$ and its complement $\Lambda_L^c = \Z^d \setminus \Lambda_L$. We choose $U_{\omega}^{\Lambda_L} = D_{\omega} S_N^{\Lambda_L}$, where $S_N^{\Lambda_L}$ is the unitary Laplacian with Neumann boundary conditions from (\ref{eq:Neumannmultidim}). In fact, the choice of boundary conditions is rather irrelevant as long as matrix elements are only affected near the boundary, e.g.\ we have
\be \label{eq:matelem1}
U_{\omega}^{\Lambda_L}(j,k) = U_{\omega}(j,k) \quad \mbox{if $j,k \in \Lambda_{L-1}$}.
\ee
Our definition of Neumann operators from Section~\ref{sec:Neumann} does not directly extend to operators on exterior domains such as $\Lambda_L^c$, where the unitary Laplacian can not be defined as a tensor product of one-dimensional Laplacians. While it is possible to define Neumann boundary conditions directly for the $d$-dimensional operator, we choose a more simplistic approach and define
\be \label{eq:extoperator}
U_{\omega}^{\Lambda_L^c} = P^{\Lambda_L^c} U_{\omega} P^{\Lambda_L^c},
\ee
viewed as an operator on $\ell^2(\Lambda_L^c)$. Here $P^{\Lambda_L^c}$ denotes the orthogonal projection onto $\ell^2(\Lambda_L^c)$. The price for our simplemindedness is that $U_{\omega}^{\Lambda_L^c}$ is not unitary. However, it is a contraction, i.e.\ $\|U_{\omega}^{(\Lambda_L^c)}\| \le 1$ and therefore $\sigma(U_{\omega}{(\Lambda_L^c)}) \subset \{z\in \C: |z|\le 1\}$, and it remains a band matrix whose entries satisfy, by definition,
\be \label{eq:matelem2}
U_{\omega}^{\Lambda_L^c}(j,k) = U_{\omega}(j,k) \quad \mbox{if $j,k \in \Lambda_L^c$}.
\ee
These properties will suffice for what we need in Section~\ref{sec:proofofbandedgelocalization}.

We will use what is often referred to as a geometric resolvent identity, relating the resolvents of $U_{\omega}$, $U_{\omega}^{\Lambda_L}$ and $U_{\omega}^{\Lambda_L^c}$. Following an argument which for the selfadjoint Anderson model is used in \cite{ASFH}, we start by defining the boundary operator $T_{\omega}^{(L)}$ through
\be
U_{\omega} = U_{\omega}^{\Lambda_L} \oplus U_{\omega}^{\Lambda_L^c} + T_{\omega}^{(L)}.
\ee
By the above construction of $U_{\omega}^{\Lambda_L}$ and $U_{\omega}^{\Lambda_L^c}$, in particular (\ref{eq:matelem1}) and (\ref{eq:matelem2}), the operator $T_{\omega}^{(L)}$ has non-vanishing matrix-elements only near the boundary of $\Lambda_L$, more specifically
\be
 T_{\omega}^{(L)} \chi_x = \chi_x T_{\omega}^{(L)} =0 \quad \mbox{if $|x|\le L-1$ or $|x| \ge L+2$}
 \ee
as well as
\be
 \chi_x T_{\omega}^{(L)} \chi_y =0 \quad \mbox{if $|x-y| \ge 2$}.
 \ee
Also, the matrix-elements of $T_{\omega}^{(L)}$ are uniformly bounded in $L$ and $\omega$.

To keep the length of the following equations under control we will drop the arguments $\omega$ and $z$ and for the rest of this section write
\be
 G := (U_{\omega}-z)^{-1}
 \ee
and
\be
 G^{(L)} := (U_{\omega}^{\Lambda_L} \oplus U_{\omega}^{\Lambda_L^c}-z)^{-1} = (U_{\omega}^{\Lambda_L}-z)^{-1} \oplus (U_{\omega}^{\Lambda_L^c}-z)^{-1}.
 \ee

We do a double-decoupling, once on $\Lambda_L$ and once on $\Lambda_{L+1}$. Using the resolvent identity twice gives
\bea \label{eq:doubleresid}
G & = & G^{(L)} - G^{(L)} T^{(L)} G \nonumber \\
& = & G^{(L)} - G^{(L)} T^{(L)} G^{(L+1)} + G^{(L)} T^{(L)} G T^{(L+1)} G^{(L+1)}.
\eea

Observe that for $y\in \Z^d$ with $|y|\ge L+2$ one has $\chi_0 G^{(L)} \chi_y =0$ and $\chi_0 G^{(L)} T^{(L)} G^{(L+1)} \chi_y =0$. Thus
\be \label{eq:resolventexp}
\chi_0 G \chi_y = \chi_0 G^{(L)} T^{(L)} G T^{(L+1)} G^{(L+1)} \chi_y,
\ee
which is the geometric resolvent identity to be used below.

\subsection{Decoupling of Fractional Moments}

The next result says that the fractional moment $\E(\|\chi_0 G \chi_y\|^s)$ can be decoupled along the boundary of $\Lambda_L$.

\begin{prop} \label{prop:decouple}
For every $s\in (0,1/3)$ there exists a constant $C = C(s) <\infty$ such that
\be \label{eq:intext}
\E(\|\chi_0 G \chi_y\|^s) \le C \sum_{u:\,|u|=L} \E(\|\chi_0 G^{(L)} \chi_u\|^s) \sum_{v':\,|v'|=L+2} \E(\|\chi_{v'} G^{(L+1)} \chi_y\|^s)
\ee
uniformly in $z$ with $1<|z|< 2$, $L\in \N$ and $y\in \Z^d$ with $|y|\ge L+2$.
\end{prop}

{\bf Proof of Proposition~\ref{prop:decouple}:}
From here on, the symbol $C$ will denote a generic constant which may change from line to line but which depends on inessential quantities only.

Define the boundary of $\Lambda_L$ by
\bea
\partial \Lambda_L & := & \{ (x,y) \in \Z^d \times \Z^d : \chi_x T^{(L)} \chi_y \not= 0\} \nonumber \\
& \subset & \{ (x,y) \in \Z^d \times \Z^d : L \le |x| \le L+1, L \le |y| \le L+1, |x-y| \le 1\}.
\eea
Expanding (\ref{eq:resolventexp}) over the boundaries of $\Lambda_L$ and $\Lambda_{L+1}$ gives
\be
\chi_0 G \chi_y = \sum_{\begin{array}{cc} (u,u') \in \partial \Lambda_L \\ (v,v') \in \partial \Lambda_{L+1} \end{array}} \chi_0 G^{(L)} \chi_u T^{(L)} \chi_{u'} G \chi_v T^{(L+1)} \chi_{v'} G^{(L+1)}\chi_y.
\ee
Taking fractional moments and also using that $T^{(L)}$ and $T^{(L+1)}$ have uniformly bounded matrix-elements we get
\be \label{eq:resolventexp2}
\E(\|\chi_0 G \chi_y\|^s) \le C \sum_{\begin{array}{cc} (u,u') \in \partial \Lambda_L \\ (v,v') \in \partial \Lambda_{L+1} \end{array}} \E \left( \|\chi_0 G^{(L)} \chi_u\|^s \|\chi_{u'} G\chi_v\|^s \|\chi_{v'} G^{(L+1)} \chi_y\|^s \right).
\ee
Notice that the first and third factor in the expectation on the right are independent. Unfortunately, they are correlated via the middle factor. In order to decouple the factors we use a {\em re-sampling argument}, following a strategy developed in \cite{AENSS} and \cite{BNSS} as a tool in the fractional moments approach to continuum Anderson-type models. For this, fix two pairs $(u,u')\in \partial \Lambda_L$ and $(v,v') \in \partial \Lambda_{L+1}$. Let $\mathcal{J} := C_u \cup C_{u'} \cup C_v \cup C_{v'}$. In the resolvents $G^{(L)}$ and $G^{(L+1)}$ we will re-sample the random variables $\theta_n$, $n\in \mathcal{J}$. For this choose i.i.d.\ random variables $\{\hat{\theta}_n\}_{n\in \mathcal{J}}$ with the same distribution as the $\theta_n$ but independent from them.

Noting that $D_{\omega} = \sum_{n\in\Z^d} e^{-i\theta_n} P_n$ (where $P_n = \langle e_n,\cdot \rangle e_n$ is the projection onto the canonical basis vector $e_n$) we define the re-sampled $D_{\omega,\hat{\omega}} := D_{\omega} - \hat{D}$, where
\be
\hat{D} := \sum_{n\in \mathcal{J}} (e^{-i\theta_n}-e^{-i\hat{\theta_n}})P_n,
\ee
i.e.\ the variables $\{\theta_n\}_{n\in \mathcal{J}}$ are replaced by the corresponding $\hat{\theta}_n$. Also define
\be
U_{\omega,\hat{\omega}}^{(L)} := D_{\omega,\hat{\omega}} S_N^{\Lambda_L} = U_{\omega}^{\Lambda_L} - \hat{D} S_N^{\Lambda_L}
\ee
where $U_{\omega}^{(L)}=U_\omega^{\Lambda_L}$, $U_0^{(L)}=S_N^{\Lambda_L}$
and
\be
\hat{G}^{(L)} := (U_{\omega,\hat{\omega}}^{(L)}-z)^{-1}.
\ee
The resolvent identity yields
\be
G^{(L)} = \hat{G}^{(L)} - \hat{G}^{(L)} \hat{D} S_N^{\Lambda_L} G^{(L)}
\ee
and
\be
G^{(L+1)} = \hat{G}^{(L+1)} - G^{(L+1)} \hat{D} S_N^{\Lambda_{L+1}} \hat{G}^{(L+1)}.
\ee
We use this to bound the terms on the right of (\ref{eq:resolventexp2}) by
\bea
\lefteqn{\E \left( \|\chi_0 G^{(L)} \chi_u\|^s \|\chi_{u'} G\chi_v\|^s \|\chi_{v'} G^{(L+1)} \chi_y\|^s \right)} \nonumber \\
& \le & \hat{\E} \E \left[ (\|\chi_0 \hat{G}^{(L)} \chi_u\|^s + \| \chi_0 \hat{G}^{(L)} \hat{D} S_N^{\Lambda_L} G^{(L)} \chi_u\|^s) \|\chi_{u'} G \chi_v\|^s \right. \label{eq:noidea} \nonumber \\
& & \left. ( \|\chi_{v'} \hat{G}^{(L+1)}\chi_y\|^s + \|\chi_{v'} G^{(L+1)} \hat{D} S_N^{\Lambda_{L+1}} \hat{G}^{(L+1)} \chi_y\|^s) \right] \nonumber \\
& =: & A_1+A_2+A_3+A_4.
\eea
Here we have argued that the above bound holds for arbitrary fixed values of the $\hat{\theta}_n$. Thus it also holds after the average over these variables, denoted by $\hat{\E}$, is taken. Of the four terms $A_1, \ldots, A_4$, found by expanding the two sums in (\ref{eq:noidea}), we will now find bounds for $A_1$, the one most easily handled, and $A_4$, the most complicated one. Corresponding bounds for the two mixed terms $A_2$ and $A_3$ can then be found by ``interpolating'' the provided arguments.

Let $\E(\ldots|\mathcal{J})$ denote the conditional expectation with respect to the $\sigma$-field generated by the family $\{\theta_k\}_{k\not\in \mathcal{J}}$. Due to independence this means that
\be
\E(X|\mathcal{J}) = \int \ldots \int X \prod_{n\in \mathcal{J}} \tau(\theta_n) d\theta_n.
\ee
The re-sampled resolvents $\hat{G}^{(L)}$ and $\hat{G}^{(L+1)}$ are independent of the variables $(\theta_n)_{n\in\mathcal{J}}$. Thus, by the rule
\be \label{eq:condexp}
\E(X)=\E(\E(X|\mathcal{J}))
\ee
for conditional expectations,
\bea \label{eq:A1}
A_1 & := & \hat{\E} \E \left[ \|\chi_0 \hat{G}^{(L)} \chi_u\|^s \|\chi_{u'} G\chi_v\|^s \|\chi_{v'} \hat{G}^{(L+1)} \chi_y\|^s \right] \nonumber \\
& = & \hat{\E} \E \left[ \|\chi_0 \hat{G}^{(L)} \chi_u\|^s \E(\|\chi_{u'} G\chi_v\|^s| \mathcal{J}) \|\chi_{v'} \hat{G}^{(L+1)} \chi_y\|^s \right] \nonumber \\
& \le & C \E(\|\chi_0 G^{(L)}\chi_u\|^s) \E(\|\chi_{v'} G^{(L+1)} \chi_y\|^s).
\eea
In the last estimate we have used the bound $\E(\|\chi_{u'} G\chi_v\|^s| \mathcal{J})\le C$, e.g.\ Theorem \ref{thm:fmbound}, that the distribution of $(\hat{\theta}_n)_{n\in\mathcal{J}}$ is identical to the distribution of $(\theta_n)_{n\in\mathcal{J}}$, and that $\chi_0 G^{(L)}\chi_u$ and $\chi_{v'}G^{(L+1)}\chi_y$ are stochastically independent. This bound for $A_1$ is of the form required in Proposition~\ref{prop:decouple}.

We continue with $A_4$, where we use (\ref{eq:condexp}) again and then apply H\"older's inequality to the conditional expectation:
\bea \label{eq:A4}
A_4 & := & \hat{\E} \E \left[ \|\chi_0 \hat{G}^{(L)} \hat{D} S_N^{\Lambda_L} G^{(L)} \chi_u\|^s \|\chi_{u'} G\chi_v\|^s \|\chi_{v'} G^{(L+1)} \hat{D} S_N^{\Lambda_{L+1}} \hat{G}^{(L+1)} \chi_y \|^s \right]  \nonumber \\
& = & \hat{\E} \E \left[ \E (\|\ldots\|^s \|\ldots\|^s \|\ldots\|^s |\mathcal{J})\right] \nonumber \\
& \le & \hat{\E} \E \left[ \E(\|\chi_0 \hat{G}^{(L)} \hat{D} S_N^{\Lambda_L} G^{(L)} \chi_u \|^{3s}| \mathcal{J})^{1/3} \right. \nonumber \\
& & \mbox{} \times \E(\|\chi_{u'} G\chi_v\|^{3s} |\mathcal{J})^{1/3} \nonumber \\
& & \left. \mbox{} \times \E( \|\chi_{v'} G^{(L+1)} \hat{D} S_N^{\Lambda_{L+1}} \hat{G}^{(L+1)} \chi_y \|^{3s}| \mathcal{J})^{1/3} \right].
\eea
We will now bound the three conditional expectations on the right separately. As $s<1/3$, we can use Theorem \ref{thm:fmbound} to bound the second factor by
\be \label{eq:secondfact}
\E(\|\chi_{u'} G\chi_v\|^{3s} |\mathcal{J})^{1/3} \le C < \infty
\ee
uniformly in $(\theta_k)_{k\not\in\mathcal{J}}$ and $z$.

To bound the first factor, we start from the definition of $\hat{D}$ to get
\be
\|\chi_0 \hat{G}^{(L)} \hat{D} S_N^{\Lambda_L} G^{(L)} \chi_u\|^{3s} \le C \sum_{\ell\in\mathcal{J}\cap \Lambda_L} \|\chi_0 \hat{G}^{(L)} P_{\ell}\|^{3s} \|P_{\ell} S_N^{\Lambda_L} G^{(L)} \chi_u\|^{3s},
\ee
note here that $\chi_0 \hat{G}^{(L)} P_{\ell}=0$ for $\ell \not\in \Lambda_L$.
The unitary diagonal operator $D_{\omega}$ commutes with $P_{\ell}$ and thus
\bea \label{eq:stuff}
\|P_{\ell} S_N^{\Lambda_L} G^{(L)} \chi_u \| & = & \|P_{\ell} S_N^{\Lambda_L} (D_{\omega}S_N^{\Lambda_L}-z)^{-1} \chi_u\| \nonumber \\
& = & \|P_{\ell} D_{\omega} S_N^{\Lambda_L} (D_{\omega}S_N^{\Lambda_L}-z)^{-1} \chi_u \| \nonumber \\
& = & \|P_{\ell} \chi_u + zP_{\ell} G^{(L)}\chi_u \| \nonumber \\
& \le & 1 + |z|\|P_{\ell}G^{(L)} \chi_u\|.
\eea
The re-sampled resolvent $\hat{G}^{(L)}$ does not depend on the variables $(\theta_n)_{n\in\mathcal{J}}$. Thus, using Theorem \ref{thm:fmbound} again, we get
\bea \label{eq:firstfact}
\lefteqn{\E(\|\chi_0 \hat{G}^{(L)} \hat{D} S_N^{\Lambda_L} G^{(L)} \chi_u\|^{3s}| \mathcal{J})} \nonumber \\
& \le & C\sum_{\ell\in \mathcal{J}\cap \Lambda_L} \|\chi_0 \hat{G}^{(L)} P_{\ell}\|^{3s} \E((1+|z|\|P_{\ell} G^{(L)} \chi_u\|)^{3s}| \mathcal{J}) \nonumber \\
& \le & C \sum_{\ell\in\mathcal{J}\cap \Lambda_L} \|\chi_0 \hat{G}^{(L)} P_{\ell}\|^{3s}.
\eea
In a similar way we find
\be
\E(\|\chi_{v'} G^{(L+1)} \hat{D} S_N^{\Lambda_{L+1}} \hat{G}^{(L+1)} \chi_y\|^{3s} |\mathcal{J})  \le C \sum_{\ell\in\mathcal{J}\cap \Lambda_{L+1}^c} \|P_{\ell} S_N^{\Lambda_{L+1}} \hat{G}^{(L+1)} \chi_y\|^{3s}.
\ee
By a calculation as in (\ref{eq:stuff}) we have $\|P_{\ell} S_N^{\Lambda_{L+1}} \hat{G}^{(L+1)} \chi_y\| = \|P_{\ell} \chi_y +z P_{\ell}\hat{G}^{(L+1)}\chi_y\| = |z|\|P_{\ell}\hat{G}^{(L+1)}\chi_y\|$. We conclude
\be \label{eq:thirdfact}
\E(\|\chi_{v'} G^{(L+1)} \hat{D} S_N^{\Lambda_{L+1}} \hat{G}^{(L+1)} \chi_y\|^{3s} |\mathcal{J}) \le C\sum_{\ell\in\mathcal{J} \cap \Lambda_{L+1}^c} \|P_{\ell} \hat{G}^{(L+1)}\chi_y\|^{3s}.
\ee
Combining the bounds (\ref{eq:secondfact}), (\ref{eq:firstfact}) and (\ref{eq:thirdfact}) into (\ref{eq:A4}) we arrive at
\bea \label{eq:A4ult}
A_4 & \le & C \sum_{\ell\in\mathcal{J} \cap \Lambda_L, \;\ell'\in\mathcal{J} \cap \Lambda_{L+1}^c} \hat{\E} \E (\|\chi_0 \hat{G}^{(L)} P_{\ell}\|^s \|P_{\ell'} \hat{G}^{(L+1)} \chi_y\|^s) \nonumber \\
& = &  C \sum_{\ell\in\mathcal{J} \cap \Lambda_L} \E(\|\chi_0 G^{(L)}P_{\ell}\|^s) \sum_{\ell'\in\mathcal{J} \cap \Lambda_{L+1}^c} \E(\|P_{\ell'} G^{(L+1)} \chi_y\|^s).
\eea
Here it was also used that $\mathcal{J}$ has a fixed finite number of elements and thus $(\sum_{j\in\mathcal{J}} x_j^{3s})^{1/3} \le C\sum_{j\in\mathcal{J}} x_j^s$. The last identity uses that $(\hat{\theta}_n)$ and $(\theta_n)$ are identically distributed and that $\chi_0 G^{(L)}P_{\ell}$ and $P_{\ell'}G^{(L+1)}\chi_y$ are stochastically independent.

The bounds (\ref{eq:A1}), (\ref{eq:A4ult}) and related bounds for the mixed terms $A_2$ and $A_3$ combine via (\ref{eq:noidea}) and (\ref{eq:resolventexp2}) to prove Proposition~\ref{prop:decouple}.
\ep

\vspace{.3cm}

\subsection{The Start of an Iteration}

We plan to use (\ref{eq:intext}) as the first step in an iterative argument, where the next step consists of applying (\ref{eq:intext}) again, but this time with $\E(\|\chi_{v'} G^{(L+1)} \chi_y\|^s)$ on the left hand side with $v'$ as the new origin. However, before doing this we need to replace $G^{(L+1)}$ with the original $G$, which can be done by reasoning similar to the decoupling argument of the previous section.

\begin{prop} \label{prop:iterationstep}
For every $s\in (0,1/3)$ there exists a constant $C = C(s) <\infty$ such that
\be \label{eq:itstep}
\E(\|\chi_0 G \chi_y\|^s) \le C L^{d-1} \sum_{|u|=L} \E(\|\chi_0 G^{(L)} \chi_u\|^s) \sum_{|x'|\in \{L+1,L+2\}} \E(\|\chi_{x'} G \chi_y\|^s)
\ee
uniformly in $z$ with $1< |z| < 2$, $L\in \N$ and $y\in \Z^d$ with $|y|\ge L+2$.
\end{prop}

{\bf Proof:}
According to Theorem~\ref{prop:decouple} we need a bound for $\E(\|\chi_{v'} G^{(L+1)} \chi_y\|^s)$ in terms of fractional moments of the full Green function $G$ for each fixed $v'$ with $\|v'\|_{\infty}=L+2$. We start from the resolvent identity
\be
G^{(L+1)} = G + G^{(L+1)} T^{(L+1)} G
\ee
and expand
\be
T^{(L+1)} = \sum_{(w,w')\in \partial \Lambda_{L+1}} \chi_w T^{(L+1)} \chi_{w'}.
\ee
Combining both yields
\bea \label{eq:onemore}
\E(\|\chi_{v'} G^{(L+1)} \chi_y\|^s) & \le & \E(\|\chi_{v'} G\chi_y\|^s) \nonumber \\
& & \mbox{} + C \sum_{(w,w')\in \partial \Lambda_{L+1}} \E(\|\chi_{v'} G^{(L+1)} \chi_w\|^s \|\chi_{w'} G\chi_y\|^s).
\eea
With the goal of factorizing the expectation on the right we fix $(w,w') \in \partial\Lambda_{L+1}$ and re-sample over the variables $\theta_n$ for $n\in \tilde{\mathcal{J}} := C_{v'} \cup C_w \cup C_{w'}$. With independent random variables $(\tilde{\theta}_n)_{n\in\tilde{\mathcal{J}}}$ independent from the $\theta_n$, but with identical distribution, define
\bea
\tilde{D} & := & \sum_{n\in \tilde{\mathcal{J}}} (e^{-i\theta_n} -e^{-i\tilde{\theta}_n})P_n, \\
D_{\omega,\tilde{\omega}} & := & D_{\omega} - \tilde{D}, \\
U_{\omega,\tilde{\omega}} & := & D_{\omega,\tilde{\omega}} S = U_{\omega} - \tilde{D} S, \\
\tilde{G} & := & (U_{\omega,\tilde{\omega}}-z)^{-1}.
\eea
The resolvent identity
\be
G = \tilde{G} - G\tilde{D}S \tilde{G}
\ee
yields
\bea \label{eq:anotherone}
\lefteqn{\E(\|\chi_{v'} G^{(L+1)} \chi_w\|^s \|\chi_{w'} G \chi_y\|^s)} \nonumber \\
& \le & \tilde{\E} \E ( \|\chi_{v'} G^{(L+1)} \chi_w\|^s \|\chi_{w'}\tilde{G}\chi_y\|^s) \nonumber \\
& & \mbox{} + \tilde{\E} \E (\|\chi_{v'} G^{(L+1)}\chi_w\|^s \|\chi_{w'} G\tilde{D} S \tilde{G}\chi_y\|^s) \nonumber \\
& =: & B_1 + B_2,
\eea
where $\tilde{\E}$ denotes averaging over the variables $\tilde{\theta}_n$. Also writing $\E(\ldots|\tilde{\mathcal{J}})$ for the conditional expectation with respect to the $\sigma$-field generated by the variables $(\theta_n)_{n\not\in\tilde{\mathcal{J}}}$ and arguing as in the previous section, one has
\bea \label{eq:B1bound}
B_1 & = & \tilde{\E} \E \left( \E(\|\chi_{v'} G^{(L+1)}\chi_w\|^s |\tilde{\mathcal{J}}) \|\chi_{w'}\tilde{G}\chi_y\|^s \right) \nonumber \\
& \le & C\E(\|\chi_{w'} G \chi_y\|^s)
\eea
H\"older's inequality yields
\bea
B_2 & \le & \tilde{\E} \E \left( \E(\|\chi_{v'} G^{(L+1)} \chi_w \|^{2s}| \tilde{\mathcal{J}} )^{1/2} \right. \nonumber \\
& & \mbox{} \times \left. \E(\|\chi_{w'} G\tilde{D} S \tilde{G} \chi_y\|^{2s}| \tilde{\mathcal{J}})^{1/2} \right).
\eea
We have $\E(\|\chi_{v'}G^{(L+1)}\chi_w\|^{2s}| \tilde{\mathcal{J}}) \le C$ and, by an argument as above,
\be
\E(\|\chi_{w'} G\tilde{D} S \tilde{G} \chi_y\|^{2s}| \tilde{\mathcal{J}}) \le C \sum_{\ell\in\tilde{\mathcal{J}}} \|P_{\ell} \tilde{G} \chi_y\|^{2s}.
\ee
This leads to the bound
\bea \label{eq:B2bound}
B_2 & \le & C \tilde{\E} \E \left( \sum_{\ell\in \tilde{\mathcal{J}}} \|P_{\ell} \tilde{G}\chi_y\|^s \right) \nonumber \\
& \le & C \sum_{x':\,C_{x'}\subset \tilde{\mathcal{J}}} \E( \|\chi_{x'} G \chi_y\|^s).
\eea
Collecting (\ref{eq:anotherone}), (\ref{eq:B1bound}) and (\ref{eq:B2bound}) into (\ref{eq:onemore}), using that $\partial \Lambda_{L+1}$ has $CL^{d-1}$ elements, and ultimately applying Proposition~\ref{prop:decouple} completes the proof of Proposition~\ref{prop:iterationstep}.
\ep

\setcounter{equation}{0}
\section{Proof of Band Edge Localization} \label{sec:proofofbandedgelocalization}

We finally have reached the point where all the main results of the previous three sections can be put together to prove Theorem~\ref{thm:bandedgelocalization}. Specifically, we will use the Combes-Thomas-type bound of Proposition~\ref{prop:CT}, the Lifshits tail estimate of Proposition~\ref{prop:lbeta} and the decoupling estimate in the form provided in Proposition~\ref{prop:iterationstep}. Also frequently enter will the a priori boundedness of fractional moments established in Theorem~\ref{thm:fmbound}.

\medskip

We will show the following fact which is equivalent to Theorem~\ref{thm:bandedgelocalization} (now in the normalization introduced at the beginning of Section~\ref{sec:lifshits} and only stated for the upper band edge): For $0<s< 1/3$ there exist $\delta>0$, $\alpha>0$ and $C<\infty$ such that
\be \label{eq:8.1}
\E (\|\chi_0 G(z) \chi_y\|^s) \le C e^{-\alpha |y|}
\ee
for all $y \in \Z^d$ and all $z\in \C$ with $1/2 < |z| < 2$, $|z|\not=1$ and arg$\,z \in [d\lambda_0-\delta, d\lambda_0]$. Here $\chi_0$ and $\chi_y$ are the characteristic functions of the cubes $C_0$ and $C_y$ introduced in Section~\ref{sec:gri}. Making the $z$-dependence explicit we write $G(z)=(U_{\omega}-z)^{-1}$ and $G^{(L)}(z) = (U_{\omega}^{\Lambda_L}-z)^{-1}$ in this section. The bound (\ref{eq:8.1}) implies $\E (\|\chi_x G(z) \chi_y\|^s) \le C e^{-\mu \|x-y\|_{\infty}}$ for arbitrary $x,y\in \Z^d$ due to ergodicity.

It suffices to prove (\ref{eq:8.1}) only for those $z$ which in addition satisfy $|z|>1$.
To see this, use the identity
\be \label{eq:8.2}
G^*(z)=-U_\omega(U_\omega -\bar z^{-1})/\bar z=-U_\omega G(1/\bar z)/\bar z,
\ee
which implies
\be \label{eq:8.2b}
\|\chi_xG(z)\chi_y\|=\|\chi_yG^*(z)\chi_x\|=\|\chi_y UG(1/\bar z)\chi_x\|/|z|.
\ee
Inserting the partition $\sum_{y'} \chi_{y'}$ and using that $\chi_y U \chi_{y'} =0$ for $|y'-y|>1$ we conclude
\be \label{eq:8.2c}
\|\chi_xG(z)\chi_y\| \le \frac{1}{|z|} \sum_{|y'-y|\le 1} \|\chi_{y'} G(1/\bar z) \chi_x\|.
\ee
This shows that (\ref{eq:8.1}) carries over to $z$ with $1/2<|z|<1$ once it has been proven for $1<|z|<2$, which is assumed for the remainder of this section.

Proposition~\ref{prop:lbeta} shows that the probability that $U^{\Lambda_{L_k}}$ has an eigenvalue close to $e^{id\lambda_0}$ is small for the sequence $L_k$ found there. Combined with the Combes-Thomas bound Proposition~\ref{prop:CT} this can be used to show smallness of the fractional moments $\E(\|\chi_0 G^{(L_k)}(z)\chi_u\|^s)$ on the right hand side of (\ref{eq:intext}) for values of $z$ close to $e^{id\lambda_0}$.

\begin{prop} \label{prop:smallfactor}
For any $s\in (0,1/3)$ there exist a sequence of integers $L_k$ with $L_k\to\infty$, $g>0$ and $C<\infty$ such that
\be \label{eq:smallfactor}
\E(\|\chi_0 G^{(L_k)}(z) \chi_u\|^s) \le Ce^{-gL_k^{d/(d+2)}}
\ee
for all $k$ sufficiently large, any $z\in \C$ such that $1<|z|<2$ and arg$\,z \in [d\lambda_0-L_k^{-2/(2+d)}, d\lambda_0]$ and any $u\in \Z^d$ with $|u|=L_k$.
\end{prop}

{\bf Proof:}\\
Let $\delta_L>0$, to be specified later. The Combes-Thomas estimate Proposition~\ref{prop:CT} states that there exists $B>0$ independent of $L$ such that
\be \label{eq:8.3}
\|\chi_0 G^{(L)}(z) \chi_u\| \le \frac{2}{\delta_L} e^{-BL\delta_L}
\ee
for all $z$ with dist$(z,\sigma(U_{\omega}^{\Lambda_L}))>\delta_L$ and all $u\in \Z^d$ with $|u|=L$.

This takes care of the realizations $\omega$ such that the values of $z$ are far enough from
$\sigma(U_{\omega}^{\Lambda_L})$. The Lifshits tail estimate takes care of the realizations where
this is not the case, in the sense that such instances are  very unlikely.

We set
\be
\Omega_G=\{\omega \ |\ \mbox{dist }(z,\sigma(U_{\omega}^{\Lambda_L}))>\delta_L\} \ \ \mbox{and}
\ \ \ \Omega_B=\Omega_G^C=\{\omega \ |\ \mbox{dist }(z,\sigma(U_{\omega}^{\Lambda_L}))\leq\delta_L\}.
\ee

Making use of (\ref{eq:8.3}), we can write by means of H\"older's inequality
\bea
\E (\|\chi_0G^{(L)}(z)\chi_u\|^s)&=&\E (\|\chi_0G^{(L)}(z)\chi_u\|^s 1_{\{\omega\in\Omega_G\}})+\E (\|\chi_0G^{(L)}(z)\chi_u\|^s1_{\{\omega\in\Omega_B\}}) \nonumber \\
&\leq &\frac{2^s}{\delta_L^s}e^{-sB L \delta_L}\E( 1_{\{\omega\in\Omega_G\}})) \nonumber \\ & & +
(\E (\|\chi_0G^{(L)}(z)\chi_u\|^{st})^{1/t}(\E(1_{\{\omega\in\Omega_B\}}))^{1/t'},
\eea
with $1<t<1/s$ and $1/t+1/t'=1$. Since $\E(1_{\{\omega\in\Omega\}})=\P(\Omega)$ and
$\E (\|\chi_0G^{(L)}(z)\chi_u\|^{st})\leq C$ for $st<1$, by Theorem \ref{thm:fmbound}, we get
\be
\E (\|\chi_0G^{(L)}(z)\chi_u\|^s)\leq C\frac{e^{-sB L \delta_L}}{\delta_L^s}+
C(\P(\mbox{dist }(z,\sigma(U^{\Lambda_L}))\leq\delta_L))^{1/t'}.
\ee
To be useful for our purpose, this last quantity need to decay as $L\ra\infty$, which requires
$L\delta_L\ra\infty$. On the other hand, we need $\delta_L\ra 0$ for the probability that $z$ is a distance $\delta_L$ only away from $\sigma(U^{\Lambda_L})$ to be very small, for suitable $z$ in a neighborhood of the band edge $e^{id\lambda_0}$. In particular, this holds for the choice $\delta_L = 1/L^{\beta}$ and any $\beta \in (0,1)$.

More specifically, with this choice of $\delta_L$ Proposition~\ref{prop:lbeta} yields the existence of a sequence $L_k$ with $L_k\to\infty$ and positive $\bar{\gamma}$ and $C$ such that
\be \label{eq:8.5}
\E (\|\chi_0G^{(L_k)}(z)\chi_u\|^s)\leq Ce^{-sB L_k^{1-\beta}}L_k^{\beta s}+CL_k^{d(1-\beta/2)/t'}e^{-\frac{\bar\gamma}{t'} L_k^{d\beta/2}}
\ee
for all $k$, $1<|z|<2$, $d\lambda_0 -1/L_k^{\beta} \le \mbox{arg}\,z \le d\lambda_0$ and $|u|=L_k$. The choice $\beta = 2/(2+d)$ leads to equal exponents of $L_k$ in the two exponentials on the right hand side of (\ref{eq:8.5}). Choosing $g< \min(sB, \bar{\gamma}/t')$ and requiring $k$ to be sufficiently large we can absorb the power terms in (\ref{eq:8.5}) into the exponentials and arrive at (\ref{eq:smallfactor})
\ep

\bigskip

We proceed with the proof of (\ref{eq:8.1}) by fixing $s\in (0,1/3)$ and choosing the sequence $L_k$ and $g$ as in Proposition~\ref{prop:smallfactor}. We now also choose $\delta_k = L_k^{-2/(2+d)}$.

Proposition~\ref{prop:iterationstep} says that
\be
\E(\|\chi_0 G(z) \chi_y\|^s) \le C L_k^{d-1} \sum_{|u|=L_k} \E(\|\chi_0 G^{(L_k)}(z) \chi_u\|^s) \sum_{|x'|\in \{L_k+1,L_k+2\}} \E(\|\chi_{x'} G(z) \chi_y\|^s)
\ee
if $|y| \ge L_k+2$. Let $1<|z|<2$ with arg$\,z\in [d\lambda_0-\delta_k,d\lambda_0]$. This along with  Proposition~\ref{prop:smallfactor} imply, for $k$ sufficiently large,
\be \label{eq:itstep2}
\E(\|\chi_0 G(z) \chi_y\|^s) \le CL_k^{2(d-1)} e^{-gL_k^{2/(2+d)}} \sum_{|x'|\in \{L_k+1,L_k+2\}} \E(\|\chi_{x'} G(z) \chi_y\|^s).
\ee
With the constant $C$ from (\ref{eq:itstep2}), fix $L=L_k$ for $k$ sufficiently large such that
\be \label{eq:bchoice}
b:= CL^{2(d-1)} e^{-gL^{2/(2+d)}} \# \{x'\in \Z^d: L+1 \le |x'| \le L+2\} < 1
\ee
and get from (\ref{eq:itstep2}) that
\be \label{eq:itstep3}
\E(\|\chi_0 G(z) \chi_y\|^s) \le b \max_{\|x'\|_{\infty} \in \{L+1,L+2\}} \E(\|\chi_{x'} G(z) \chi_y\|^s).
\ee
Note that $\E(\|\chi_{x'} G(z) \chi_y\|^s) = \E(\|\chi_0 G(z) \chi_{y-x'}\|^s)$, which allows to iterate (\ref{eq:itstep3}). If $x', x^{(2)}, x^{(3)}, \ldots$ is one of the chains of sites obtained in this way, then the iteration may be continued as long as $|x^{(j)}-y| \ge L+2$, i.e.\ at least $\frac{|y|}{L+2}-1$ times. For the last entry in the chain we use Theorem \ref{thm:fmbound}  to bound $\E(\|\chi_{x^{(j)}} G(z) \chi_y\|^s)$ by $\tilde{C}$. In (\ref{eq:itstep3}) this leads to the bound
\be \label{eq:grandfinale}
\E(\|\chi_0 G(z) \chi_y\|^s) \le \tilde{C} b^{\frac{|y|}{L+2}-1} = \frac{\tilde{C}}{b} e^{\frac{\log b}{L+2}|y|}.
\ee
Thus we have proven (\ref{eq:8.1}) with $C= \frac{\tilde{C}}{b}$ and $\alpha = \frac{|\log b|}{L+2}$.


\end{document}